\documentclass[%
  superscriptaddress,
  twocolumn,
  floatfix,
  10pt,
  aps,
  pre,
  nofootinbib,
  final,letterpaper
]{revtex4-2}

\usepackage{graphicx}
\usepackage{hyperref}
\usepackage[utf8]{inputenc}

\usepackage{dcolumn}
\usepackage{amsmath,amsfonts,amssymb}
\usepackage{physics}

\usepackage{color}

\begin{document}

\title{Directed Percolation in Random Temporal Network Models with Heterogeneities}

\author{Arash Badie-Modiri}
\affiliation{Department of Computer Science, School of Science, Aalto University, FI-0007, Finland}
\author{Abbas K.~Rizi}
\affiliation{Department of Computer Science, School of Science, Aalto University, FI-0007, Finland} 
\author{Márton Karsai}
\affiliation{Department of Network and Data Science
Central European University, 1100 Vienna, Austria}
\affiliation{Alfr\'ed R\'enyi Institute of Mathematics, 1053 Budapest, Hungary}
\author{Mikko Kivelä}
\affiliation{Department of Computer Science, School of Science, Aalto University, FI-0007, Finland} 

\date{\today}

\begin{abstract}
The event graph representation of temporal networks suggests that the connectivity of temporal structures can be mapped to a directed percolation problem. However, similar to percolation theory on static networks, this mapping is valid under the approximation that the structure and interaction dynamics of the temporal network are determined by its local properties, and otherwise, it is maximally random. We challenge these conditions and demonstrate the robustness of this mapping in case of more complicated systems. We systematically analyze random and regular network topologies and heterogeneous link-activation processes driven by bursty renewal or self-exciting processes using numerical simulation and finite-size scaling methods. We find that the critical percolation exponents characterizing the temporal network are not sensitive to many structural and dynamical network heterogeneities, while they recover known scaling exponents characterizing directed percolation on low dimensional lattices. While it is not possible to demonstrate the validity of this mapping for all temporal network models, our results establish the first batch of evidence supporting the robustness of the scaling relationships in the limited-time reachability of temporal networks.
\end{abstract}

\keywords{directed percolation, temporal networks, spreading processes, weighted event graph} 
\maketitle


\section{Introduction}\label{sec:introduction}
Connectivity is an essential characteristic of complex networks as it determines how far information or influence can spread in a network structure. Consequently, it governs the emergence and scale of any macroscopic phenomena often modelled on networks such as disease spreading, transportation, or information diffusion, to mention a few examples. Percolation theory provides a comprehensive understanding that characterizes network connectivity with various mathematical and algorithmic tools primarily developed for complex networks. For example, percolation can be mapped to late-stage results of specific epidemic processes \cite{newman2002spread, kenah2007second, kenah2011epidemic, pastor2015epidemic, rizi2021epidemic}, such that the size of percolating components determine the final size of the epidemic. Meanwhile, the percolation transition and its related critical behavior explain the disease outcome close to the epidemic threshold. 

However, these theoretical descriptions commonly assume that the network is static, with links and nodes always present, ignoring the typical character of several complex structures where links may vary in time. Since information, disease or other effects can pass between two nodes in a network only at the time of their interactions, the temporal alternation of links may crucially influence the critical behavior and final outcome of any ongoing spreading processes \cite{liu2013contagion, karimi2013threshold, backlund2014effects, delvenne2015diffusion, bazzi2016community, aslak2018constrained, koher2019spreading}. To characterize these processes, one needs to measure connectivity in temporal networks across time, where components are defined in terms of network nodes and links and the temporal distribution of interactions. Consequently, beyond the well-studied structural heterogeneities of static networks, like in their node degrees, the effects of temporal correlations leading to temporal heterogeneities in the interaction dynamics, like burstiness, become important \cite{karsai2011small, rocha2011simulated, horvath2014spreading, ubaldi2017burstiness, hiraoka2018correlated, karsai2018bursty}. This is especially the case for so-called limited-waiting-time processes, where an effect or information, e.g.~a disease or a meme \cite{kuhn2014inheritance}, arriving at a node can pass over to another node only if an interaction appears within a time window $\delta t$. Otherwise, the pathogen times out, e.g., the patient recovers or the meme becomes irrelevant, making it impossible to reach other nodes.

Similar to static networks, the connectivity of temporal networks passes through a phase-transition. However, close to this critical threshold, temporal networks exhibit different critical behavior as compared to static structures~\cite{kivela2018mapping,parshani2010dynamic,shortpaper}.
For limited-waiting-time connectivity, where the control parameter is $\delta t$, this phase-transition can be theoretically understood under some simplifying assumptions about the homogeneous dynamics of connectivity \cite{shortpaper}. Since there is an embedded direction (or flow) of time, the microscopic dynamics can be fundamentally irreversible with a broken detailed balance and non-equilibrium steady-state. These results suggest that the dynamics of percolation on temporal networks are generically the same as any other system belonging to the Directed Percolation (DP) universality class, which is characterized by a one-component order parameter without additional symmetries and unconventional features such as quenched disorder \cite{henkel2008non}.

The homogeneity approximations used for the derivations presented in \cite{shortpaper}, however, become less grounded when the underlying structure deviates from a random graph or if the interaction dynamics become inhomogeneous. In this paper, our goal is to build on the theory laid down in \cite{shortpaper} to investigate further the relation between temporal networks and directed percolation. In other words, the primary objective of this manuscript is as follows: to show empirically that diverse classes of temporal networks, with various degrees of temporal and spatial heterogeneity, combined with the very general notion of limited-waiting-time reachability, will show an absorbing phase transition in connectivity that belongs to the directed percolation universality class.

In its epidemic interpretation, directed percolation can be one of the most basic non-equilibrium second-order phase transitions from fluctuating states into so-called absorbing states, which exhibit universal features, determined by symmetry properties and conservation laws. We demonstrate the precision of this mapping using extensive numerical simulations and provide further theoretical calculations to study synthetic temporal networks as directed percolation processes with a range of temporal and spatial inhomogeneities.

The remainder of Sec.~\ref{sec:introduction} will be dedicated to laying the groundwork and presenting the context in which this manuscript is set: in Sec.~\ref{sec:event-graph} we will discuss connectivity on temporal networks, the event graph representation and modeling spreading processes and Sec~\ref{sec:directed-percolation} will introduce directed percolation and its characteristics.

The next section, Sec.~\ref{sec:methods}, is dedicated to an overview of our contributions. Section \ref{sec:dp-analogous-in-temporal-networks} will describe our mapping of concepts of directed percolation and temporal networks. Section \ref{sec:mean-field-solution-dp-in-temporal-network} provides an overview of the theoretical results from Ref.~\cite{shortpaper}, which will be extended further in Sec.~\ref{sec:random-regular-solutions} by explicitly deriving some critical exponents and scaling relations. Section \ref{sec:estimation-tricks} will lay down the algorithmic techniques that make large-scale simulations of spreading processes on temporal networks possible.

Finally, in Sec.~\ref{sec:results} we will describe the experimental setup and provide numerical evidence for validity of our hypothesis by application of the methods described previously, while Sec.~\ref{sec:discussion} provides an overview of the implications of the results and the limitations of our study.

\subsection{Temporal networks and the event graph}\label{sec:event-graph}
A temporal network $G = (\mathcal{V}, \mathcal{E})$ provides representations of a dynamically changing complex system as a set of timed interactions known as events $\mathcal{E}$ between a set of entities $\mathcal{V} = \{v_1, v_2,\ldots, v_n\}$ known as nodes or vertices during an observation period $\mathcal{T}$. Each event indicates a time-dependent interaction between two nodes, e.g.~physical contact or communication between two people or trade between two commercial entities \cite{holme2019temporal}, i.e.~$e=(u, v, t_\text{start}, t_\text{end})$ such that $u, v \in \mathcal{V}$ between times of $t_\text{start}, t_\text{end} \in \mathcal{T}$ ($t_\text{start} < t_\text{end}$). Note that this definition can be easily extended to directed events and to directed or undirected temporal hypergraphs.

Two events $e, e' \in \mathcal{E}$ are \emph{adjacent} if they share at least one endpoint node in common, $\{u, v\} \cap \{u', v'\} \neq \emptyset$, and they follow each other in time such that $ \Delta t(e, e') = t'_\text{start} - t_\text{end} > 0$. Therefore, any temporal network can be represented as a higher-order static directed acyclic weighted graph known as the \textit{event graph} $D = (\mathcal{E}, E_D, \Delta t(e, e'))$~\cite{kivela2018mapping,Mellor}. Nodes of the event graph are the events of the original temporal network and the weight of a link between two connected nodes (adjacent events) is then defined as the time difference $\Delta t$ between the corresponding events.

Every path on the event graph constitutes a \emph{causal chain} as, by definition, a path constitutes a list of events where every two consecutive events are adjacent. Paths in event graphs are, therefore, equivalent to time-respecting paths in the corresponding temporal network representation \cite{saramaki2019weighted}. Therefore, calculating time-respecting reachability on a temporal network is equivalent to connectivity on its corresponding (static) event graph representation. The weakly connected components on an event graph determine \emph{causal domains}, disjoint sets of events where there can be no causal connections whatsoever between events if they belong to two different weakly connected components. In addition, as compared to reachability, the size and distribution of weakly connected components are quantities, which are much easier to measure for temporal networks and they characterize a percolation transition if we assume an undirected network. Moreover, the sizes of these components put an upper bound on how much an effect can spread starting from one of the events in that component \cite{kivela2018mapping}.

Temporal networks preserve the dynamic properties of the represented complex system, unlike aggregated static networks where this information is lost. 
Through the studies of time-varying networks, several new phenomena in human dynamics have been explored over the last decades, such as node and link burstiness~\cite{karsai2018bursty,jo2020burst,urena2020estimating}, causal, temporal motifs \cite{kovanen2011temporal}, or the cyclic activation patterns of human interaction activities \cite{aledavood2018social}, to mention a few. 
As opposed to systems governed by homogeneous and independent processes, these correlations and the induced temporal dynamics may have significant effects on various dynamical processes evolving on temporal networks such as spreading \cite{takaguchi2013bursty, sajjadi2020impact}, reachability \cite{holme2005network, lee2019concurrency}, diffusion \cite{moody2016interdependent, delvenne2015diffusion}, and opinion formation \cite{garimella2018polarization}.

The different dynamics of a temporal network are often straightforward to study through simulations. For example, in the case of spreading processes, transmission can be modeled by temporal network events \cite{karsai2011small, rocha2011simulated, horvath2014spreading, ubaldi2017burstiness, hiraoka2018correlated}. More concretely, in a physical interaction network, where nodes represent people and events represent two people coming to close proximity, 
each of these contact events will have a probability of transmitting the disease. The disease then spreads to all the nodes that can be reached via such infecting events from the initially infected nodes. Similarly, in a network where events represent communication of information at a specific time, such as mobile phone calls or email exchanges, it is straightforward to model the spreading of information by keeping track of the information nodes have access to at each point in time.

Many dynamics evolving on top of networks, such as some spreading processes \cite{lambiotte2016guide, holme2012temporal, holme2015modern}, social contagion \cite{daley1964epidemics, castellano2009statistical,unicomb2021dynamics} ad-hoc message passing by mobile agents \cite{tripp2016special} or routing processes \cite{nassir2016utility}, can have a limited memory thus can only use paths constrained by limited waiting times. Limited waiting-time reachability can be modeled using the event graph, $D$, that contains a superposition of all temporal paths~\cite{kivela2018mapping, saramaki2019weighted, badiemodiri2020efficient}. In a limited waiting-time spreading process unfolding over a temporal network, either the spreading agent (e.g.~the pathogen in the disease spreading) must be transmitted onward from a node within some time $\delta t$ or the infection has to be renewed before that time. In other words, the node must participate in a possibly disease-carrying event in $\delta t$ time, or the process stops and the node reverts to susceptible. Therefore, all the spreading paths in the network are $\delta t$-constrained time-respecting paths. Let's call two adjacent events $e$ and $e'$ as $\delta t$-adjacent if $\Delta t(e, e') \leq \delta t$. A subset of the event graph $D$ with an upper threshold of weights no greater than $\delta t$, i.e., where directed links indicate $\delta t$-adjacency, enables us to calculate reachability for $\delta t$ limited-time spreading process for the corresponding temporal network. Therefore, the event graph encapsulates a complete set of $\delta t$-constrained time-respecting paths for all values of $\delta t$ simultaneously.


\subsection{Directed Percolation}\label{sec:directed-percolation}
The waiting-time limit $\delta t$ can be regarded as the control parameter of a continuous phase-transition, where connectivity in the event graph is determined by $\delta t$-connected paths of events. As the value of maximum waiting time decreases, more and more of the links of the event graph get removed, where each deleted link corresponding to an adjacency relationship between two events that are temporally more than $\delta t$ apart. This leads to a drop in connectivity in the event graph, which is exactly equivalent to the drop in connectivity on the temporal network. In order to characterize these phase transitions, unlike characterizing the superficially similar phase transitions that take place when removing links in static (undirected) networks, we need to consider a percolation framework that can explicitly model the one-way flow of time.

Directed percolation is a paradigmatic example of dynamical phase transitions into absorbing states with a well-defined set of universal critical exponents and is often used to model phenomena with inherent directionality, such as fluids passing through porous media \cite{hinrichsen2000non, odor2004universality, hinrichsen2006non, henkel2008non}. Originally introduced as a model for directed random connectivity \cite{broadbent1957percolation}, directed percolation attracted scrutiny in percolation theory in the late seventies \cite{blease1977directed}. Since then, a considerable body of work has been devoted to this approach of interpretation in the literature since the critical behavior of many stochastic many-particle non-equilibrium processes can be shown to belong to the directed percolation universality class. Directed percolation has applications in various domains at multiple scales ranging from galaxies to semiconductors \cite{schlogl1972chemical,gerola1980theory,van1981hopping,bak1993punctuated}.

As the simplest model exhibiting a transition between active and absorbing phases \cite{hinrichsen2000non}, it is straightforward to define and implement models governed by directed percolation, e.g. in the case of lattice models \cite{domany1984equivalence, kinzel1985phase, harris1974contact, jensen1993critical, mendes1994generalized, ziff1986kinetic, dhar1987collapse}. Directed percolation, however, does not appear to be an integrable model and its critical behavior is highly non-trivial. Moreover, it seems that the basic features of directed percolation, such as non-fluctuating states, are quite difficult to realize in nature \cite{hinrichsen2000possible}. Another fundamental problem is quenched disorder due to microscopic inhomogeneities of the system \cite{henkel2008non}. One of the earliest unambiguous and robust experimental realizations of a system exhibiting critical behavior in the directed percolation class was for the rather specific case of liquid crystal electrohydrodynamic convection \cite{takeuchi2007directed}. Another experimental evidence was reported in 2016 in the case of transition to turbulence \cite{lemoult2016directed}. Due to the simplicity and robustness of directed percolation, it seems to be a good model for explaining ubiquitous phase-transitions in many real-world phenomena, especially in the so-called contact processes \cite{barrat2008dynamical, pastor2015epidemic, kempe2002connectivity, moody2002importance, holme2005network, pan2011path, scholtes2014causality, stehle2011high, dai2020temporal, aleta2020data} in the realm of temporal networks \cite{holme2019temporal}.

Before presenting the mapping between reachability in temporal networks and the concepts in directed percolation, for the remainder of this section we will review these concepts for the case of the simple infinite lattice. Let us take the example of a spreading process across time in an infinitely large $d$-dimensional square lattice: assume that each infected (or occupied) node can infect any of its neighbors independently with probability $p$ at each tick of a discrete timer. Let us also assume that an infected node recovers (becomes unoccupied) in one tick of the clock after infection unless it is re-infected by a neighbor. This configuration is denoted in many sources as a $d+1$-dimensional lattice, substituting the temporal axis with another discrete spatial dimension with the only difference that, unlike the other $d$ dimensions, this one has an inherent directionality. Throughout the rest of this section, we will continue to use the space and time analogy to facilitate a better transition to modeling phenomena on temporal networks.

The dynamics of this spreading process is defined by the topology and dimensionality of the medium of percolation and competition between two processes: the probability that an infected node infects each of its neighbors in a single tick of the clock, or ``reproduction'' from the perspective of the spreading agent, and the time it takes for each infected node to recover, or ``self-annihilation'' or ``death'' of the spreading agent. In the many classic representations of directed percolation, the reproduction probability is often denoted by the parameter $p$ and the ``self-annihilation'' is set to happen in exactly one tick of the clock. For large enough values of $p$, the system will forever stay in an ``active state'' where there is a non-vanishing density of nodes infected (occupied) at all times. Conversely, if the annihilation process has the upper hand, the system eventually transitions irreversibly into an ``absorbing phase'' where no occupied nodes are left in the lattice and the spreading agent is extinct.

More generally, let us say the reproduction and self-annihilation process respectively happen at rates $\mu_p$ and $\mu_r$. Let us assume that at $t=0$, nodes are uniformly occupied with density $\rho_0$. To write a mean-field rate equation for occupation density $\rho(t)$, we need to take into account how often more than one spreading agent (pathogens) simultaneously occupies (infects) the same node, in which case only one new node is occupied. Let us only consider the rate $\mu_c$ at which two other nodes simultaneously infect a single node and assume the probabilities of three or more simultaneous infections are small. In this case, the rate equation is of the form
\begin{equation}\label{eq:general-mean-field-rate-equation}
    \pderivative{t} \rho(t) = \tau \rho(t) - g \rho(t)^2\,,
\end{equation}
where the \emph{control parameter} $\tau = \mu_p - \mu_r$ is the manifestation of the competition between reproduction and death as described above and coupling constant $g = \mu_c$ describes the events of infecting a node already infected by another neighbour \cite{henkel2008non}. This equation has a steady-state at $\lim_{t\rightarrow\infty} \rho(t) = \rho_\text{stat}(\tau) = 0$ which corresponds to the aforementioned absorbing phase. Furthermore for $\tau > 0$ the value of $\rho(t)$ approaches a stationary occupation density of $\lim_{t\rightarrow\infty} \rho(t) = \rho_\text{stat}(\tau) = \tau/g$, which is identified as the order parameter of the directed percolation process. At exactly $\tau = 0$, occupation density decays algebraically with time $\rho(t) \sim (\rho_0^{-1} + gt)^{-1}$. Naturally, for values of $\tau < 0$ the system eventually arrives at the absorbing phase $\rho(t) \rightarrow 0$ in finite time.

More generally, starting from a homogeneously occupied initial condition, order parameter $\rho_\text{stat}(\tau)$ of a system in the directed percolation universality scales as $\rho_\text{stat}(\tau) \sim \tau^\beta$, when control parameter $\tau$ is close to $\tau_c = 0$. For $\tau > 0$, density decays algebraically as $\rho(t) \sim t^{-\alpha}$ where in the mean-field regime (i.e.~$d\geq4$), $\beta = \alpha = 1$. In the case of a spreading process controlled by a percolation probability $p$ introduced at the beginning of this section, it can be shown that $\tau \propto p - p_c$ where critical percolation probability $p_c$ is a function of topology and dimensionality of the percolation medium \cite{saberi2015recent, li2021percolation}.

Alternatively, we can focus on the ramifications of starting from a single seed of infection, as opposed to a homogeneous initial distribution of occupied nodes. A characteristic property of this scenario is survival probability $P(t)$: the probability that a spreading process starting from a single seed would still be in the active phase ($\rho(t) > 0$) at time $t$. Similar to occupation density $\rho(t)$, at criticality $\tau = \tau_c = 0$ survival probability also decays algebraically with time $P(t) \sim t^{-\delta}$. A second alternative for order parameter is the ultimate probability of survival $P_\text{surv}(\tau) = \lim_{t \rightarrow \infty} P(t)$. When the control parameter is close to the critical threshold $\tau \rightarrow 0^-$, the ultimate probability of survival scales algebraically as $P_\text{surv}(\tau) \sim \tau^{\beta'}$.

Continuous phase transitions in models with time-like dimensions generally have the same system of two separate order parameters, controlled by two different critical exponents $\beta$ and $\beta'$. For the case of directed percolation, however, ``rapidity-reversal symmetry'', an invariance property under time-reversal, ensures the two exponents have the same value $\beta = \beta'$ \cite{grassberger1979reggeon} which implies that $P(t)$ and $\rho(t)$ are at least asymptotically proportional as $t \rightarrow \infty$, and in some cases exactly equal $P(t) = \rho(t)$ \cite{henkel2008non}. Rapidity-reversal symmetry limits the number of independent critical exponents to three \cite{1982ZPhyB..47..365G, 1981ZPhyB..42..151J}.

\subsubsection{Characteristic quantities of the directed percolation}
The single-source initial condition also allows us to define additional interesting characteristic quantities in the absorbing phase, which might lend themselves to experimental observation. Let's define \emph{pair-connectedness function} $c(\Vec{r_1}, t_1, \Vec{r_2}, t_2)$ as the probability that a path exists from a node with spatial coordinates $\Vec{r_1}$ at time $t_1$ and another in $\Vec{r_2}$ at time $t_2$. Note that the definition of spatial coordinates for nodes as a $d$-dimensional vector $\Vec{r_i}$ implies that the percolation medium and node $i$ is embedded in a $d$-dimensional space, e.g.~a $d$-dimensional lattice. Assuming that the percolation medium is invariant with respect to translations across time and space, we can simplify the pair-connectedness function by fixing the origin on the source node and denote the pair-connectedness function as $c(\Vec{r}, t)$. \emph{Mean cluster mass} $M$ is defined as the integration of the pair-connectedness function across time and space:
\begin{equation}\label{eq:dp-mean-component-mass}
    M = \int_0^\infty \dd{t} \int \dd{\Vec{r}} c(\Vec{r}, t) \,,
\end{equation}
which, with control parameter close to the critical threshold $\tau_c = 0$ scales like $M \sim (-\tau)^{-\gamma}$ where $\gamma = \nu_\parallel + d \nu_\perp - \beta - \beta'$. Similarly, \emph{mean spatial volume} $V$ can be defined as the number of unique nodes that will ever get infected in a single-source spreading scenario. As with the case of the cluster mass $M$, spatial volume scales through a power relationship $V \sim (-\tau)^{-\upsilon}$ close to the critical threshold where $\upsilon = d \nu_\perp - \beta'$. It is possible to think of spatial volume $V$ as the size of the projection of the percolation cluster over the $d$-dimensional spatial plane, i.e., over the original $d$-dimensional lattice. Projection of the same cluster on the temporal dimension will define the survival time of the cluster, which is distributed according to the probability of survival $P(t)$.

The homogeneous, fully-occupied initial condition, on the other hand, allows us to study the response of a system to an \emph{external field} $h$ on the order parameter static density $\rho_\text{stat}$. For the case of directed percolation, an external field can be implemented as the spontaneous occupation of nodes at a rate $h$. A positive external field deprives the system of the possibility of ever transitioning into an absorbing phase. \emph{Susceptibility} $\chi$ is defined as the magnitude of the response generated by a minuscule disturbance in the external field
\begin{equation}\label{eq:susceptibility}
    \chi(\tau, h)=\frac{\partial}{\partial h} \rho_{stat}(\tau, h)\,,
\end{equation}
which diverges algebraically as the control parameter $\tau$ converges to the critical threshold $\tau_c = 0$, $\chi \sim \abs{\tau}^{-\gamma}$ where $\gamma$ is the same exponent as the mean cluster mass $M$. For the rest of this paper, when not specified, susceptibility $\chi$ is studied at minuscule values of external field ($h=0$) as $\tau$ converges to the critical threshold $\tau_c = 0$. In practical terms, susceptibility is a useful tool for finding the transition point, as unlike the order parameters, we do not need to define an arbitrary threshold for what constitutes a small or large value for a quantity such as $M(t)$ or $V(t)$ close to the transition point in a finite system. Instead, the susceptibility will typically show a peak even in finite systems, which are discussed in more detail in Sec.~\ref{sec:finite-size-scaling}.

\subsubsection{Finite size scaling properties of the system}\label{sec:finite-size-scaling}

While the dynamics described previously explain the behavior of an infinitely large system, measuring properties of infinitely large systems is a rather involved task. Verifying that the behavior of a system at criticality is explained by a specific set of critical exponents is often easier performed by studying the finite-size scaling properties of the system. This can be carried out by measuring a set of quantities for realizations at different scales and plotting the universal scaling function of each quantity as a function of scale-invariant ratios. If the exponents used are correct, all the scaling functions of different linear system sizes for the same quantity should collapse on top of each other.

The effect of the finite size of the system manifest themselves as deviations from the scaling laws as described before and their effects are measurable after some characteristic size-dependent amount of time has elapsed since the beginning of the simulation. For example, while in an infinitely large system in active phase $\tau > 0$ the system will forever stay in an active phase, a finite system will always have a non-vanishing probability of transitioning to the absorbing state due to fluctuation of the order parameter. These finite-size effects take place at a characteristic time $t_f$ that scales as $t_f \sim l^{z}$ where $z = \nu_\parallel / \nu_\perp$ is the so-called \emph{dynamical exponent} and $l$ is the lateral (or linear) size of the system as opposed to system size $N$ measured in number of nodes, where $N \propto l^d$.

In phenomenological scaling theory \emph{simple scaling} is assumed for absorbing phase transitions. This means that large-scale properties of the system are invariant under scale transformations with the control parameter close to the critical threshold. A multiplicative transformation, or ``concentration'', of the control parameter $\tau$ by a factor of $\lambda$, $\tau \mapsto \lambda \tau$ would result in re-scaling of other quantities as
\begin{equation}
    \begin{aligned}
        &t \mapsto \lambda^{-\nu_\parallel} t \; &&l \mapsto \lambda^{-\nu_\perp} l\\
        &\rho \mapsto \lambda^\beta \rho \; &&P \mapsto \lambda^{\beta'} P \\
        &h \mapsto \lambda^\sigma h \; &&\chi \mapsto \lambda^{-\gamma} \chi\,,
    \end{aligned}
\end{equation}
where $t$ and $l$ stand for time-like and length-like quantities respectively.

More specifically, scale-invariance mandates very specifically how a quantity will change under multiplicative scale change. As an example, let's study changes of $\rho(t, l)$ 
\begin{equation}
    \rho = f(t, l) \mapsto \lambda^\beta \rho = f(\lambda^{-\nu_\parallel} t, \lambda^{-\nu_\perp} l)\,,
\end{equation}
where $t$ is time from initial infection seed and $l$ is the linear system size.

Since this relationship is valid for all values of $\lambda$, we can remove one parameter of the function by selecting a special value $\lambda = l^{1/{\nu_\perp}}$
\begin{equation}\label{eq:scaling-for-rho}
    l^{\beta/\nu_\perp} \rho = f(l^{-\nu_\parallel/\nu_\perp} t, 1) = F(l^{-\nu_\parallel/\nu_\perp} t)\,,
\end{equation}
where the function $F(x)$ is referred to as the ``(universal) scaling function'' of its corresponding quantity, in this case, density $\rho$. The parameter to this function $l^{-\nu_\parallel/\nu_\perp} t$ is in itself invariant to scale transformations. This parameter and those similarly derived for other quantities are oftentimes known as ``scale-invariant ratios''. The function $F(x)$ is universal, meaning that if measured to sufficient accuracy, we obtain exactly the same type of scaling function for systems with similar boundary conditions and shape for any phenomena in the directed percolation universality class \cite{henkel2008non}.

The value of each exponent is only a function of a few large-scale properties of the system, such as the number of spatial dimensions of the system. There exists an upper critical dimension $d_c$ where systems with spacial dimensionality $d \geq d_c$ all follow the same set of values for critical exponents, which are exactly equal to those derived through mean-field estimation. For the case of the directed percolation universality class the upper critical dimension has a value of $d_c = 4$ \cite{henkel2008non}.

\section{Methods}\label{sec:methods}
\subsection{Directed percolation in temporal networks}\label{sec:dp-analogous-in-temporal-networks}
Let us now take the case of $\delta t$ limited-time spreading from a single source on a temporal network. Similar to the classic directed percolation single-source spreading process, each temporal network node can participate in the spreading process by becoming infected, infecting others and recovering multiple times. Temporal networks are different from the archetypal directed percolation systems presented in Sec.~\ref{sec:directed-percolation} in that they do not present a regular lattice or metric space in the spatial dimension. Furthermore, there is typically no discrete structure in the temporal dimension, which is usually modeled as a continuous axis. Nevertheless, if the various concepts such as order parameter, control parameter and cluster sizes are defined carefully, temporal networks and limited waiting-time connectivity can be mapped to directed percolation \cite{shortpaper}.

To put it in the same reference frame as with other absorbing phase transitions, changing the parameter $\delta t$, in this scenario, controls the relative occurrence of ``annihilation'' and ``multiplication'' processes. A small enough value of $\delta t$, will lead to a situation where spreading scenarios will eventually die out, at which point the system enters an absorbing phase. Similarly, as $\delta t$ grows, a spreading agent will be able to avoid extinction for longer time, until after some threshold $\delta t > \delta t_c$ a random spreading scenario will not die out (in an infinitely large network). As discussed in Section~\ref{sec:event-graph}, such spreading scenarios are closely related to various properties of the $\delta t$ thresholded limited waiting-time event graph $D$.

\begin{figure*}[ht]
    \centering
    \includegraphics[width=\linewidth]{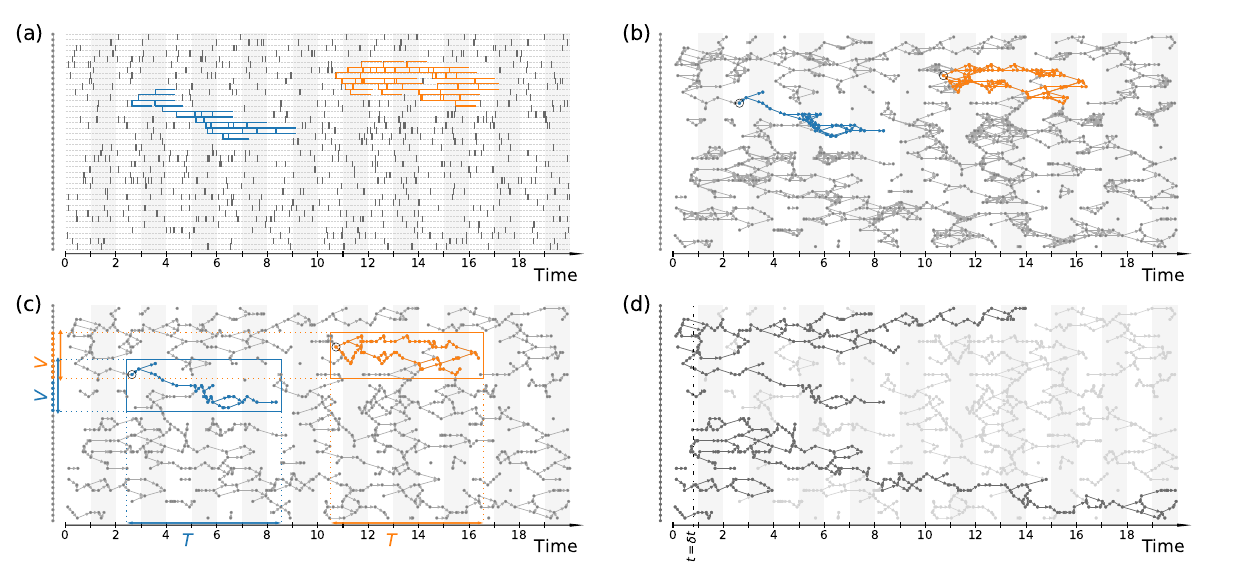}
    \caption{Two spreading scenarios starting from random events (marked with black circles on (b) and (c)) represented over (a) temporal network (b) $\delta t$-limited event graph and (c) reduced event graph of a temporal network built from a one-dimensional grid of 40 nodes (displayed on the left side) with Poisson activation of events with mean inter-event time 1 unit of time, simulated for 20 units of time. The adjacency relations have a maximum waiting time $\delta t = 0.8$ unit of time. Spatial volume $V$ can be visualized as the mean size of projection of a spreading cluster on the spatial plain, i.e.~the static base network on the vertical axis, whereas survival time $T$ is equivalent to the mean size of projection of the cluster on time (horizontal) axis. While measuring a direct analogue to component mass $M$, integrating the pair-connectedness function across time and space equivalent to the mean sum of lengths of colored horizontal lines in (a) is not straightforward with the event graph representation. It is possible to show that the mean number of uniquely counted events involved in the spreading process, corresponding to the cardinality of the out-component of the initial event here represented by the total number of colored nodes in the event graph, show the same scaling behavior. (d) Homogeneous, fully-occupied initial condition with the occupied events shown in a darker shade than unoccupied events shows the decline and eventual stabilization of the occupation density as time grows. In this scenario, all nodes are considered occupied for time $-\infty < t < 0$, which translates to the occupation of all events in period $0 \leq t < \delta t$ and all events in their out-components.}
    \label{fig:dag-schematic}
\end{figure*}

As illustrated in Fig.\ref{fig:dag-schematic}, the projection of the spreading cluster over the spatial plane amounts to a subset of temporal network nodes $\mathcal{V}$ that has ever participated in the spreading process. This can be measured by calculating the mean number of unique temporal network nodes involved in the out-components of the event graph. The (ensemble) average number of unique nodes participating in single random source spreading processes is analogous to mean spatial volume $V$. The projection of the spreading cluster over the temporal axis is equal to the time window from the beginning of the spreading process to its end. The ensemble average of this time duration is analogous to mean survival time $T$. The sum of the duration of infectiousness for all the nodes, i.e.~the integration of the pair-connectedness function, would therefore be analogous to spatial and temporal integration of the pair-connectedness function or mean component mass $M$. Note that the duration of the infectiousness is equal in all of the events, therefore, we use the number of reachable (i.e.~possibly infection-carrying) events as a proxy for $M$, ignoring the overlaps. The above-defined quantities can be measured as features of the event graph. The average number of events in the out-component of a node in the event graph (equivalent to an event in the temporal network) measures the number of reachable events. The survival probability $P(t)$ can be similarly defined over an ensemble of single-source spreading instances based on the distribution of the lifetime of each spreading scenario, accounting for the finite temporal window of the simulation of the temporal network using a Kaplan--Meier estimator \cite{kaplan1958nonparametric}.

Another scenario is the simulation of the spreading process from homogeneous, fully occupied, initial conditions. Translating this from classic directed percolation poses a new problem; a homogeneous initial condition cannot translate to a ``full row'' of occupied nodes since we are dealing with continuous-time as opposed to the typical directed percolation case of discretized time presented in Sec.~\ref{sec:directed-percolation}. Rather, a better translation of the fully-occupied initial condition to continuous time is to assume all nodes to be occupied at the beginning of the observation period $t = min(\mathcal{T})$, or more accurately by assuming all nodes to be occupied for all values of $t$ where $t \leq min(\mathcal{T})$. Occupation density $\rho(t)$ is defined as the fraction of infected nodes at time $t$. Stationary density $\rho_\text{stat}(\tau)$ is therefore defined as occupation density after the system had enough time to reach a stationary state. We can also emulate the effects of an external field $h$ in this scenario: In continuous-time, this is equivalent to each node spontaneously becoming occupied through an independent Poisson point process with a rate of $h$. Susceptibility $\chi(\tau, h)$ can then be measured, from Eq.~\eqref{eq:susceptibility}, by the rate of change in stationary density as external field changes.

\subsection{Empirical methods for estimation of characteristic quantities}\label{sec:estimation-tricks}
In practice, we can estimate $M$, $V$, $T$ and $P(t)$ on the event graph by finding all the \emph{out-components}, i.e., every reachable event starting from every event (see, Figure \ref{fig:dag-schematic}(c)). Calculating the exact set of out-components for every event in the event graph is time and memory-intensive. However, if we are only interested number of events or number of unique nodes that participate in those events, as opposed to the full set of events in the out-components, we can use probabilistic cardinality estimation data-structures to estimate out-component sizes with arbitrary precision in $\mathcal{O}(\abs{E} \log \abs{E})$ time, as opposed to $\mathcal{O}(\abs{E}^2)$ time required for exact calculation \cite{badiemodiri2020efficient}. Minimum and maximum time of all events in the out-component can be exactly calculated in $\mathcal{O}(\abs{E} \log \abs{E})$ time. Calculating properties of the \emph{in-component} of an event is possible through a simple reversal of direction of all links in the event graph and applying the same algorithms.

Similarly, in the homogeneous fully-occupied initial condition scenario, we do not need to directly estimate occupation density $\rho(t)$, stationary occupation density $\rho_\text{stat}(\tau)$ and susceptibility $\chi(\tau, h)$ via naive algorithms, which would explicitly compute these measures by simulating propagation. The properties of homogeneous, fully occupied, $\delta t$-constrained reachability 
can be estimated by marking as occupied any event that is in the out-component of at least one event with time $-\infty < t < t_0$. This can be accomplished by running the in-component size estimation algorithm \cite{badiemodiri2020efficient} once over the whole network, recording minimum observed time in in-component of each event and marking those with minimum in-component time smaller than $t_0$ as occupied. In practice, temporal networks are only recorded or generated for a finite window of time $t_\text{min} < t < t_\text{max}$. As there are no adjacency relationship between events more than $\delta t$ apart temporally, any event that has at least one event in its in-component with time $t_\text{min} < t < t_\text{min} + \delta t$ can be considered occupied. Figure \ref{fig:dag-schematic}(d) shows all occupied events (dark grey) with the initial condition that assumes all nodes are occupied from  $-\infty < t < 0$. The density of occupied events, which corresponds to particle density $\rho(t)$, can be estimated from the event graph representation by the number of occupied nodes in a band of time divided by the area covered by the band, i.e.~number of nodes multiplied by the width of the band.

Normally, calculating the effects of an external field $h$ would require simulating a fully occupied initial condition, marking some nodes randomly selected with rate $h$ as occupied, computing their out-components, and measuring how many new events got occupied. As we are interested in the effects of a minuscule positive external field, indicated by susceptibility $\chi(\tau, 0)$, we can instead calculate the effects of spontaneously marking exactly one random event in the whole network as occupied using probabilistic counting and in-components of all events (i.e., looking back in time). If the number of events in the in-component of an event $e$ is denoted as $|E^\text{in}(e)|$ and the minimum time among all events in its in component as $t^\text{in}_\text{min}(e) = \min_{(u,v,t) \in E^\text{in}(e)} t$, the expected number of spontaneously occupied events when a minuscule external field $h$ is applied can be estimated as $\sum_{e \in \mathcal{E}} P_\text{occupied}(e)$ where
\begin{equation}
   P_\text{occupied}(e) = \begin{cases}
     1 & \text{if}\ t^\text{in}_\text{min}(e) < t_0\\
     \frac{|E^\text{in}(e)|}{|\mathcal{E}|} & \text{otherwise}\,.
   \end{cases}
\end{equation}
In this scenario, the respective value for the external field that would spontaneously occupy on average one event is proportional to $h \propto 1/|\mathcal{E}|$.
We approximate $\rho(t)$ by number of occupied events within a $\delta t$ time window divided by spatio-temporal hyper-volume of the time window $\delta t \times \left| \mathcal{V} \right|$. The estimate for $\rho(t)$ can in turn be used to approximate quantities like stationary density $\rho_{stat}(\tau)$ and susceptibility $\chi(\tau, h)$.

\section{Results}\label{sec:results}
\subsection{Experimental setup}\label{sec:experimental-setup}
In this section, we focus on validating and exploring the limits to our hypothesis that $\delta t$ limited-time spreading in many forms of temporal networks belongs to the directed percolation universality class. We do this by performing single-seed and homogeneous initial-condition spreading simulations following the method defined in Sec.~\ref{sec:dp-analogous-in-temporal-networks} and explained in detail in Sec.~\ref{sec:estimation-tricks}. By measuring various observables for networks of different sizes as described in Sec.~\ref{sec:finite-size-scaling}, we can verify whether for each quantity the corresponding universal scaling functions collapse for systems of different finite sizes when using the same values of critical exponents $\beta$, $\beta'$, $\nu_\parallel$ and $\nu_\perp$ as that of DP corresponding to the dimensions of the system as a previous mean-field approximation and experimental setups for the directed percolation.

The experiments are performed on a variety of synthetic temporal networks. The generation procedure consists of generating a static \emph{base network} corresponding to the aggregate network and generating events, i.e.~activations or timestamps, for each link based on some temporal dynamic. In total, we analyzed 26 combinations of base networks and link-activation processes. In order to perform the finite-size scaling analysis, we computed all the statistics for ten network sizes, starting from $N=2^8$ nodes and increasing the size by a factor of two until we reached $N=2^{17}$ nodes. For the case of $d$-dimensional square grids where $d \in \{ 2, 3, 4 \}$, however, closest powers of $d$ to the powers of two from $2^8$ to $2^{17}$ was used with a periodic boundary condition, to provide spatial translational invariance. Each statistic was calculated as the average of at least 256 (up to 4096) realizations and each realization of the largest configuration consists of around $3.7\times10^7$ events. No sampling of spreading scenarios was required for each network's realization, as the effect of starting a spreading process from any possible combinations of nodes and times could be gathered in one pass as described in Sec.~\ref{sec:estimation-tricks}. See supplementary material for a more detailed overview of the experimental setup.

Static base networks are either (a) one to four-dimensional square lattice grids with periodic boundary conditions (b) random regular graphs with specified average degree \cite{kim2003generating,steger1999generating} or (c) Erdős–Rényi $G(n, p)$ random networks with specified expected average degree \cite{batagelj2005efficient}. For the random networks, we chose the average degrees 8 for the Erdős–Rényi graphs and 9 for the random regular graphs (such that both networks have the same expected excess degree). The higher degrees of random networks ensure that the probability of generating networks with large isolated components remains negligible and that even locally, the network would be of high enough dimensionality to be in the mean-field regime above the upper critical dimension $d_c = 4$.

\begin{figure}
    \centering
    \includegraphics[width=\linewidth]{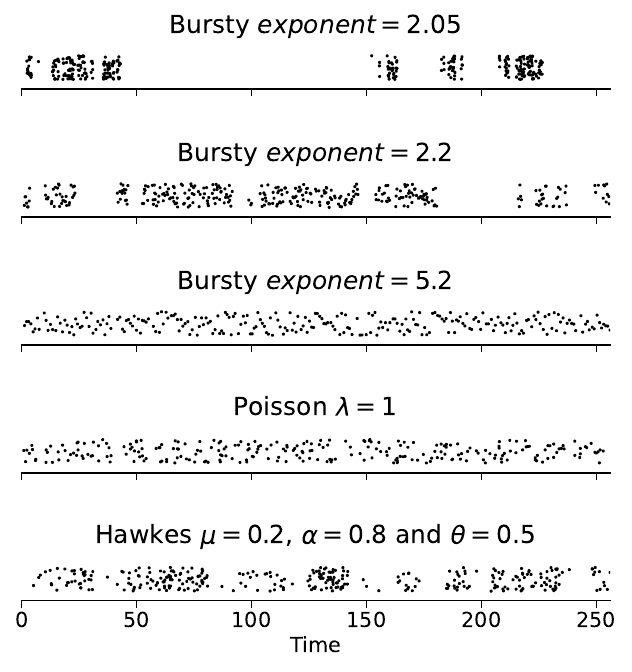}
    \caption{Sample timestamps from a single realization (activations of a single link) with different temporal dynamics. Each point represents a single activation at a specific time. The points are scattered over the vertical axis to avoid overlaps in the visualization. All timestamps were generated for 256 units of time with parameters or minimum cutoffs that would result in an expected inter-event time of 1. Equation \eqref{eq:hawkes-process-intensity} defines the parameters and the intensity function of the Hawkes univariate exponential self-exciting process.}
    \label{fig:timestamps}
\end{figure}

Temporal dynamics of the links are either governed by (a) Poisson processes, i.e.~exponential inter-event times, (b) bursty processes, i.e.~renewal processes with power-law inter-event time distributed as $\propto \Delta t^-\gamma$ with exponents $\gamma \in \{ 2.05, 2.2, 2.8, 5.2 \}$ and minimum interval cutoff set so that the expected inter-event would be equal to one and (c) Hawkes independent self-exciting processes with different parameter sets. The Hawkes univariate exponential self-exciting process \cite{hawkes1971point}, is defined by the conditional intensity function
\begin{equation}\label{eq:hawkes-process-intensity}
    \lambda^*(t) = \mu + \alpha \theta \sum_{t_i < t} \mathrm{e}^{-\theta (t - t_i)}\,.
\end{equation}
The parameters of this formulation of the Hawkes process are (1) background (or exogenous) intensity of events $\mu$ indicating the random probability of events happening without being caused through self-excitement, (2) the infectivity factor $\alpha$, which can be interpreted as the expected number of induced self-exciting events per each event, and (3) the rate parameter of the delay $\theta$. Based on the properties of exponential kernel used in defining Eq.~\ref{eq:hawkes-process-intensity}, $1/\theta$ is the expected inter-event time between an event (e.g.~a coincidental social interaction) and its corresponding induced self-exciting event (e.g.~the follow-up social interactions) \cite{laub2015hawkes}.

As the unit of time is arbitrary, temporal processes are scaled, without loss of generalization, so that they produce timestamps with a mean inter-event time equal to one. The processes are initialized in their stationary state, and in practice, the first timestamp for each event is generated through residual time distribution of each process, except for the case of Hawkes process where the process is allowed a burn-in time equal to the simulation time window before the first timestamp is recorded. The temporal processes of pairs of links are simulated independently of each other. Figure \ref{fig:timestamps} shows a visualization of the different methods of generating event activations. Temporal networks were simulated for a time window of at least $T = 64$ and up to $T = 8192$ units of time. See supplementary material for the exact experimental setup for each system size. The difference in system sizes and time windows for the simulations were necessitated by the limitations and optimal utilization of the computational facilities.

\subsection{Estimating the critical threshold \texorpdfstring{$\delta t_c$}{dt-c} and the critical exponents \texorpdfstring{$\beta$}{beta}, \texorpdfstring{$\beta'$}{beta'}, \texorpdfstring{$\nu_\parallel$}{nu-parallel} and \texorpdfstring{$\nu_\perp$}{nu-perpendicular}}\label{sec:estimating-dtc}

Best estimate of the critical exponents $\beta$, $\beta'$, $\nu_\parallel$, $\nu_\perp$ and critical threshold $\delta t_c$ can be determined by finding the values of these exponents that would produce the best data collapse for the universal scaling functions corresponding to $\rho(t)$, $\hat{P}(t)$, $M(t)$, and $V(t)$. The quality of collapse, in turn, can be assessed by comparing the deviation of the scaling function curves for different system sizes from the average trajectory. Here, for each of the quantities $\hat{P}(t)$, $\rho(t)$, $M(t)$ and $V(t)$, we calculated one trajectory for finite-size scaling function for each system size, as defined for example for the case of $\rho(t)$ by Eq.~\eqref{eq:scaling-for-rho}. As the tested value of critical exponents and $\delta t_c$ gets closer to the actual critical threshold, the curves for different sizes should more closely collapse on top of each other. Plotted with the correct values of critical exponents and critical threshold, we expect to see all trajectories collapse into one with the possible exception of very small values of $t$. To quantify the quality of a collapse, we measure the mean curve in the area where all system sizes have defined values for the scaling function and measure the root mean square difference of all points from all system sizes to the mean curve. The errors were measured after logarithmically scaling the values to account for the power-law nature of the scaling functions. Sum of errors for the collapse of $\hat{P}(t)$, $\rho(t)$, $M(t)$ and $V(t)$ was used in evaluating each set of parameters.

\begin{figure}
    \centering
    \includegraphics[width=\linewidth]{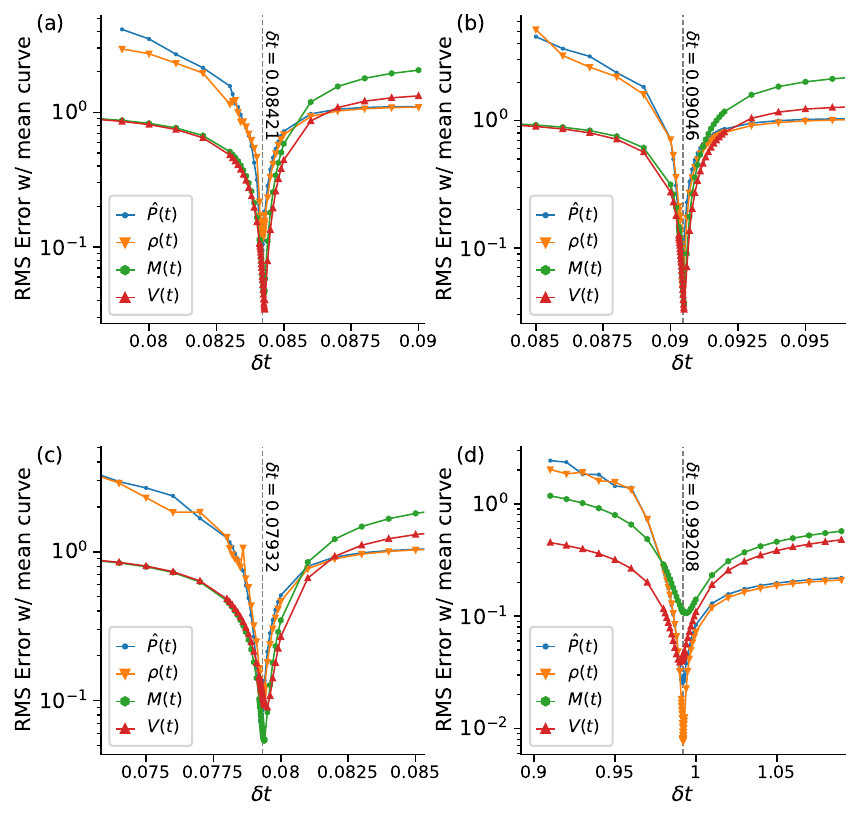}
    \caption{Root mean square (logarithmic) deviation of scaling-corrected functions of probability of survival $\hat{P}(t)$, density $\rho(t)$, mass $M(t)$ and volume $V(t)$ for different system sizes from average trajectory shows a sharp drop at $\delta t_c$ due to data collapse. Each instance of the network is made through realizations of (a) Erdős–Rényi static network $\langle k \rangle = 8$ and Poisson process $\lambda=1$ activations, (b) random 9-regular networks with bursty (power-law with minimum cutoff) inter-event time distribution with mean 1 and exponent $\gamma = 2.8$, (c) Erdős–Rényi static network $\langle k \rangle = 8$ and Hawkes univariate exponential self-exciting process with parameters $\mu=0.2$, $\alpha=0.8$ and $\theta=1.0$ and (d) one-dimensional grid with periodic boundary conditions (a circle) and Poisson process $\lambda=1$ link activations. Refer to Sec.~\ref{sec:experimental-setup} for the definitions of the parameters.
    }
    \label{fig:dtc-optimisation}
\end{figure}

In order to assess collapse of the universal scaling functions, we first determine a value for $\delta t_c$ for each network configuration. That is, the best candidate for $\delta t_c$ is selected based on the least total error for collapse of $\hat{P}(t)$, $\rho(t)$, $M(t)$ and $V(t)$ assuming DP critical exponents. Figure \ref{fig:dtc-optimisation} shows this total error of collapse for two network configurations. This shows is a clear minimum for each configuration indicating the critical value $\delta t_c$, which is consistent across $\hat{P}(t)$, $\rho(t)$, $M(t)$, and $V(t)$ trajectories.

\begin{figure}[hb!]
    \centering
    \includegraphics[width=\linewidth]{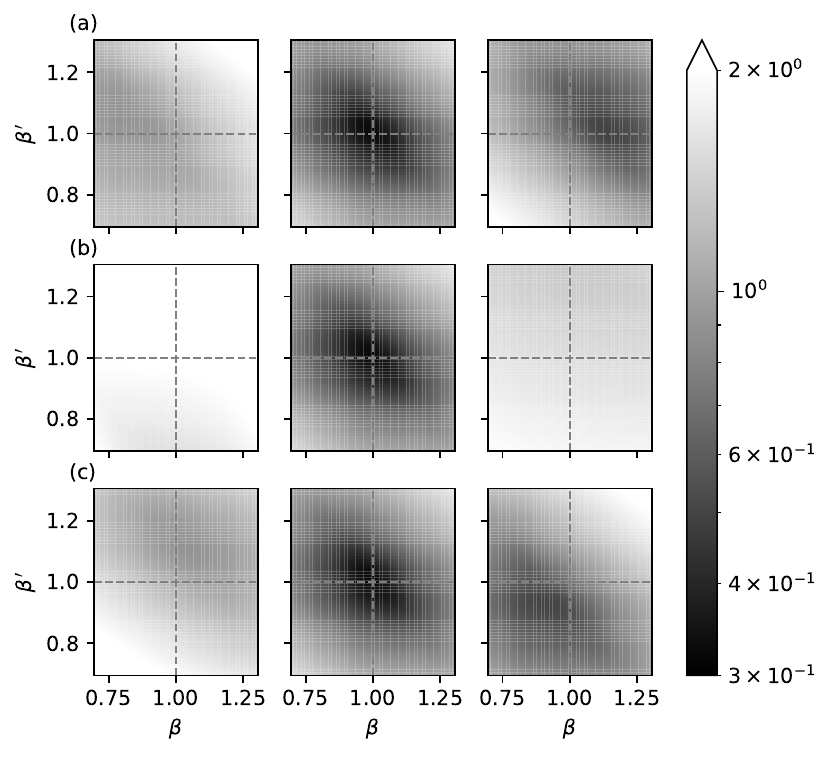}
    \caption{Total error of collapse of universal scaling functions of $M(t)$, $V(t)$, $\hat{P}(t)$ and $\rho(t)$ for Erdős–Rényi networks $\langle k \rangle = 8$ and Poisson process activation $\lambda = 1$ as a function of $\beta$ and $\beta'$. In these visualizations we set $\nu_\perp=0.5$, $\nu_\parallel=1$ and $\delta t_c=0.08421$, and vary one of these parameters such that the three panels from left to right correspond to values (a) $\nu_\perp \in \{0.34, 0.5, 0.66\}$, (b) $\nu_\parallel \in \{0.84, 1, 1.16\}$ and (c) $\delta t_c \in \{0.0840,0.08421,0.0844\}$. Note that the center panel is repeated across the rows and always has parameter values $\nu_\perp=0.5$, $\nu_\parallel=1$ and $\delta t_c=0.08421$. We see that there is a minimum in the error close to $\beta = \beta' = \nu_\parallel = 1$ and $\nu_\perp=0.5$ within this five-dimensional space.}
    \label{fig:grid-search-beta-beta-prime}
\end{figure}

\begin{figure*}[ht!]
    \centering
    \includegraphics[width=\linewidth]{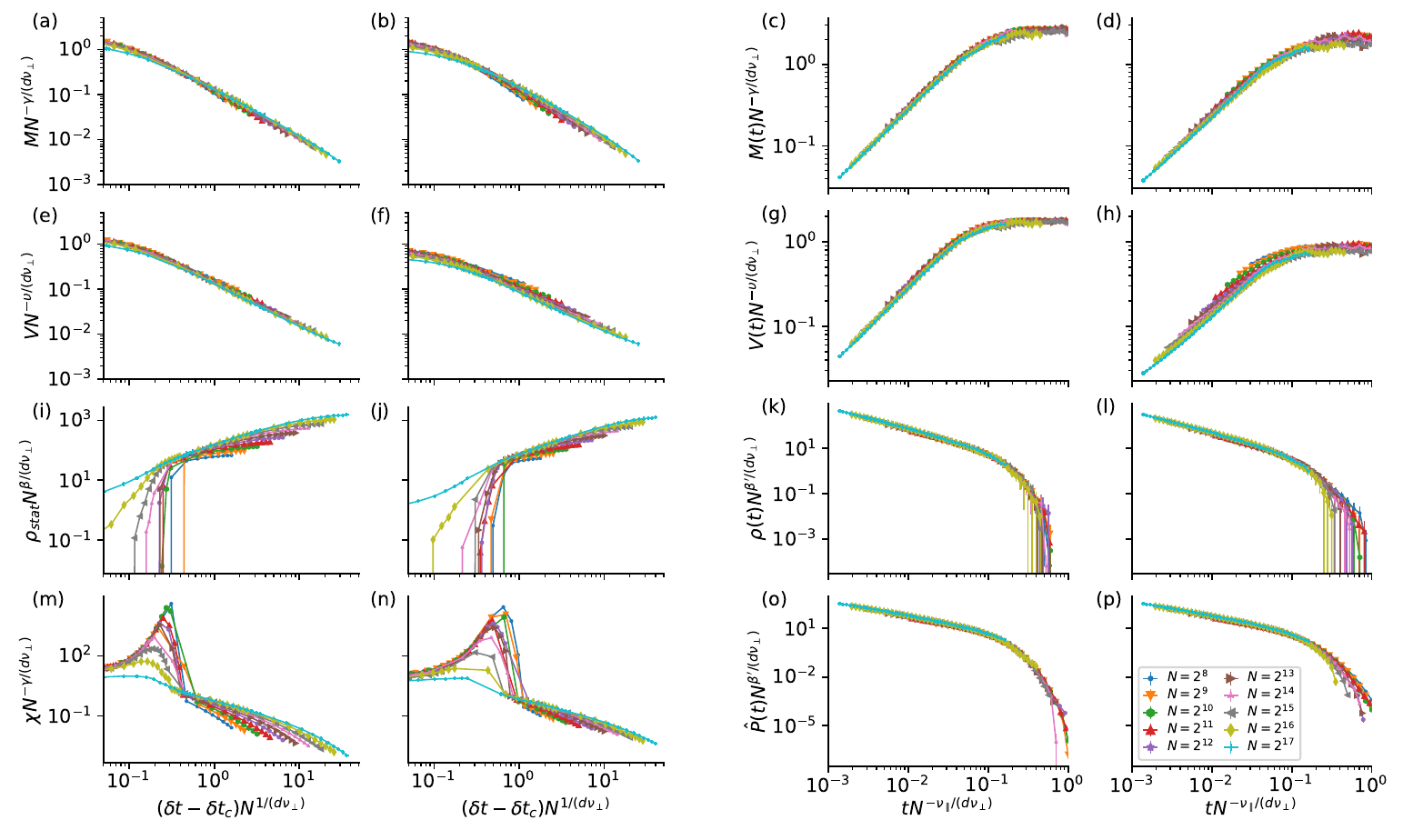}
    \caption{Universal scaling functions for $\delta t$ limited-time reachability over (a,c,e,g,i,k,m,o) random 9-regular network with bursty (heavy-tail with minimum value cutoff) link activation with mean inter-event time of 1 and exponent $\gamma=2.8$  and (b,d,f,h,j,l,n,p) random Erdős–Rényi network $\langle k \rangle = 8$ with Hawkes univariate exponential self-exciting process link activation with parameters $\mu = 0.2$, $\alpha = 0.8$ and $\theta = 1.0$. The finite size scaling is performed for the following single-source scenarios: (a,b) the mean component mass $M$ as function of $\delta t$ close to critical point and (c,d) as function of time $t$ at the critical point, (e,f) the mean component volume $V$ as function of $\delta t$ close to critical point and (g,h) as function of time $t$ at the critical point and (o,p) Survival probability $\hat{P}$ as function of time $t$ at the critical point. For fully-occupied initial conditions the finite-size scaling is performed for (k,l) the occupation density $\rho$ as function of time $t$ at the criticality, and both (i,j) the static density $\rho_{\text{stat}}$ and (o,p) susceptibility $\chi$ as function of $\delta t$ close to the critical point. The collapse of the universal scaling functions validates the hypothesis that these systems are governed by the same critical exponents as in directed percolation in the mean-field regime. 
    See Sec.~\ref{sec:experimental-setup} for the full definitions of the parameters.}
    \label{fig:regular-bursty-scaling}
\end{figure*}

The resulting estimates for $\delta t_c$ can be used to visually verify directed percolation critical exponents and our selected optimal value of $\delta t_c$ for each system by plotting the finite-size universal scaling functions of different system sizes. In total, we produce collapses for eight characteristic quantities measured in a single source or homogeneous initial conditions. Figure~\ref{fig:regular-bursty-scaling} shows these collapses measured for regular networks with bursty dynamics (renewal process with power-law inter-event times) and Erdős–Rényi networks with a Hawkes self-exciting process dynamics. The full set of plots for all of the 26 configurations are shown in Supplementary Materials. In all cases, a satisfactory collapse can be observed for at least probability of survival $\hat{P}(t)$ and density $\rho(t)$ and in most cases, other quantities show a good collapse as well. It is important to note that quantities that depend on measuring values as time approaches infinity, e.g., $\rho_\text{stat}(\delta t)$ and $\chi(\delta t)$ have generally lower quality of measurement and collapse since the time to reach a stable value for these increases substantially close to criticality \cite{hinrichsen2000non}.

Table \ref{tab:best-fits-and-errors} (column ``Est. $\delta t_c$'') shows our best estimate of the critical threshold $\delta t_c$ for each configuration using the method described above. As the systems become rapidly more and more connected after the critical threshold, a lower value for the critical threshold $\delta t_c$  indicates higher, or more robust, spatio-temporal connectivity, meaning that the same $\delta t$ limited-time spreading would result in a larger number of reachable nodes, $V(\delta t)$, or larger number of reachable events, $M(\delta t)$. When modeling infectious disease spreading as directed percolation on temporal networks, larger values for $V(\delta t)$ and $M(\delta t)$ may indicate larger epidemic sizes and the total number of human hours of infection in the population, respectively.

\begin{table*}[h!]
\caption{\label{tab:best-fits-and-errors}
Column ``Est. $\delta t_c$'' shows the best candidate for critical threshold $\delta t_c$ selected by minimizing the collapse error of the universal scaling functions for probability of survival $\hat{P}(t)$, density $\rho(t)$, mass $M(t)$ and volume $V(t)$ derived for different system sizes, assuming DP exponents. The collapse error is measured by the sum of root mean squared deviation of logarithmically scaled trajectories for all four scaling functions. For the best estimate for exponents $\alpha=\beta/\nu_\parallel$ and $\delta=\beta'/\nu_\parallel$, reported respectively in columns ``Est. $\alpha$'' and ``Est. $\delta$'', a power-law was fitted to the head of values of probability of survival $\hat{P}(t)$ and density $\rho(t)$ respectively for the largest system size simulated for the time period between $2 < t < 0.04 \times N^{\nu_\parallel/{d \nu_\perp}}$, where both functions are expected to be still mostly behaving, similar to an infinite system, according to power relations $t^{-\alpha}$ and $t^{-\delta}$ respectively. Directed percolation mean-field values for these critical exponents are $\delta = \alpha = 1$, which is close to the value estimated for random, high-dimensional networks. Furthermore the value of these critical exponents in a DP system are expected to be close to $\alpha=\delta=0.15946$ for 1+1 dimensional, $\alpha=\delta=0.450$ for 2+1 dimensional, $\alpha=\delta=0.732$ for 3+1 dimensional and equal to the mean-field estimates systems $\alpha=\delta=1$ for 4+1 dimensional \cite{hinrichsen2000non}, which is close to values estimated for 1 dimensional lattice and 2 to 4 dimensional square lattices.
}
\begin{ruledtabular}
\begin{tabular}{lddddddd}
\textrm{Configuration}&
\multicolumn{1}{c}{\textrm{Est. $\delta t_c$}}&
\multicolumn{1}{c}{\textrm{$\beta$ Error}}&
\multicolumn{1}{c}{\textrm{$\beta'$ Error}}&
\multicolumn{1}{c}{\textrm{$\nu_\parallel$ Error}}&
\multicolumn{1}{c}{\textrm{$\nu_\perp$ Error}}&
\multicolumn{1}{c}{\textrm{Est. $\alpha$}}&
\multicolumn{1}{c}{\textrm{Est. $\delta$}}\\
\colrule
Erdős--Rényi $\langle k \rangle = 8$\\
\hspace{1em}Poisson & 0.08421 & 0.01 & 0.01 & 0.06 & 0.03 & 1.0702 & 1.0338\\
\hspace{1em}Bursty\\
\hspace{2em}$\gamma=2.05$ & 0.06231 & 0.06 & 0.17 & 0.08 & 0.09 & 1.0110 & 0.9816\\
\hspace{2em}$\gamma=2.2$ & 0.08013 & 0.02 & 0.05 & 0.04 & 0.03 & 1.0320 & 1.0285\\
\hspace{2em}$\gamma=2.8$ & 0.08649 & 0.01 & 0.01 & 0.05 & 0.01 & 1.0625 & 1.0368\\
\hspace{2em}$\gamma=5.2$ & 0.08655 & 0.01 & 0.01 & 0.06 & 0.02 & 1.0540 & 1.0499\\
\hspace{1em}Hawkes self-exciting\\
\hspace{2em}$\mu=0.2\ \alpha=0.8\ \theta=0.5$ & 0.0815 & 0.01 & 0.04 & 0.07 & 0.03 & 0.9929 & 1.0015\\
\hspace{2em}$\mu=0.2\ \alpha=0.8\ \theta=1.0$ & 0.07932 & 0.02 & 0.06 & 0.06 & 0.04 & 1.0185 & 0.9747\\
\hspace{2em}$\mu=0.5\ \alpha=0.5\ \theta=0.5$ & 0.08339 & 0.01 & 0.03 & 0.05 & 0.02 & 1.0791 & 1.0328\\
\hspace{2em}$\mu=0.5\ \alpha=0.5\ \theta=1.0$ & 0.08281 & 0.01 & 0.04 & 0.07 & 0.03 & 1.0311 & 1.0116\\
\hspace{2em}$\mu=0.8\ \alpha=0.2\ \theta=0.5$ & 0.08397 & 0.01 & 0.01 & 0.07 & 0.02 & 1.0542 & 1.0246\\
\hspace{2em}$\mu=0.8\ \alpha=0.2\ \theta=1.0$ & 0.08383 & 0.01 & 0.02 & 0.07 & 0.02 & 1.0251 & 1.0087\\
Random 9-regular\\
\hspace{1em}Poisson & 0.08808 & 0.03 & 0.05 & 0.05 & 0.02 & 1.0096 & 0.9947\\
\hspace{1em}Bursty\\
\hspace{2em}$\gamma=2.05$ & 0.06484 & 0.08 & 0.17 & 0.11 & 0.08 & 0.9752 & 0.9660\\
\hspace{2em}$\gamma=2.2$ & 0.08413 & 0.04 & 0.05 & 0.05 & 0.03 & 1.0044 & 0.9825\\
\hspace{2em}$\gamma=2.8$ & 0.09046 & 0.02 & 0.03 & 0.05 & 0.02 & 1.0190 & 0.9874\\
\hspace{2em}$\gamma=5.2$ & 0.09049 & 0.02 & 0.02 & 0.07 & 0.01 & 0.9886 & 0.9755\\
\hspace{1em}Hawkes self-exciting\\
\hspace{2em}$\mu=0.2\ \alpha=0.8\ \theta=0.5$ & 0.0853 & 0.05 & 0.06 & 0.06 & 0.03 & 0.9982 & 0.9686\\
\hspace{2em}$\mu=0.2\ \alpha=0.8\ \theta=1.0$ & 0.08303 & 0.02 & 0.06 & 0.08 & 0.04 & 0.9680 & 0.9564\\
\hspace{2em}$\mu=0.5\ \alpha=0.5\ \theta=0.5$ & 0.08728 & 0.02 & 0.03 & 0.06 & 0.02 & 1.0094 & 0.9702\\
\hspace{2em}$\mu=0.5\ \alpha=0.5\ \theta=1.0$ & 0.08663 & 0.02 & 0.05 & 0.09 & 0.03 & 0.9861 & 0.9664\\
\hspace{2em}$\mu=0.8\ \alpha=0.2\ \theta=0.5$ & 0.0879 & 0.05 & 0.01 & 0.12 & 0.01 & 0.9901 & 0.9563\\
\hspace{2em}$\mu=0.8\ \alpha=0.2\ \theta=1.0$ & 0.08769 & 0.04 & 0.05 & 0.06 & 0.03 & 0.9936 & 0.9796\\
1D lattice\\
\hspace{1em}Poisson & 0.9919 & 0.01 & 0.03 & 0.01 & 0.03 & 0.1583 & 0.1456\\
2D square lattice\\
\hspace{1em}Poisson & 0.28428 & 0.01 & 0.08 & 0.03 & 0.01 & 0.4109 & 0.3922\\
3D square lattice\\
\hspace{1em}Poisson & 0.15375 & 0.01 & 0.06 & 0.02 & 0.01 & 0.7229 & 0.6899\\
4D square lattice\\
\hspace{1em}Poisson & 0.1045 & 0.02 & 0.03 & 0.03 & 0.02 & 1.0077 & 0.9870\\

\end{tabular}
\end{ruledtabular}
\end{table*}

These results indicate that within each spatial configuration, increased burstiness (as indicated by lower value for the power-law exponent $\gamma$) generally leads to a lower value for $\delta t_c$ threshold and higher connectivity. Furthermore, for the case of the self-exciting process, increasing the expected number of self-induced events, as indicated by $\alpha$, generally results in a lower value for $\delta t_c$ (higher connectivity). While it was previously understood that a wide range of temporal inhomogeneities slows down spreading processes over temporal networks \cite{karsai2011small}, these results demonstrate that certain temporal inhomogeneities, e.g.~a highly bursty or self-exciting temporal dynamic, can enable a \emph{more limited} spreading agent (expressed in terms of a maximum waiting time) to spread to a wider set of nodes. For example, spreading processes with maximum waiting time between $0.063 < \delta t < 0.084$ over an Erdős–Rényi networks $\langle k \rangle = 8$ will spread to a much larger set of nodes and span a longer span of time if the link activations are highly bursty ($\gamma = 2.05$) compared to a Poisson process with the same mean inter-event time, as the latter will be spreading in the sub-critical regime compared to the super-critical regime for the former.

It is also interesting to note that while the random spatial configurations, namely random 9-regular networks and the Erdős–Rényi networks $\langle k \rangle = 8$, both result in networks with the same expected excess degree value, the Erdős–Rényi networks with higher levels of spatial inhomogeneity, which manifests as a wider spread degree distribution, can be observed to have a lower $\delta t_c$ critical threshold. While testing on a wider range of spatial (structural) inhomogeneities would be required before a conclusion is reached, these results might hint at a similar behavior as with temporal inhomogeneities, namely that introducing certain spatial inhomogeneities might result in higher connectivity in the sense that the same limited-time spreading agent can eventually spread to a wider share of the network.

Additionally, we present a method to assess the quality of a collapse for a range of different values of critical exponents ($\beta$, $\beta'$, $\nu_\parallel$ and $\nu_\perp$) and $\delta t_c$. A five-dimensional grid search for optimal values for critical exponents and $\delta t_c$ based on the quality of collapse for $P(t)$, $\rho(t)$, $M(t)$ and $V(t)$ shows that the total error declines around critical exponent values close to that of directed percolation, i.e.~$\beta=\beta'=\nu_\parallel=1$ and $\nu_\perp = 0.5$ for mean-field regimes and their respective DP values for lower dimensionality square grid networks. Figure~\ref{fig:grid-search-beta-beta-prime} shows for Erdős–Rényi static networks with $\langle k \rangle = 8$ and Poisson process link activation, the $\beta\cross\beta'$ plane from the five-dimensional grid search with two sandwiching parallel planes along each of the $\nu_\parallel$, $\nu_\perp$ and $\delta t_c$ dimensions. This verifies that there is a minimum close to $\beta=\beta'=\nu_\parallel=1$, $\nu_\perp = 0.5$ and $\delta t_c = 0.08421$ for total error of collapse of $P(t)$, $\rho(t)$, $M(t)$ and $V(t)$. Similar plots for some other network configurations (along with a different two-dimensional slice, $\nu_\parallel \cross \nu_\perp$) can be viewed in the Supplementary Material. It is important to note that while other combinations of parameters in the grid might lead to other local optima, visual inspection of the resulting collapse show that to be mainly numerical artifacts where the total error changes rapidly close to extreme values of the parameters (i.e., critical exponents and $\delta t_c$) where only a very small fraction of the trajectories for different finite sizes actually overlap.

Furthermore, for each of the critical exponents, we can measure an estimation error based on this five-dimensional parameter grid. For each exponent, we find a range of values where, assuming that all other exponents are fixed at their DP values, would produce collapses of higher or equal quality compared to the DP value of that exponent. The sizes of these ranges, which by definition includes the DP value for all exponents, provides a confidence interval for the range of possible exponent values that are able to explain the behavior of the system with at least the same quality as that of directed percolation. As shown in Table~\ref{tab:best-fits-and-errors}, these errors are in most cases only a few percent, with a notable exception of the highly bursty renewal processes with $\gamma = 2.05$. Simulating power-law distributions becomes a much harder problem as the magnitude of the exponent approaches 2. Close to this exponent, it takes a larger and larger number of realizations for the properties of the population, e.g.~average inter-event time for bursty temporal dynamics, to converge. It is also possible that the large estimation error is an indicator that the system is approaching a breakdown of one of the key symmetries, with the most likely candidate being rapidity-reversal symmetry based on the fact that the estimation error for $\beta'$ is much larger than that of the other exponents.

\subsection{Estimating critical exponents by simulating very large systems}\label{sec:estimating-alpha-delta}

\begin{figure}
    \centering
    \includegraphics[width=\linewidth]{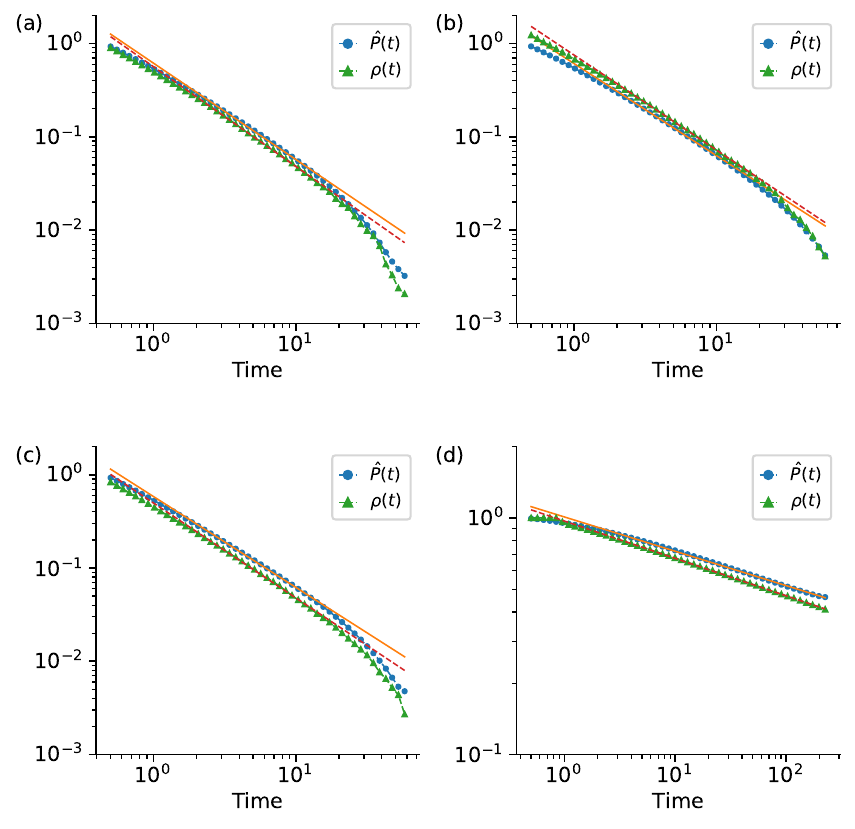}
    \caption{An example of fitting power-law functions on empirical $\rho(t)$ and $\hat{P}(t)$ results on finite networks for deriving critical exponents $\alpha = \beta/\nu_\parallel$ and $\delta = \beta'/\nu_\parallel$. Power-law functions were fitted on experimental results of spreading over Erdős–Rényi networks with $\langle k \rangle = 8$ and $N=2^{17}$ nodes and bursty (power-law with minimum cutoff) inter-event time distribution with mean 1 and exponent $\gamma = 2.8$. The fitting was performed on values in range $1 < t < 0.04 \times N^{\nu_\parallel/{d \nu_\perp}}$ (i.e.~$1 < t < 14.48$) to limit the interference of finite-size effects with the scaling behavior.}
    \label{fig:exponent-fit}
\end{figure}

As discussed before in Sec.~\ref{sec:directed-percolation}, the effects of the finite size of the system manifest at characteristic times $t_f \propto N^{\nu_\parallel/{d \nu_\perp}}$ in the form of fluctuations that causes the transition of the system to the absorbing phase. At times much smaller than $t_f$ the system  shows approximately the scaling behavior of an infinitely large system where at criticality, $\rho(t) \sim t^{-\alpha}$ and $\hat{P}(t) \sim t^{-\delta}$ where $\alpha = \beta/\nu_\parallel$ and $\delta = \beta'/\nu_\parallel$. On the other hand, the power-law scaling behavior becomes visible at times comparable to the mean inter-event time of the dynamic process but not up to arbitrarily infinitesimal values of $t$. Given these properties, we fitted two power-law functions using the least-squares method to the results of experiment with the largest system size for the range of time $2 < t < 0.04 \times N^{\nu_\parallel/{d \nu_\perp}}$ on $\rho(t)$ and $\hat{P}(t)$ to derive exponent $\alpha$ and $\delta$. Figure \ref{fig:exponent-fit} shows one such fitting for a system made from Erdős–Rényi networks with $\langle k \rangle = 8$ and $N=2^{17}$ nodes and bursty (power-law with minimum cutoff) inter-event time distribution with mean 1 and exponent $\gamma = 2.8$. Table \ref{tab:best-fits-and-errors} (columns Est. $\alpha$ and Est. $\delta$) shows the best estimates of these exponents, which as expected are very close to respective directed percolation critical exponents of 1 (for the mean-field regime $d \geq 4$) for the case of random networks and 0.159, 0.450 and 0.732 for one-, two- and three dimensional lattices respectively \cite{hinrichsen2000non}.

\section{Discussions}\label{sec:discussion}
Through combining multiple methods of empirical and theoretical verification, we are able to confidently state that limited waiting-time connectivity percolation over a wide range of synthetic temporal networks incorporating a range of temporal and topological inhomogeneities show behavior compatible with the directed percolation universality class. It is of utmost importance to discuss the limitations of our method: chief among them, that our empirical finite-size simulation method, as described in Sec.~\ref{sec:estimation-tricks}, is not able to measure quantities which are defined at $t \rightarrow \infty$, such as the ultimate probability of survival $P_\text{surv}$ and static density $\rho_{stat}$ (and therefore susceptibility $\chi$) to the same standard of accuracy as the other quantities due to the finite size of the synthetic networks used for analysis. This is exacerbated close to the critical threshold where the equilibration time, the time required for the network to reach a stationary state, grows rapidly while the memory and computational cost of simulating a temporally larger temporal network grow linearly and log-linearly, respectively, with the increased simulated time \cite{hinrichsen2000non}. This is visible in Fig.~\ref{fig:regular-bursty-scaling}e,f,k,l as a worse collapse as compared to other quantities.

Also, while it is computationally much more feasible to measure susceptibility $\chi$ by inducing occupation of exactly one existing event in the temporal network (described in Sec.~\ref{sec:estimation-tricks}) as compared to inducing occupation of nodes at random times (as described in Sec.~\ref{sec:directed-percolation}), the latter method might be more robust, especially when dealing with a temporal network with a high degree of temporal inhomogeneity. Although our experiments with this alternative method were limited to smaller system sizes, we could not observe any significant difference between the two methods for the network configurations presented in this manuscript.

While a wide range of temporal dynamics and network structures with different levels of inhomogeneity are studied here, there is still a wide variety of systems that present computational and theoretical challenges. First, the effects of event-event correlations between links are not studied. It has been shown that event-event correlations, among other forms of inhomogeneity, can affect the rapidity of the spreading process on temporal networks \cite{karsai2011small}. Conceptually, local event-event correlations such as temporal motifs \cite{kovanen2011temporal}, are close to temporal event graphs, which are in practice computed using isomorphisms on slightly modified temporal event graphs. Thus, incorporating temporal motifs to the framework at the level of analytical computations is an interesting future direction, as that corresponds to modifying the frequency of appearance of structural motifs in the event graphs. Second, the effect of static base networks with heavy-tail degree distributions and other more complicated network topologies are absent from this study. Here, of especial interest are the networks with heavy-tailed degree distribution with static network reachability percolation threshold at zero occupied links, e.g.~$p(k) \propto k^{-2}$. While initial results did not support the conclusion that a $\delta t$ limited waiting time over this class of synthetic temporal networks would be in the directed percolation universality class, due to limitations on computational resources, we were not able to perform the analysis on the larger system sizes comparable to the other types of networks.

Depending on the physical mechanism involved in the modeled connectivity phenomenon or spreading process, alternative methods of defining the adjacency relationship might be more suitable than the one used here. For example, for the case of disease spreading over a physical contact network, the currently used definition of event graph causes a ``re-infection'' of the infected party, manifested as a restart of their $\delta t$ duration of disease. This can be resolved by substituting each undirected event in the temporal network with two simultaneous directed events. Similarly, for a disease spreading scenario over transportation networks, such as an airplane traffic network, the time between two events (the value that is compared to the maximum duration of disease $\delta t$ to determine whether two flights are adjacent) should be calculated from the departure of one flight to departure of the possibly adjacent flight and not, as it is currently presented, from the arrival of the latter to the departure of the former. This might be an important factor when dealing with scenarios in which the reasonable values for $\delta t$ are comparable to the delay or the duration of the events, e.g.~the time from the departure of a flight to the arrival in a spreading process over an air transport network.

For some spreading mechanisms, it might also be more suitable to replace the hard $\delta t$ limited-time cutoff of adjacency requirement used in this work with a probabilistic process by measuring quantities over an ensemble of event graphs. For example, using a Poisson process instead of a $\delta t$ limited-time cutoff would produce dynamics similar to simulations of SIS processes over networks while simulating results of the simulation starting at every possible starting point in one pass. Viewed this way, normal $\delta t$ limited-time cutoff can be seen as a probabilistic process where the probability of adjacency is a step function at $\Delta t = \delta t$. It is also possible to combine an occupation probability similar to classic directed percolation (see Sec.~\ref{sec:directed-percolation}) with a $\delta t$ limited-time cutoff (or a Poisson process cutoff or other forms of temporal locality constraint) to construct a two-dimensional phase diagram for each temporal network.

It would also be possible to define connectivity in the event graphs in a way that mimics the SIR process. In this case, one would need to prune some of the temporal paths in the event graph such that temporal network nodes are not repeated. This distinction is equivalent to paths and simple paths (or walks and paths, respectively) in static graphs. The algorithmic techniques employed in this work are not directly applicable to this case, and in fact, it has been recently shown that algorithmic problems in such settings can be computationally difficult. For example, in the SIR interpretation of the event graph, finding if it is possible for a node infected at a specific time to infect a given node is an NP-hard problem \cite{casteigts2019computational}. In any case, averaging over explicit simulations of spreading scenarios is always an alternative option to the algorithms that take advantage of the redundancies in computing reachability.

Connectivity, which encapsulates several important phenomena on complex systems such as spreading processes \cite{lambiotte2016guide, holme2012temporal, holme2015modern} and routing dynamics \cite{nassir2016utility}, has not yet undergone the same level of development on temporal networks as the static networks. It has been previously suggested that connectivity on temporal networks, or other adjacent representations such as dynamic networks, might show the same properties as any other directed percolation system \cite{parshani2010dynamic,kivela2018mapping}, a class of percolation models with built-in directionality which has enjoyed abundant attention in the past decades. In Ref.~\cite{shortpaper}, we laid formal foundations by providing one-to-one analogues between concepts from directed percolation and temporal network connectivity and provided theoretical evidence supporting this hypothesis. In this work, we presented multiple accounts of empirical evidence showing that connectivity on many model temporal networks belongs to the directed percolation universality class and that this hypothesis is robust for a range of temporal and spatial heterogeneities.

This work focused mainly on establishing the vocabulary and developing the required tools in the hopes of rendering studies of connectivity in temporal networks ripe for future analysis, especially from a critical phenomena perspective. It is important to note that this work has only scratched the surface of the analytical study of connectivity on temporal networks and still, a vast body of analytical and phenomenological topics, some of which were eluded to in the previous paragraphs, remains open for future study.

\section{Acknowledgement}
We would like to thank János Kertész and Géza Ódor for their helpful comments and suggestions. The authors wish to acknowledge CSC -- IT Center for Science, Finland, and Aalto University ``Science-IT'' project for generous computational resources. Márton Karsai acknowledges support from the DataRedux ANR project (ANR-19-CE46-0008) and the SoBigData++ H2020 project (H2020-871042).

\bibliography{citations}

\section{Appendix: Mean-field solution for directed percolation in temporal networks}\label{sec:mean-field-solution-dp-in-temporal-network}

The event graph representation contains many redundant adjacency relationships, e.g.~triangles, or more generally feed-forward loops, that can be removed without changing reachability of nodes, producing a \emph{reduced event graph} \cite{shortpaper}. Assuming the probability of two or more adjacent events happening at exactly the same time is negligible, the reduced event graph, a subset of the event graph with exactly the same reachability properties, has a maximum in-~and out-degree of 2 \cite{Mellor,shortpaper}. If we make the simplifying assumption that the reduced event graph representation of $\delta t$ limited waiting-time spreading process on a specific temporal network is indistinguishable from a random directed network with the same joint in- and out-degree distribution $P(k_{\text{in}}, k_{\text{out}})$, a mean-field solution to order parameter occupation density $\rho_\text{stat}$ for a $\delta t$ limited-time spreading process over temporal networks, as defined in Section \ref{sec:dp-analogous-in-temporal-networks}, can be derived in the form
\begin{equation}\label{eq:rate-equation}
    \frac{\partial}{\partial t}\rho(t) =(\langle Q_\text{out} \rangle - 1) \rho(t) -\langle Q_\text{out} \rangle \rho(t)^2 \,,
\end{equation}
where $\langle Q_\text{out} \rangle$ is the mean excess out-degree of the reduced event graph \cite{shortpaper}. This rate equation has the same form as Eq.~\eqref{eq:general-mean-field-rate-equation}. The solution to this equation shows a phase transition at $\tau_c = 0$ and other behavior consistent with $\tau = \langle Q_\text{out} \rangle - 1$ being the control parameter of directed percolation. As with Eq.~\eqref{eq:general-mean-field-rate-equation} this sets two of the four critical exponents in the mean-field regime to the same values as those of mean-field DP, $\alpha = \beta = 1$.

Under the same assumption, the probability-generating function representation of the out-degree distribution is $G^\text{out}_0(y) = G(1, y)$ where $G(x, y)$ is the joint in- and out-degree distribution probability-generating function. Similarly, the excess out-degree distribution probability-generating function can be defined as
\begin{equation}
G^\text{out}_1(y) = \frac{1}{\langle k_{EG} \rangle} \left.\frac{\partial}{\partial x} G(x, y)\right|_{x=1}\,,
\end{equation}
where $\langle k_{EG} \rangle = \left.\frac{\partial}{\partial x} G(x, y)\right|_{x=y=1} = \left.\frac{\partial}{\partial y} G(x, y)\right|_{x=y=1}$ is the mean in- or out-degree on the event graph. This can be used to derive the out excess-degree distribution as $Q^{\text{out}}_i = \frac{\partial^i}{i! \partial y^i} G^\text{out}_1(y)|_{y=0}$.

Making the same assumption as above, namely that the event graph representation is indistinguishable from a random directed network with the same joint in- and out-degree distribution $P(k_{\text{in}}, k_{\text{out}})$, we can derive the mean cluster mass, which as discussed in Sec.~\ref{sec:dp-analogous-in-temporal-networks} can be calculated as the number of reachable events or mean out-component size on the event graph, as
\begin{equation}
    M = 1 + \langle k_{EG} \rangle (-\tau)^{-1} = \frac{\langle k_{EG} \rangle - \tau}{-\tau} \,,
    \label{eq:m}
\end{equation}
which has a power-law asymptote at $\tau_c = 0$ of the form $M \sim -\tau^{-1}$ which confirms the mean-field DP exponent $\gamma = 1$ \cite{shortpaper}.

Deriving a closed-form solution for $\tau$ becomes prohibitively complex for many types of synthetic temporal networks that involve even the slightest traces of spatial or temporal inhomogeneities and require many simplifying approximations of the structure of networks. As the nature of the assumption are similar to the ones we used while showing the critical exponents in the mean-field regime, this alone would not be productive as a mean to validate or refute the previous theoretical claims for networks with heterogeneous structure or dynamics. Therefore, we complemented these analytical derivations of $\tau$ (from Sec.~\ref{sec:random-regular-solutions}) and the critical exponents (from the mean-field approach of Ref.~\cite{shortpaper} and the current section) with measurements derived from simulations. While it would be possible to measure $\tau$ from the simulated event graphs, we elected to use $\delta t - \delta t_c$ as a stand-in for control parameter $\tau$, similar to how $p - p_c$ was used in Sec.~\ref{sec:directed-percolation} for lattices. Very close to the critical threshold $\tau \rightarrow \tau_c$, $\delta t - \delta t_c$ linearly approximates the control parameter $\tau$, which would preserve the power-law relationships mentioned before at least for some neighbourhood of $\tau = \tau_c = 0$. $\delta t$ is simply a parameter of the simulation and $\delta t_c$ can be derived empirically for each configuration, either by trial and error or through the finite-size scaling method described in Sec.~\ref{sec:finite-size-scaling}. This means that, by virtue of not relying on the methods and assumptions presented previously, we can provide a clean separation between the empirical validation and our theoretical assumptions.

It is possible to find a closed-form solution for $\tau$ for very simple systems, such as the case of random $k$-regular networks with Poisson process link activations. This, however, entails making simplifying assumptions about the structure of the event graph. The results of this derivation and the comparison with empirical measurements follows in Sec.~\ref{sec:random-regular-solutions}.

\subsubsection{Solution for random \texorpdfstring{$k$}{k}-regular static base networks with Poisson link activation}\label{sec:random-regular-solutions}

\begin{figure}
    \centering
    \includegraphics[width=\linewidth]{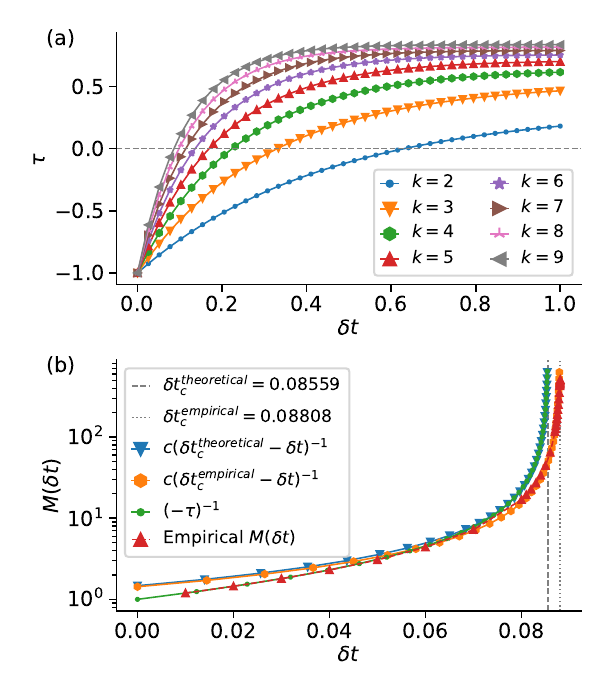}
    \caption{(a) Theoretically derived value of control parameter $\tau$ as a function of $\delta t$ as given in Eq.~\eqref{eq:tau_theory_kreg} for random k-regular networks with Poisson processes link activation with $\lambda = 1$. The intersection with the horizontal line at $\tau=0$ indicates the predicted critical value $\delta t_c$. (b) The analytical solutions for mean out-component size $M = (-\tau)^{-1}$ as a function of $\delta t$ compared to empirical measurement of $M(\delta t)$ over 256 realizations of large ($N=2^{17}$) finite network for random 9-regular networks in the absorbing phase $\delta t < \delta t_c$. Also visualised is the effect of using $\delta t - \delta t_c$ as an approximation of control parameter $\tau$, which shows similar behavior close to $\delta t_c$.
    }
    \label{fig:regular-tau-vs-dt-analytical}
\end{figure}

For the case of random $k$-regular static base networks and Poisson process activation of links with mean inter-event time $\lambda$, we were able to analytically derive a closed-form solution of the control parameter $\tau$ as a function of $\delta t$, $k$ and $\lambda$. To this end, it is necessary to derive the joint degree distribution probability-generating function $G(x, y)$ of the event graph based on the excess degree distribution of the base random $k$-regular network and the Poisson process \cite{shortpaper}. This leads to a formulation of out-degree and excess out-degree distribution probability-generating functions of the form
\begin{equation}\label{eq:regular-out-degree-generating-function}
\begin{aligned}
    G_0^\text{out}(y) = G_1^\text{out}(y) = -\frac{2 (k-1) (y-1) y e^{\delta t (-k) \lambda }}{k} +\\
    \frac{(y-1) (2 k (y-1)-2 y+1) e^{\delta t (1-2 k) \lambda }}{2 k-1} + \\
    \frac{y \left(2 (k-1)^2 y+3 k-2\right)}{k (2 k-1)}\,.
\end{aligned}
\end{equation}
This in turn, based on relation $\tau = \langle Q_\text{out} \rangle -1 = G^{'\text{out}}_1(1) - 1$, produces
\begin{equation}
    \tau = \frac{\left(4 k^2-6 k+2\right) e^{\delta t (-k) \lambda }+k e^{\delta t (1-2 k) \lambda }-2 (k-1)^2}{(1-2 k) k}\,.
    \label{eq:tau_theory_kreg}
\end{equation}

Figure \ref{fig:regular-tau-vs-dt-analytical}(a) shows the relationship between the theoretically derived value of the control parameter $\tau$ from Eq.~\eqref{eq:tau_theory_kreg} for different random $k$-regular networks with a Poisson process with mean inter-event time fixed to 1. As expected, a denser network has a lower onset of criticality in terms of the maximum waiting time $\delta t$. Furthermore, a linear approximation of $\tau \propto \delta t - \delta t_c$ works quite well for these systems for the neighborhood close to $\tau = 0$ given the lower curvature for at least the immediate surrounding of $\tau_c$.

Given that, for the event graph representation of an infinite random $k$-regular networks with a Poisson process activation configuration the out-degree and the excess out-degree distributions are equal, as derived in Eq.~\eqref{eq:regular-out-degree-generating-function}  (i.e.~$G^\text{out}_0(x) = G^\text{out}_1(x)$), Eq.~\eqref{eq:m} simplifies to $M = -\tau^{-1}$ for $\tau < 0$. Figure \ref{fig:regular-tau-vs-dt-analytical}(b) compares this analytical solution of mean out-component size (calculated with the assumption of the randomness of the event graph) with empirical measurements of a large network. Note that, for $k=9$, our best empirical estimate for $\delta t_c$, $\delta t^\text{empirical}_c = 0.08808$, compared to the estimate from the analytical method, $\delta t^\text{theoretical}_c = 0.08559$, have a difference of around 3\%. This is also visible when comparing empirical measurements of mean cluster mass $M(\delta t)$ and the theoretical estimations for the system in Fig.~\ref{fig:regular-tau-vs-dt-analytical}(b). This can be attributable to the fact that the rate equation Eq.~\eqref{eq:rate-equation} is constructed for temporal networks under the assumption that the event graph is indistinguishable from a random directed network with the same joint in- and out-degree distribution. This difference seems to suggest that certain local structures in the event graph are very slightly over-represented compared to a random directed graph with the same degree distribution. Also indicated by Fig.~\ref{fig:regular-tau-vs-dt-analytical}(b) is the fact that the power-law behavior of the empirical trajectory with a critical exponent of $\gamma = -1$ can quite easily be validated by using an empirical estimation of $\delta t_c$.
\end{document}


\title{Supplementary Material for Evidence for Directed Percolation in Random Temporal Network Models}
\author{Arash Badie-Modiri}
\affiliation{Department of Computer Science, School of Science, Aalto University, FI-0007, Finland}
\author{Abbas K.~Rizi}
\affiliation{Department of Computer Science, School of Science, Aalto University, FI-0007, Finland} 
\author{Márton Karsai}
\affiliation{Department of Network and Data Science
Central European University, 1100 Vienna, Austria}
\affiliation{Alfr\'ed R\'enyi Institute of Mathematics, 1053 Budapest, Hungary}
\author{Mikko Kivelä}
\affiliation{Department of Computer Science, School of Science, Aalto University, FI-0007, Finland} 

\date{\today}

\maketitle

\section{Directed percolation critical exponents}
For the purposes of this paper we used the following values for the critical exponents of the directed percolation universality class based on values presented in \cite{henkel2008non} collected from various original sources \cite{jensen1999low,voigt1997epidemic,grassberger1996self,jensen1992critical}. The implementation (namely the grid-search step) requires all exponents to be truncated to 3 digits after decimal point.

\begin{table}[h!]
\caption{\label{tab:dp-exponents} Values used for the critical exponents of the directed percolation universality class. The values \cite{henkel2008non,jensen1999low,voigt1997epidemic,grassberger1996self,jensen1992critical} were all truncated to three places after the decimal point.}
\begin{ruledtabular}
\begin{tabular}{crrrr}
\multicolumn{1}{c}{\textrm{exponent}} &
\multicolumn{1}{c}{\textrm{$d=1$}} &
\multicolumn{1}{c}{\textrm{$d=2$}} &
\multicolumn{1}{c}{\textrm{$d=3$}} &
\multicolumn{1}{c}{\textrm{Mean-field}}\\
\colrule
$\beta=\beta'$ & 0.276 & 0.583 & 0.814 & 1.000 \\
$\nu_\parallel$ & 1.734 & 1.295 & 1.110 & 1.000 \\
$\nu_\perp$ & 1.097 & 0.733 & 0.584 & 0.500
\end{tabular}
\end{ruledtabular}
\end{table}

Additionally the other ``dependent'' exponents, used throughout the paper and the implementation, can be derived from the exponents in Tab.~\ref{tab:dp-exponents} as follows:

\begin{equation}\label{eq:dp-exponents-relations}
    \begin{aligned}
        z &= \nu_\parallel / \nu_\perp \\
        \alpha &= \beta/\nu_\parallel \\
        \delta &= \beta'/\nu_\parallel \\
        \gamma &= \nu_\parallel + d \nu_\perp - \beta - \beta' \\
        \upsilon &= d \nu_\perp - \beta'
    \end{aligned}
\end{equation}

Although the values used for the exponents not presented in Tab.~\ref{tab:dp-exponents} are calculated and used as double precision floating points in the implementation, their representative values up to three digits after decimal point is presented in Tab.~\ref{tab:dp-other-exponents} for the sake of comparison and quick reference.
\begin{table}[h!]
\caption{\label{tab:dp-other-exponents} Values used for the other critical exponents of the directed percolation universality class. The values presented here, calculated from Tab.~\ref{tab:dp-exponents} and Eq.\ref{eq:dp-exponents-relations} are truncated to three places after the decimal point for the purposes of this table.}
\begin{ruledtabular}
\begin{tabular}{crrrr}
\multicolumn{1}{c}{\textrm{exponent}} &
\multicolumn{1}{c}{\textrm{$d=1$}} &
\multicolumn{1}{c}{\textrm{$d=2$}} &
\multicolumn{1}{c}{\textrm{$d=3$}} &
\multicolumn{1}{c}{\textrm{Mean-field}}\\
\colrule
$z$ & 1.581 & 1.767 & 1.901 & 2.000 \\
$\alpha=\delta$ & 0.159 & 0.45 & 0.733 & 1.000 \\
$\gamma$ & 2.279 & 1.595 & 1.234 & 1.000 \\
$\upsilon$ & 0.821 & 0.883 & 0.938 & 1.000 \\
\end{tabular}
\end{ruledtabular}
\end{table}

\section{Implementation}
The experiments of this paper are implemented using mainly C++ and are made available online \cite{badie2021implementation}. The implementation and the paper itself makes use of the work of various other software projects \cite{tange2018gnu,yoo2003slurm,hagberg2008exploring,wolfram2021mathematica} as well as generous computational resources provided by the Aalto University Science-IT project and CSC -- IT Center for Science, Finland.

The different types of random networks are generated on the fly in the \verb|networks/| directory by the scripts provided in the \verb|scripts/| directory. Of the real-world datasets, analyzed in \cite{shortpaper}, the data for the US air transportation network \cite{bts2017air} and Helsinki public transportation network \cite{kujala2018collection} are provided along with the implementation.

The code can be compiled using the command \verb|make all|. Various C++17 features are used extensively throughout the code and it is only tested to compile with GCC 9.3.0, though it is expected to work with a more recent version of GCC as well.

You can generate a random temporal network using the \verb|random_network| executable:
\begin{verbatim}
$ ./random_network --seed 0 --nodes 512 --average-degree 9 --static-model regular \
       --temporal-model poisson --mean-dt 1 --max-time 2048 > example/example-small.events
\end{verbatim}
or by passing a pre-generated static network to the \verb|--static-model|, instead of specifying one of the existing a static network model. To estimate quantities for source source limited-time spreading process on a synthetic temporal network with $\delta t = 0.8$, you can use a variation of the following command:
\begin{verbatim}
$ ./random_large_single --seed 0 --dt 0.8 --network example/example-small.events \
      --summary /dev/stdout --window-min 0 --window-max 2048 \
      --time-bins 16 --time-bins-min 0.25 --time-bins-max 2048
\end{verbatim}
which prints the results, including statistics on cluster mass, volume and lifetime and some basic information on the temporal network, in JSON format on standard output. The quantities are presented as average over all events, as well as binned logarithmically based on the time of the event and the time from the event to the end of the observation window. The lifetime and temporal distribution of events that produce ``censored'' percolation clusters, clusters that might span longer than the window of observation of the temporal network, are reported separately as well to facilitate estimating the probability of survival using Kaplan--Meier estimators \cite{kaplan1958nonparametric}. Similarly, if the same set of parameters is used with either of the executables \verb|random_large_homogeneous| or \verb|random_large_homogeneous_alternative|, it provides information about the distribution of occupied events in case of starting from homogeneous, fully occupied initial condition. The former accepts an additional parameter \verb|--field h| to simulate an external field of rate $h$ by spontaneously occupying the events, while the latter uses the much faster algorithm presented in this paper which calculates the average effect of occupying exactly one random event. The values reported for \verb|dt-band-homogeneous-occupied| and \verb|dt-band-external-field-occupied| corresponds to the number of normally occupied events in a $\delta t$ band before the end of the observation time window, and the number of occupied events in the presence of a minor external field that spontaneously occupies exactly one random event. The aggregation script \verb|aggregate_data.py| can provide more details on how the output of multiple runs of these executables results are aggregated and transformed into quantities that are reported in this manuscript.

Results for the synthetic networks were generated through the following iterative system: We started at an initial estimation for $\delta t_c$ and precision of $p=1$, where precision indicates the number of digits of certainty for $\delta t_c$ after the decimal point. At each step we calculated all the characteristic quantities for all possible values of $\delta t$ with one extra digit after decimal point ($\delta t_c - 9\times10^{-(p-1)} \geq \delta t \geq  \delta t_c + 10\times10^{-(p-1)}$) and calculate total error of collapse for $P(t)$ $\rho(t)$ $V(t)$ and $M(t)$ trajectories assuming that value is the critical threshold $\delta t_c$. If one of the $\delta t_c$ candidates has a significantly higher quality of collapse than the current value, $\delta t_c$ is updated, precision is incremented and the process is repeated. After $\delta t_c$ is determined up to the desired level of precision or after determining that the precision cannot be increased further with the current method, system sizes or the number of realizations, we use all the data produced at every step to plot quantities as a function of $\delta t - \delta t_c$ and as a function of time at $\delta t = \delta t_c$. Please note that the main goal of the current manuscript is not to provide the best estimate of $\delta t_c$ for various synthetic temporal networks, but to validate the hypothesis that limited-time spreading on these networks belong to the same universality class as any directed percolation system. It might be possible to design more efficient numerical methods of estimating $\delta t_c$, but the current method and the estimates presented would suffice for the purposes of this manuscript.

\clearpage
\section{Full suit of empirical results for random \texorpdfstring{Erdős–Rényi}{Erdos–Renyi} based temporal networks}

\begin{table}[htbp!]
\caption{\label{tab:experiment-parameters-random} Experimental setup for Erdős–Rényi networks with $\langle k \rangle = 8$ and random 9-regular networks.}
\begin{ruledtabular}
\begin{tabular}{rrr}
\multicolumn{1}{c}{\textrm{Size $N$}} &
\multicolumn{1}{c}{\textrm{Time window $T$}} &
\multicolumn{1}{c}{\textrm{Realisations}}\\
\colrule
256 & 2048 & 4096 \\
512 & 2048 & 4096 \\
1024 & 2048 & 4096 \\
2048 & 1024 & 4096 \\
4096 & 1024 & 1024 \\
8192 & 1024 & 512 \\
16384 & 512 & 512 \\
32768 & 256 & 512 \\
65536 & 128 & 256 \\
131072 & 64 & 256 \\
\end{tabular}
\end{ruledtabular}
\end{table}
\begin{figure}[htbp!]
    \centering
    \includegraphics[width=0.65\linewidth]{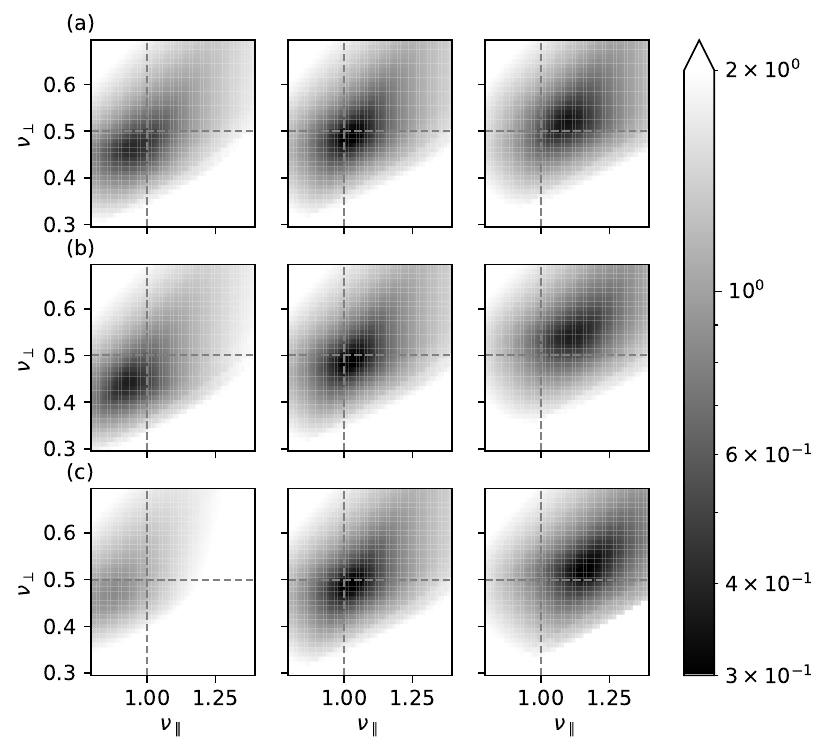}
    \caption{Total error of collapse of $M(t)$, $V(t)$, $\hat{P}(t)$ and $\rho(t)$ universal scaling functions for Erdős–Rényi networks $\langle k \rangle = 8$ and Poisson process activation $\lambda = 1$ with $\nu_\perp$ and $\nu_\parallel$ values assuming (a) $\beta \in \{0.84, 1, 1.16\}$, (b) $\beta' \in \{0.84, 1, 1.16\}$ and (c) $\delta t_c \in \{0.0840,0.08421,0.0844\}$ which shows a minimum around $\beta = \beta' = \nu_\parallel = 1$ and $\nu_\perp=0.5$}
    \label{fig:erdos-grid-search-nu-para-nu-perp}
\end{figure}
\begin{figure}[htbp!]
    \centering
    \includegraphics[width=0.65\linewidth]{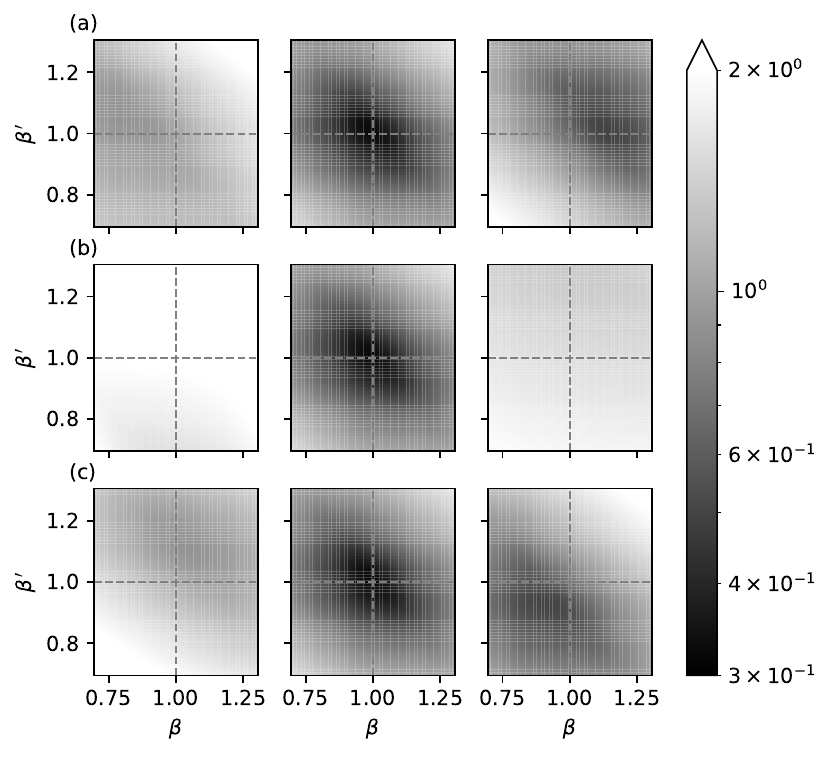}
    \caption{Total error of collapse of $M(t)$, $V(t)$, $\hat{P}(t)$ and $\rho(t)$ universal scaling functions for Erdős–Rényi networks $\langle k \rangle = 8$ and Poisson process activation $\lambda = 1$ with $\beta$ and $\beta'$ values assuming (a) $\nu_\perp \in \{0.34, 0.5, 0.66\}$, (b) $\nu_\parallel \in \{0.84, 1, 1.16\}$ and (c) $\delta t_c \in \{0.0840,0.08421,0.0844\}$ which shows a minimum around $\beta = \beta' = \nu_\parallel = 1$ and $\nu_\perp=0.5$}
    \label{fig:erdos-grid-search-beta-beta-prime}
\end{figure}

\begin{figure}[htbp!]
    \centering
    \includegraphics[width=\linewidth]{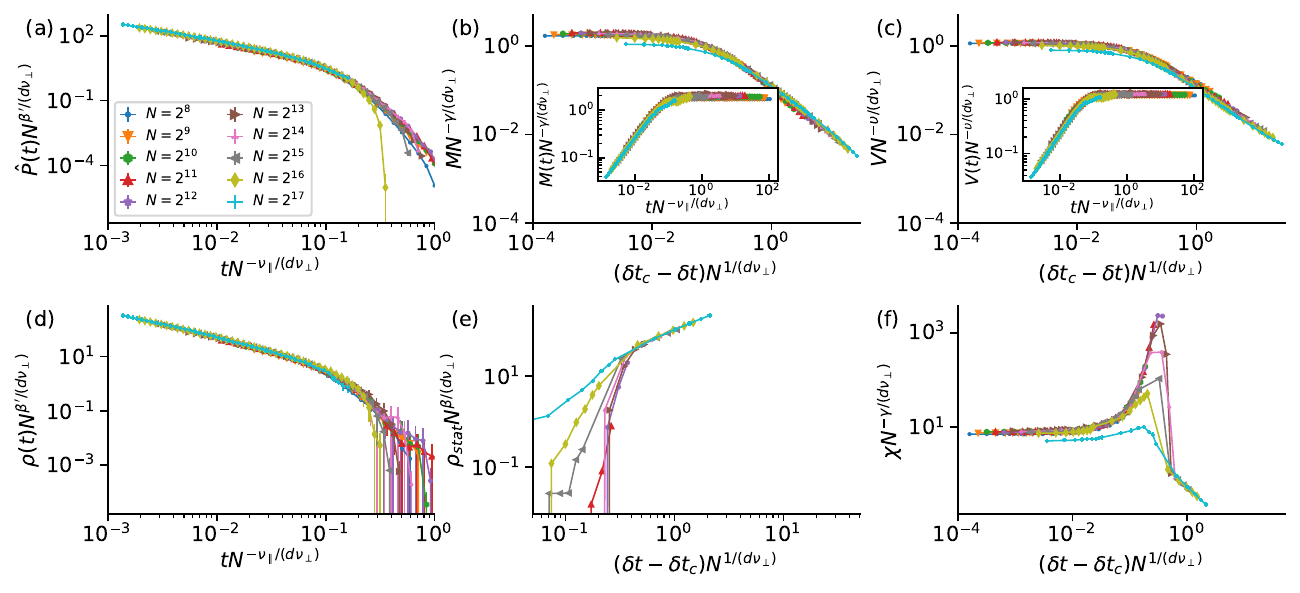}
    \caption{Random Erdős–Rényi network $\langle k \rangle = 8$ with Poisson link activation with mean inter-event time of 1.}
    \label{fig:erdos-poisson-scaling}
\end{figure}

\begin{figure}[htbp!]
    \centering
    \includegraphics[width=\linewidth]{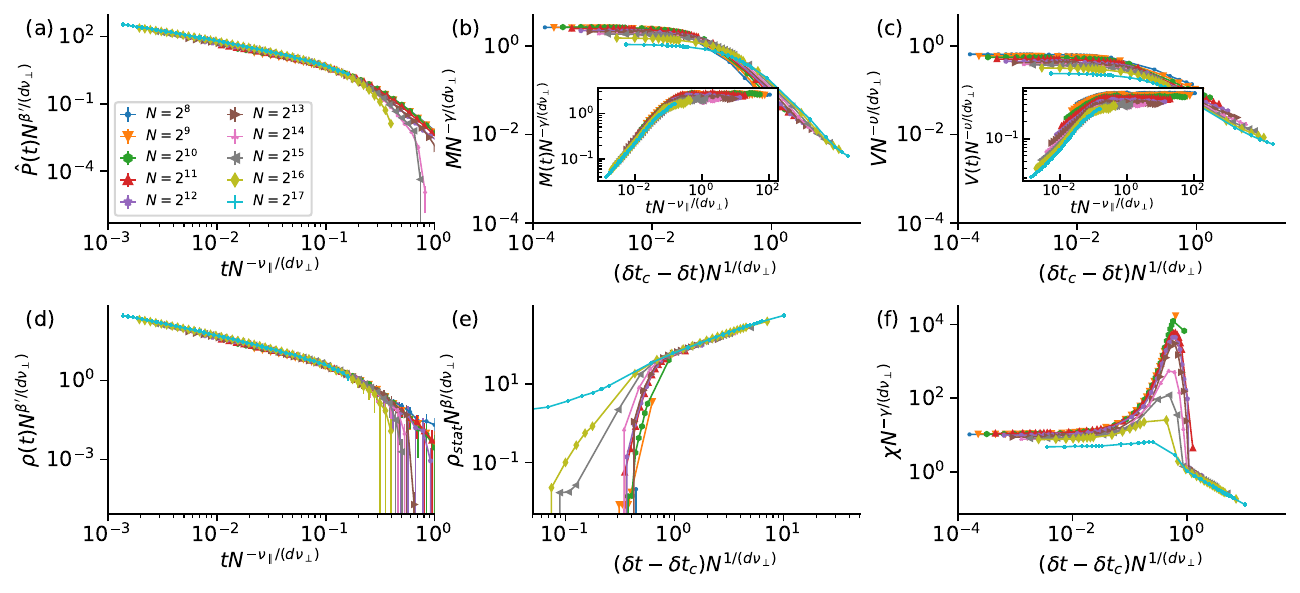}
    \caption{Random Erdős–Rényi network $\langle k \rangle = 8$ with bursty link activation (i.e. power-law inter-event time distribution with minimum cutoff) with mean inter-event time of 1 and inter-event time exponents of 2.05.}
    \label{fig:erdos-bursty-scaling-2.05}
\end{figure}
\begin{figure}[htbp!]
    \centering
    \includegraphics[width=\linewidth]{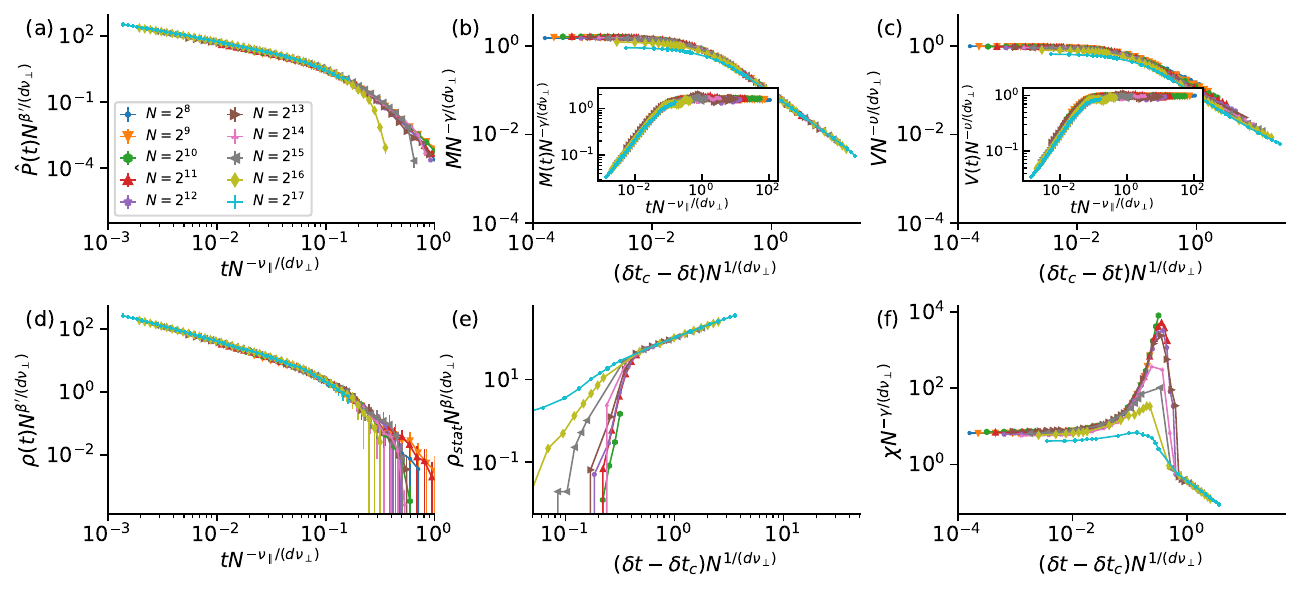}
    \caption{Random Erdős–Rényi network $\langle k \rangle = 8$ with bursty link activation (i.e. power-law inter-event time distribution with minimum cutoff) with mean inter-event time of 1 and inter-event time exponents of 2.20.}
    \label{fig:erdos-bursty-scaling-2.2}
\end{figure}
\begin{figure}[htbp!]
    \centering
    \includegraphics[width=\linewidth]{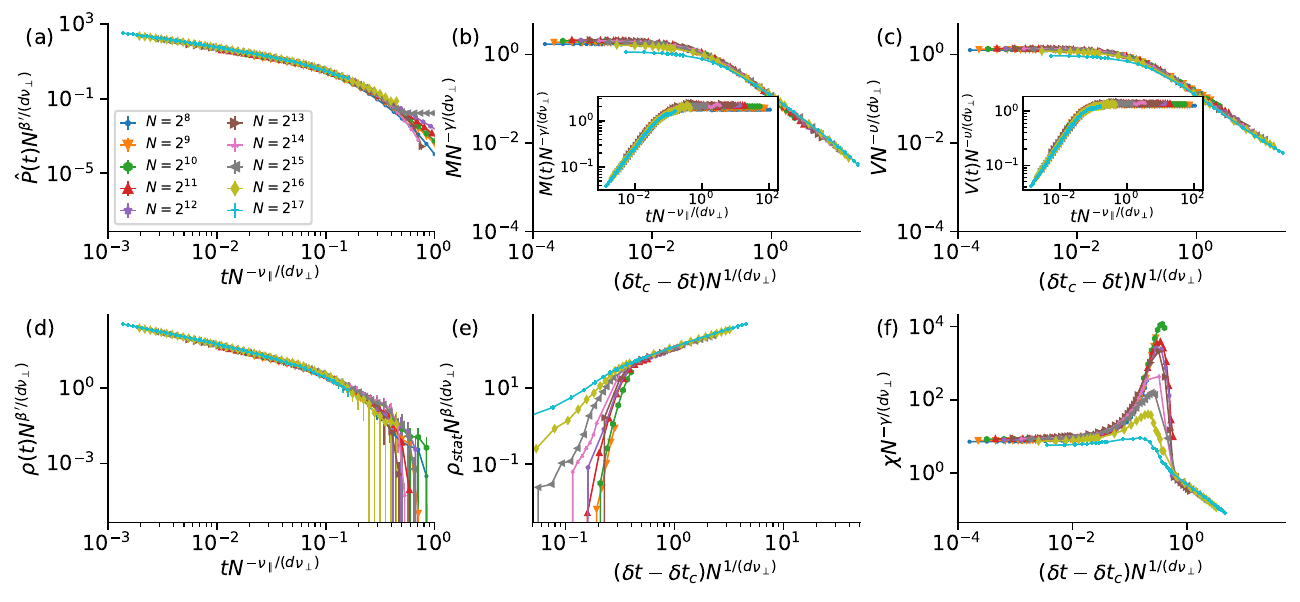}
    \caption{Random Erdős–Rényi network $\langle k \rangle = 8$ with bursty link activation (i.e. power-law inter-event time distribution with minimum cutoff) with mean inter-event time of 1 and inter-event time exponents of 2.80.}
    \label{fig:erdos-bursty-scaling-2.8}
\end{figure}
\begin{figure}[htbp!]
    \centering
    \includegraphics[width=\linewidth]{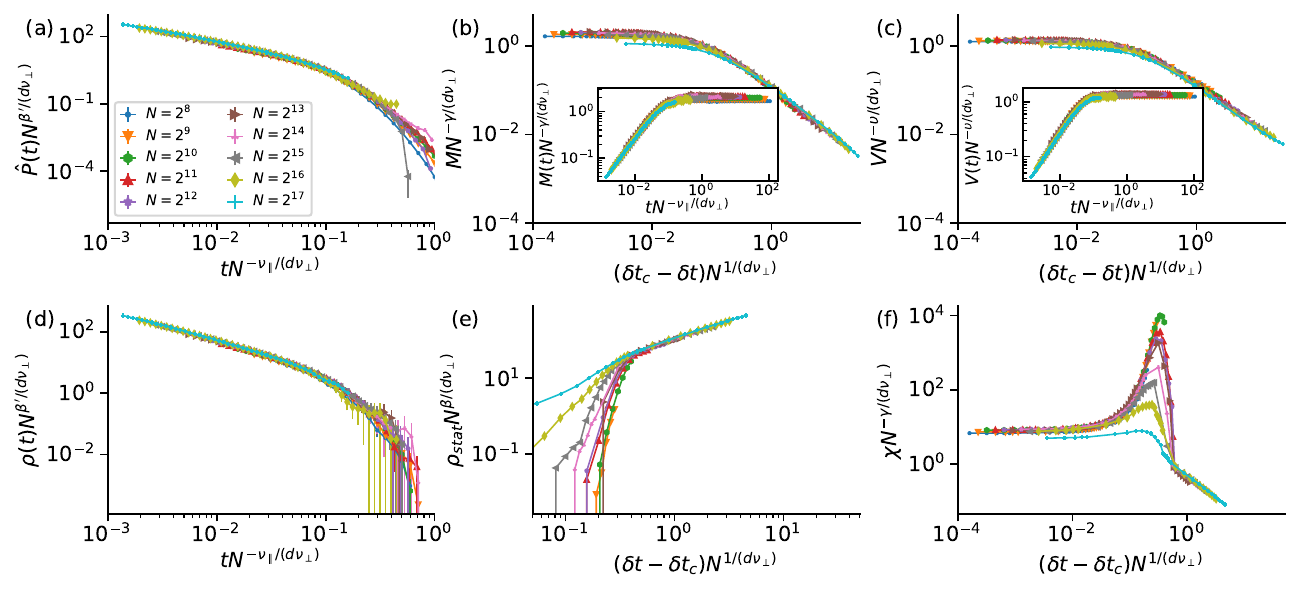}
    \caption{Random Erdős–Rényi network $\langle k \rangle = 8$ with bursty link activation (i.e. power-law inter-event time distribution with minimum cutoff) with mean inter-event time of 1 and inter-event time exponents of 5.2.}
    \label{fig:erdos-bursty-scaling-5.2}
\end{figure}
\begin{figure}[htbp!]
    \centering
    \includegraphics[width=\linewidth]{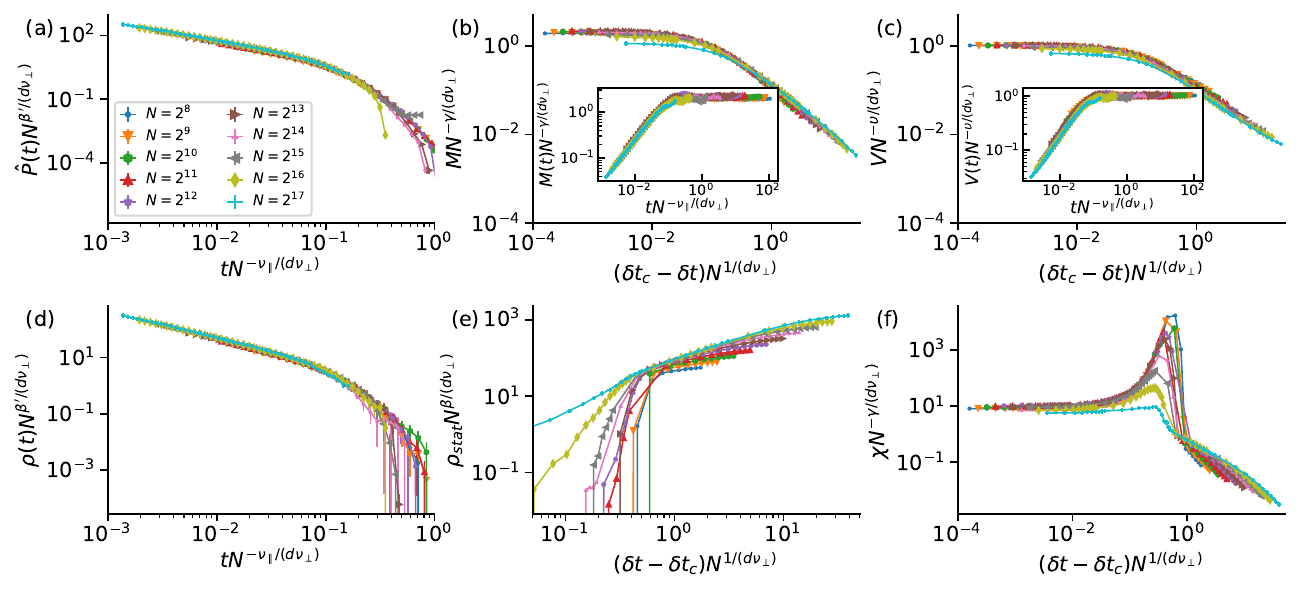}
    \caption{Random Erdős–Rényi network with self-exciting Hawkes process with parameters $\mu=0.2$, $\alpha=0.8$ and $\theta=0.5$.}
    \label{fig:erdos-self-exciting-scaling-0.2-0.8-0.5}
\end{figure}
\begin{figure}[htbp!]
    \centering
    \includegraphics[width=\linewidth]{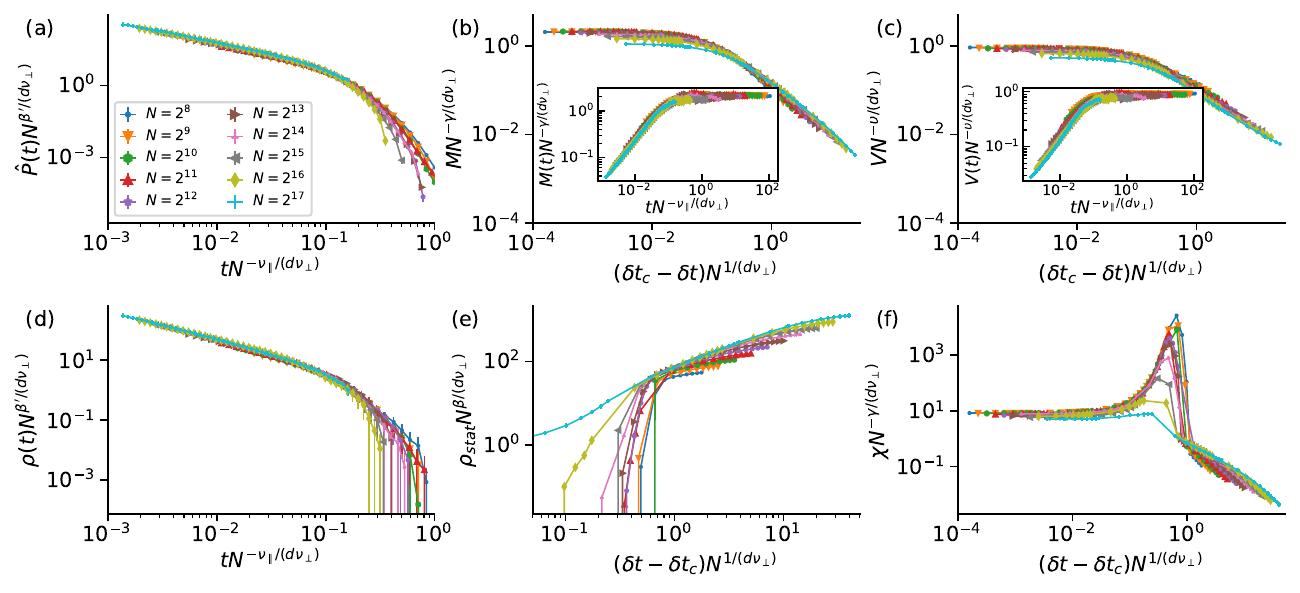}
    \caption{Random Erdős–Rényi network with self-exciting Hawkes process with parameters $\mu=0.2$, $\alpha=0.8$ and $\theta=1.0$.}
    \label{fig:erdos-self-exciting-scaling-0.2-0.8-1.0}
\end{figure}
\begin{figure}[htbp!]
    \centering
    \includegraphics[width=\linewidth]{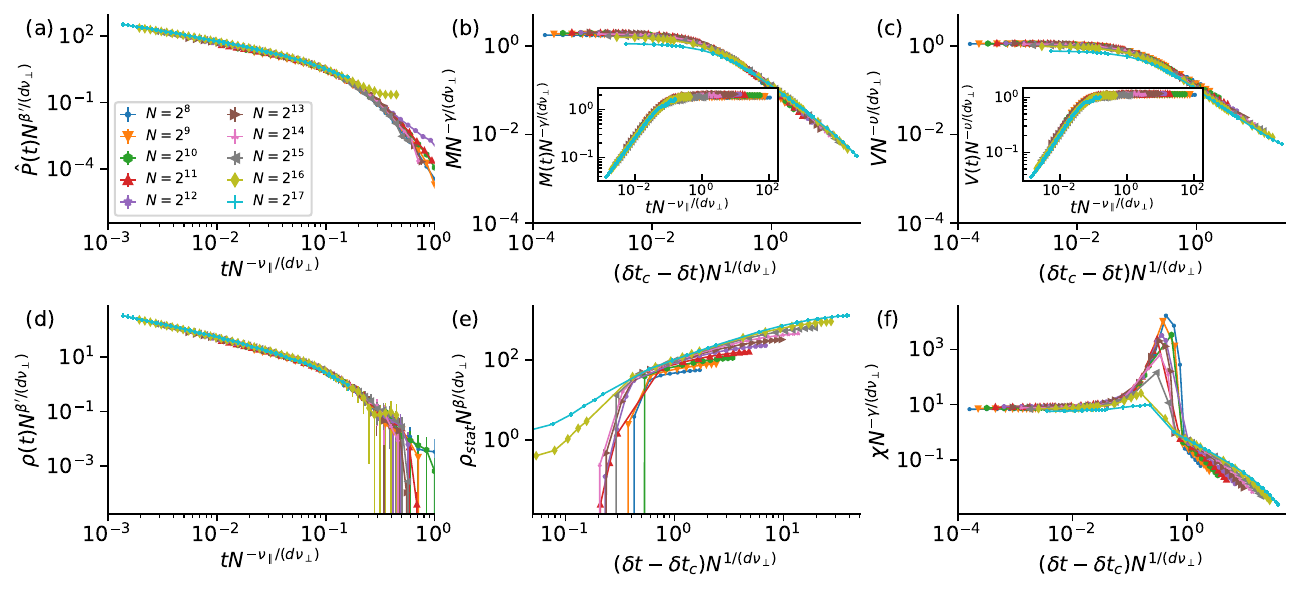}
    \caption{Random Erdős–Rényi network with self-exciting Hawkes process with parameters $\mu=0.5$, $\alpha=0.5$ and $\theta=0.5$.}
    \label{fig:erdos-self-exciting-scaling-0.5-0.5-0.5}
\end{figure}
\begin{figure}[htbp!]
    \centering
    \includegraphics[width=\linewidth]{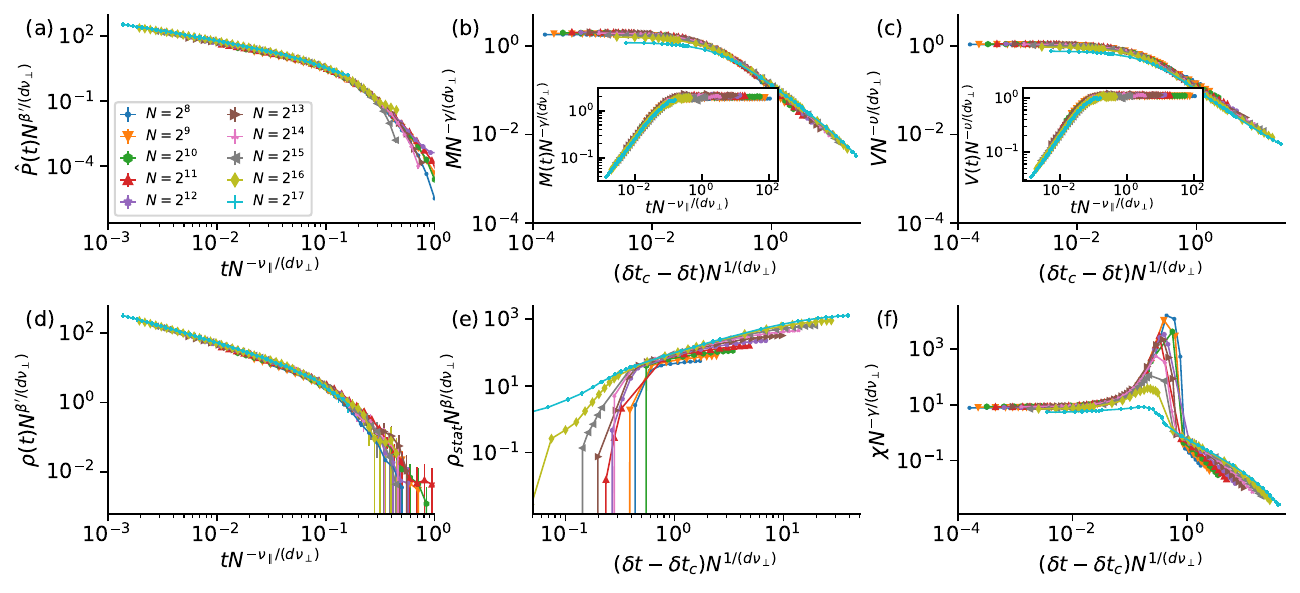}
    \caption{Random Erdős–Rényi network with self-exciting Hawkes process with parameters $\mu=0.5$, $\alpha=0.5$ and $\theta=1.0$.}
    \label{fig:erdos-self-exciting-scaling-0.5-0.5-1.0}
\end{figure}
\begin{figure}[htbp!]
    \centering
    \includegraphics[width=\linewidth]{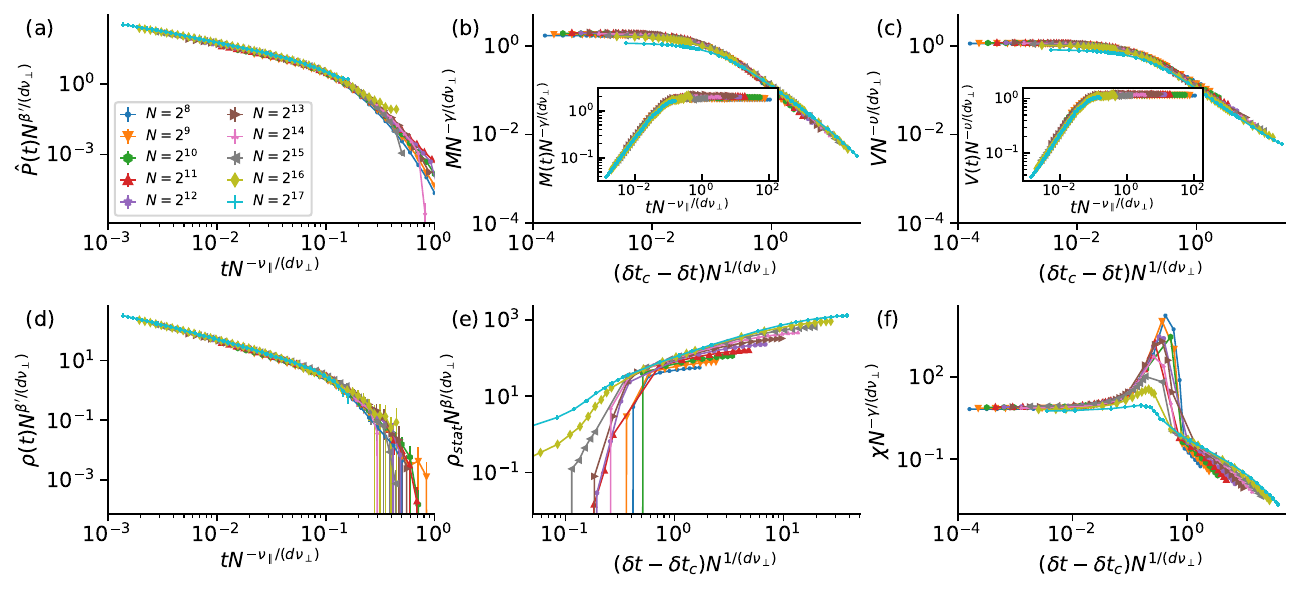}
    \caption{Random Erdős–Rényi network with self-exciting Hawkes process with parameters $\mu=0.8$, $\alpha=0.2$ and $\theta=0.5$.}
    \label{fig:erdos-self-exciting-scaling-0.8-0.2-0.5}
\end{figure}
\begin{figure}[htbp!]
    \centering
    \includegraphics[width=\linewidth]{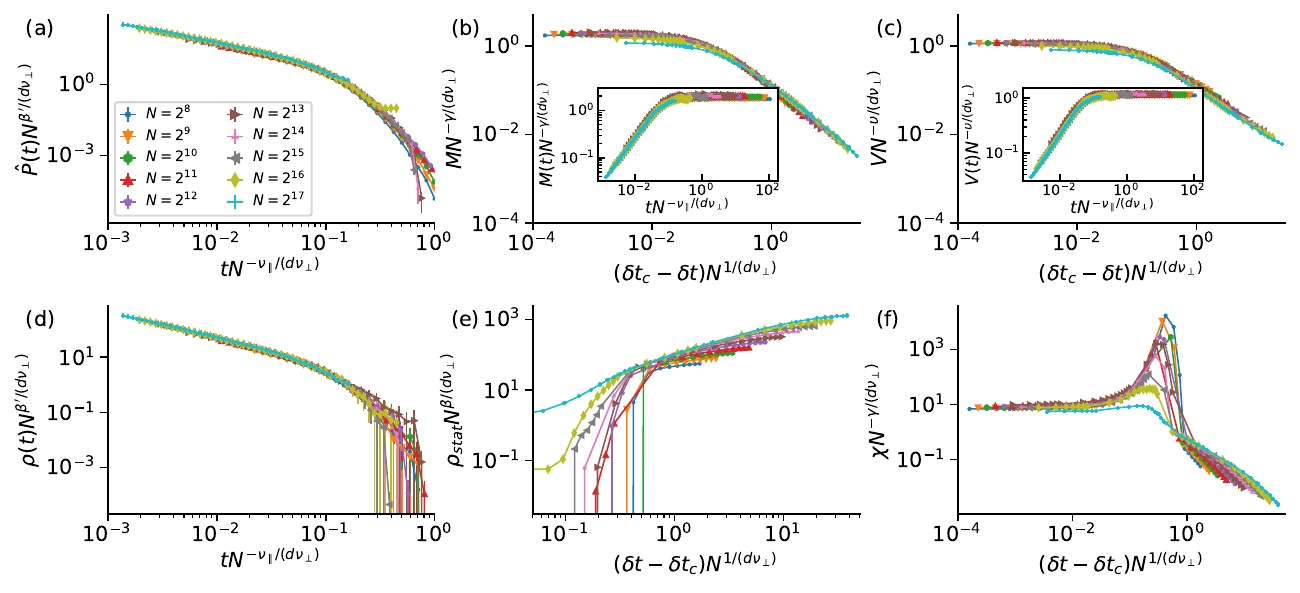}
    \caption{Random Erdős–Rényi network with self-exciting Hawkes process with parameters $\mu=0.8$, $\alpha=0.2$ and $\theta=1.0$.}
    \label{fig:erdos-self-exciting-scaling-0.8-0.2-1.0}
\end{figure}

\section{Full suit of empirical results for random regular based temporal networks}

\begin{figure}[htbp!]
    \centering
    \includegraphics[width=0.65\linewidth]{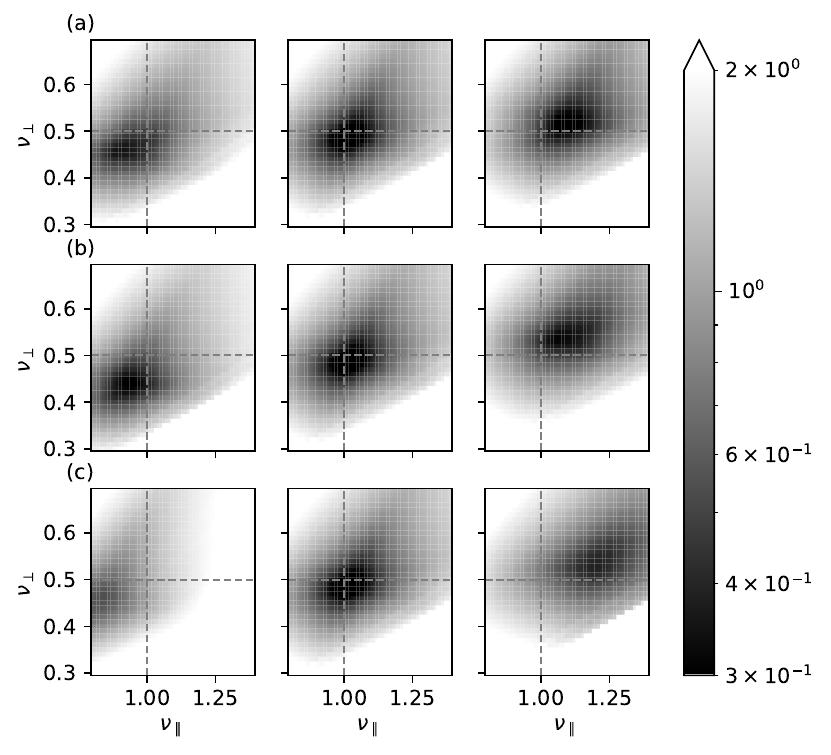}
    \caption{Total error of collapse of $M(t)$, $V(t)$, $\hat{P}(t)$ and $\rho(t)$ universal scaling functions for random 9-regular networks and Poisson process activation $\lambda = 1$ with $\nu_\perp$ and $\nu_\parallel$ values assuming (a) $\beta \in \{0.84, 1, 1.16\}$, (b) $\beta' \in \{0.84, 1, 1.16\}$ and (c) $\delta t_c \in \{0.0878,0.08808,0.0883\}$ which shows a minimum around $\beta = \beta' = \nu_\parallel = 1$ and $\nu_\perp=0.5$}
    \label{fig:regular-grid-search-nu-para-nu-perp}
\end{figure}
\begin{figure}[htbp!]
    \centering
    \includegraphics[width=0.65\linewidth]{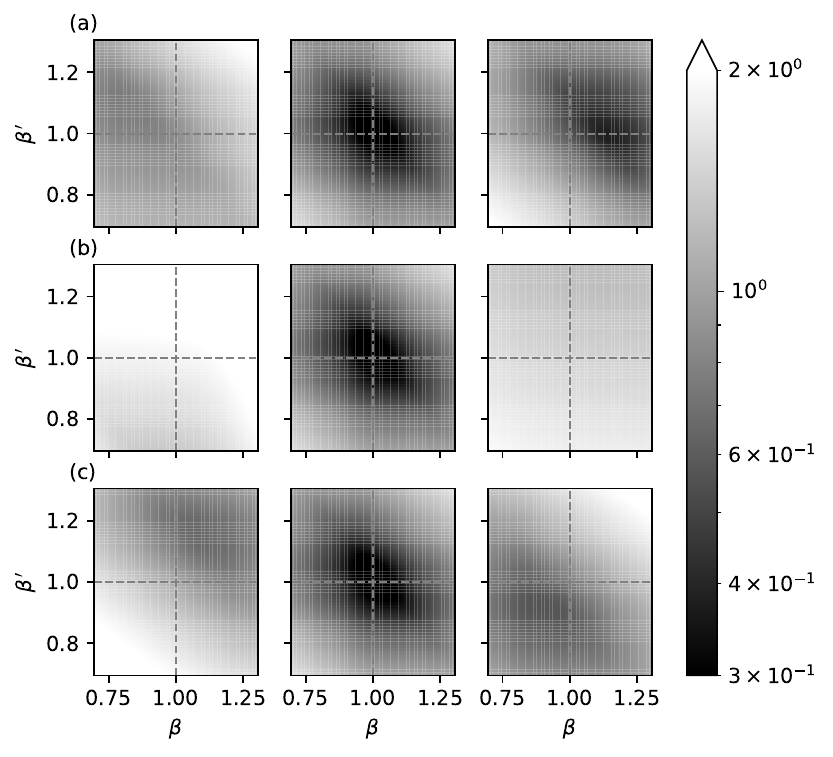}
    \caption{Total error of collapse of $M(t)$, $V(t)$, $\hat{P}(t)$ and $\rho(t)$ universal scaling functions for random 9-regular networks and Poisson process activation $\lambda = 1$ with $\beta$ and $\beta'$ values assuming (a) $\nu_\perp \in \{0.34, 0.5, 0.66\}$, (b) $\nu_\parallel \in \{0.84, 1, 1.16\}$ and (c) $\delta t_c \in \{0.0878,0.08808,0.0883\}$ which shows a minimum around $\beta = \beta' = \nu_\parallel = 1$ and $\nu_\perp=0.5$}
    \label{fig:regular-grid-search-beta-beta-prime}
\end{figure}

\begin{figure}[htbp!]
    \centering
    \includegraphics[width=\linewidth]{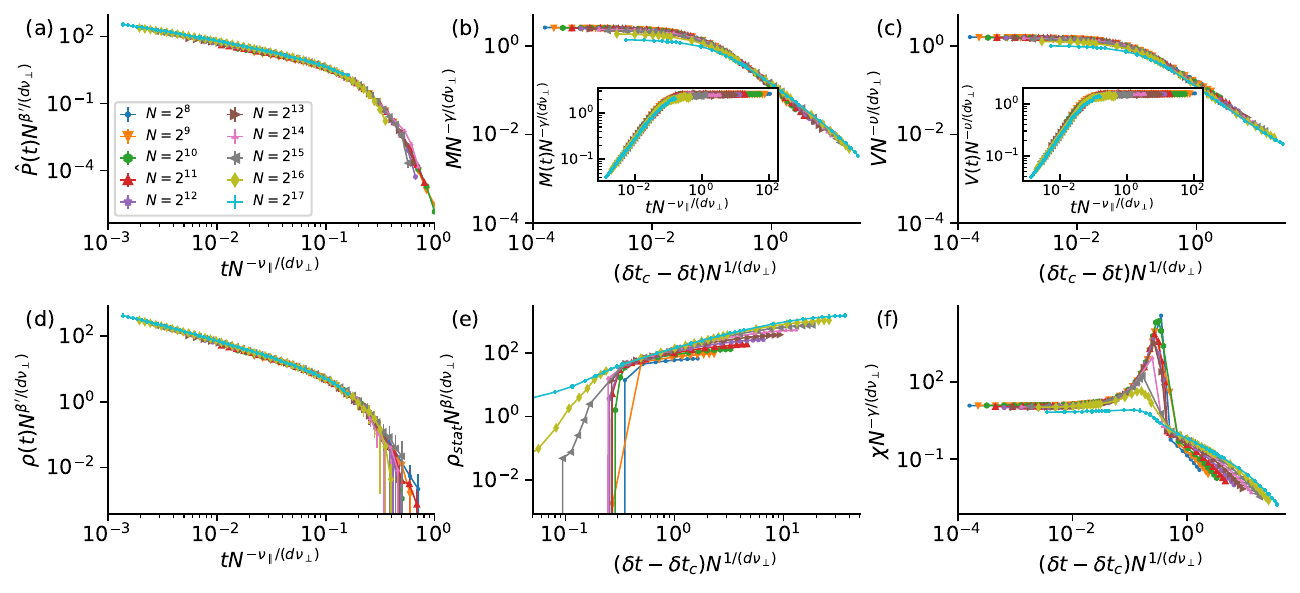}
    \caption{Random 9-regular network with Poisson link activation with mean inter-event time of 1.}
    \label{fig:regular-poisson-scaling}
\end{figure}

\begin{figure*}[htbp!]
    \centering
    \includegraphics[width=1\linewidth]{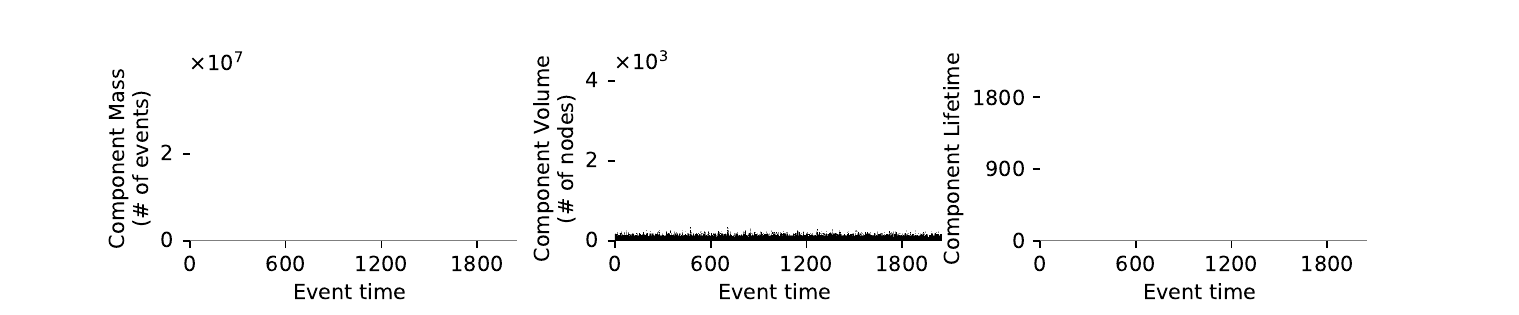}
    \includegraphics[width=1\linewidth]{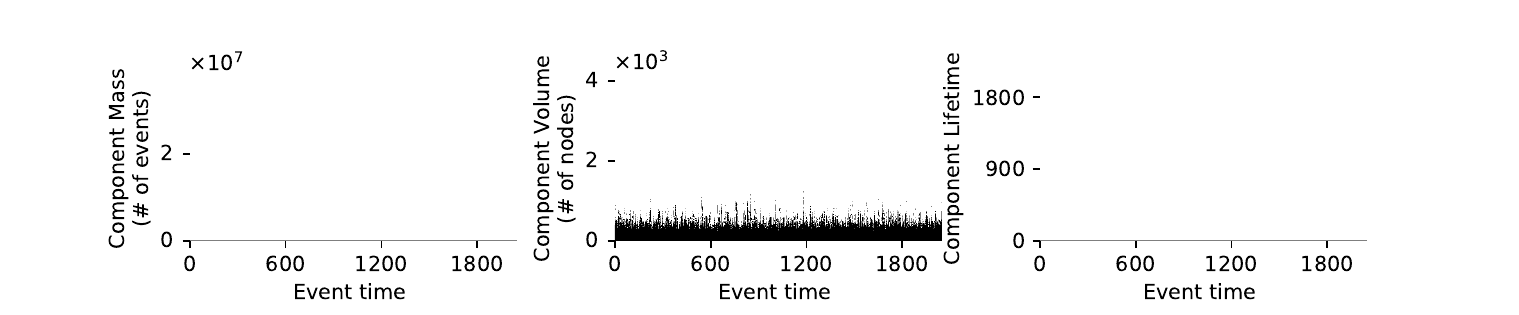}
    \includegraphics[width=1\linewidth]{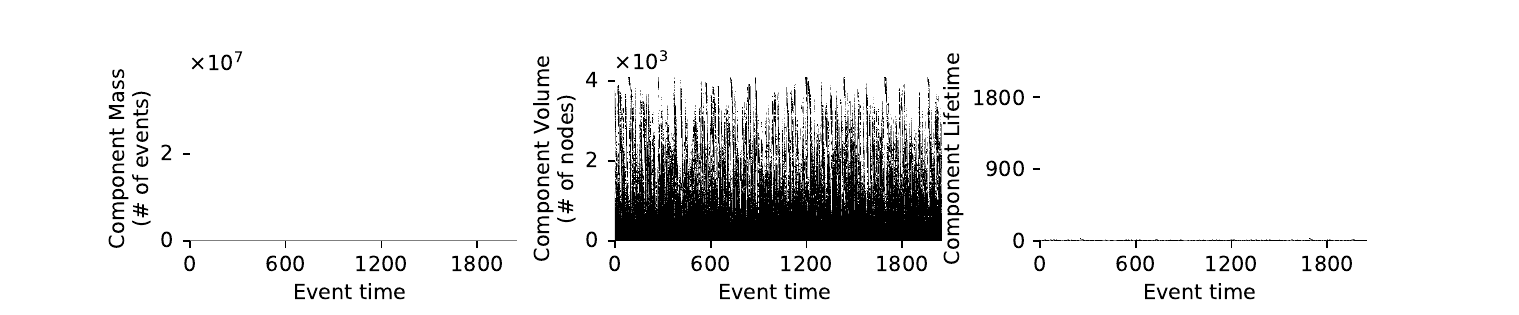}
    \includegraphics[width=1\linewidth]{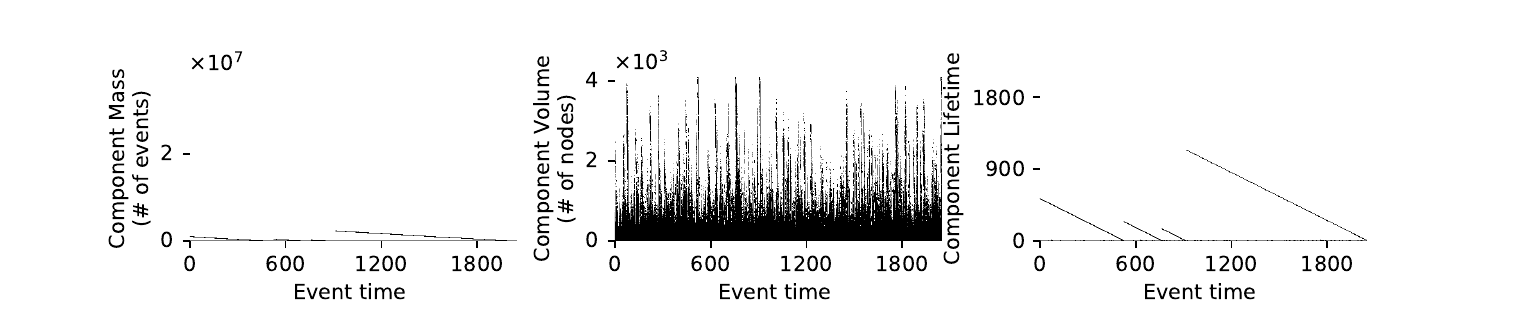}
    \includegraphics[width=1\linewidth]{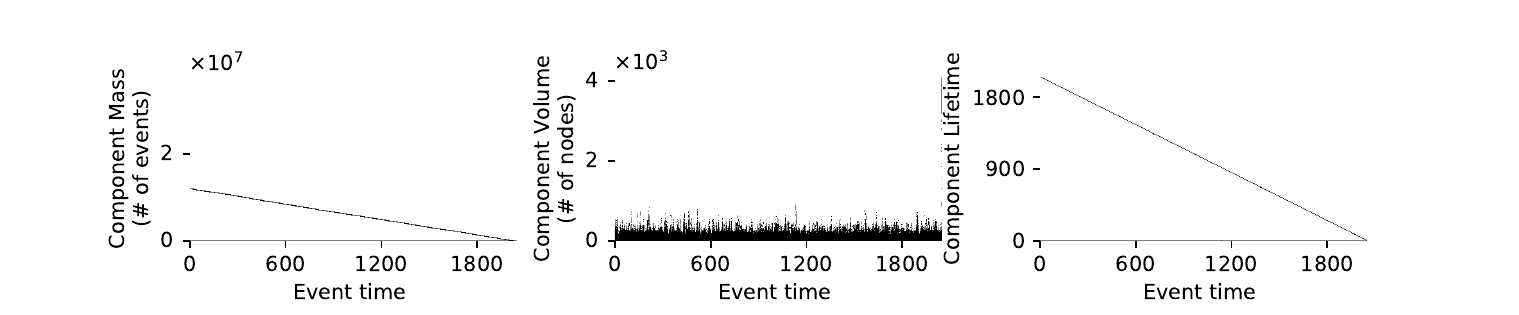}
    \caption{Out-component size estimates of all events of a 9-regular network with Poisson process link activations $\lambda=1$ for (a) $\delta t=0.07$, (b) $\delta t=0.08$, (c) $\delta t=0.08802$, (d) $\delta t=0.092$ and (e) $\delta t=0.1$.}
    \label{fig:regular-out-component-sizes}
\end{figure*}

\begin{figure}[htbp!]
    \centering
    \includegraphics[width=\linewidth]{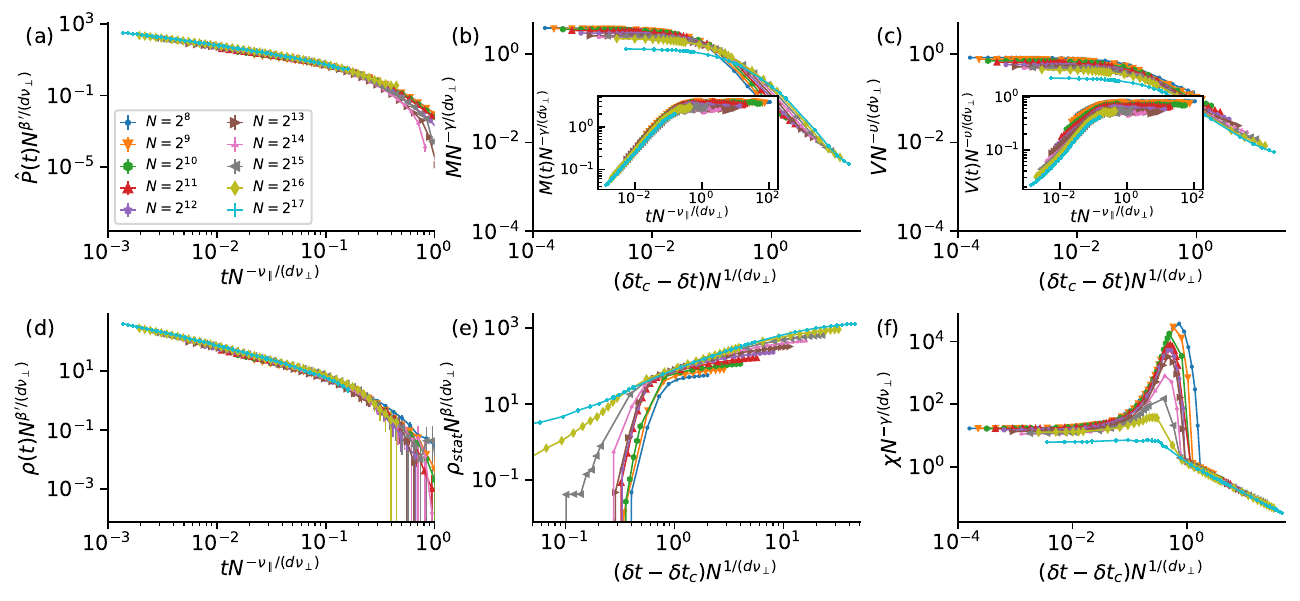}
    \caption{Random 9-regular network with bursty link activation (i.e. power-law inter-event time distribution with minimum cutoff) with mean inter-event time of 1 and inter-event time exponents of 2.05.}
    \label{fig:regular-bursty-scaling-2.05}
\end{figure}
    
\begin{figure}[htbp!]
    \centering
    \includegraphics[width=\linewidth]{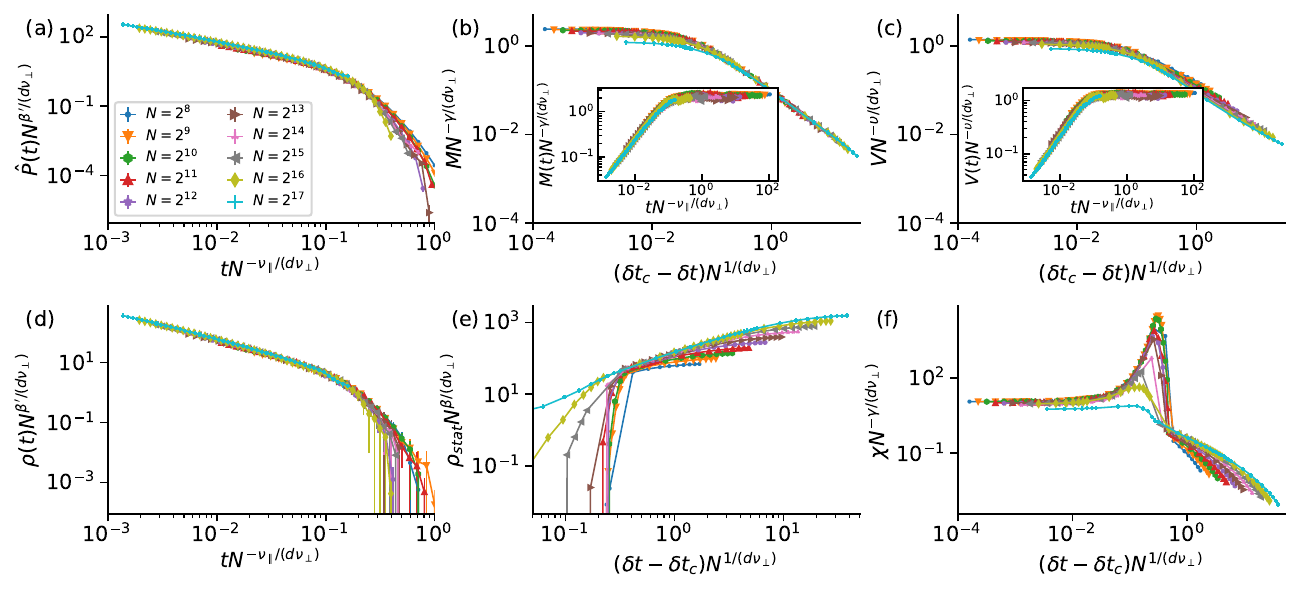}
    \caption{Random 9-regular network with bursty link activation (i.e. power-law inter-event time distribution with minimum cutoff) with mean inter-event time of 1 and inter-event time exponents of 2.2.}
    \label{fig:regular-bursty-scaling-2.2}
\end{figure}

\begin{figure}[htbp!]
    \centering
    \includegraphics[width=\linewidth]{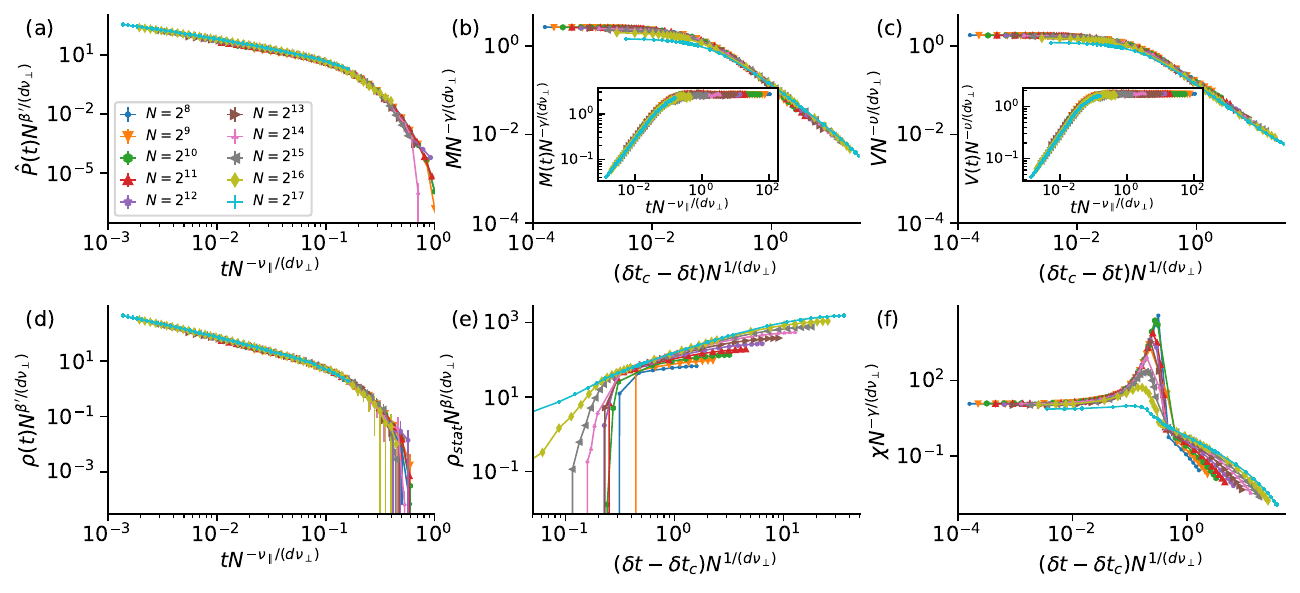}
    \caption{Random 9-regular network with bursty link activation (i.e. power-law inter-event time distribution with minimum cutoff) with mean inter-event time of 1 and inter-event time exponents of 2.8.}
    \label{fig:regular-bursty-scaling-2.8}
\end{figure}

\begin{figure}[htbp!]
    \centering
    \includegraphics[width=\linewidth]{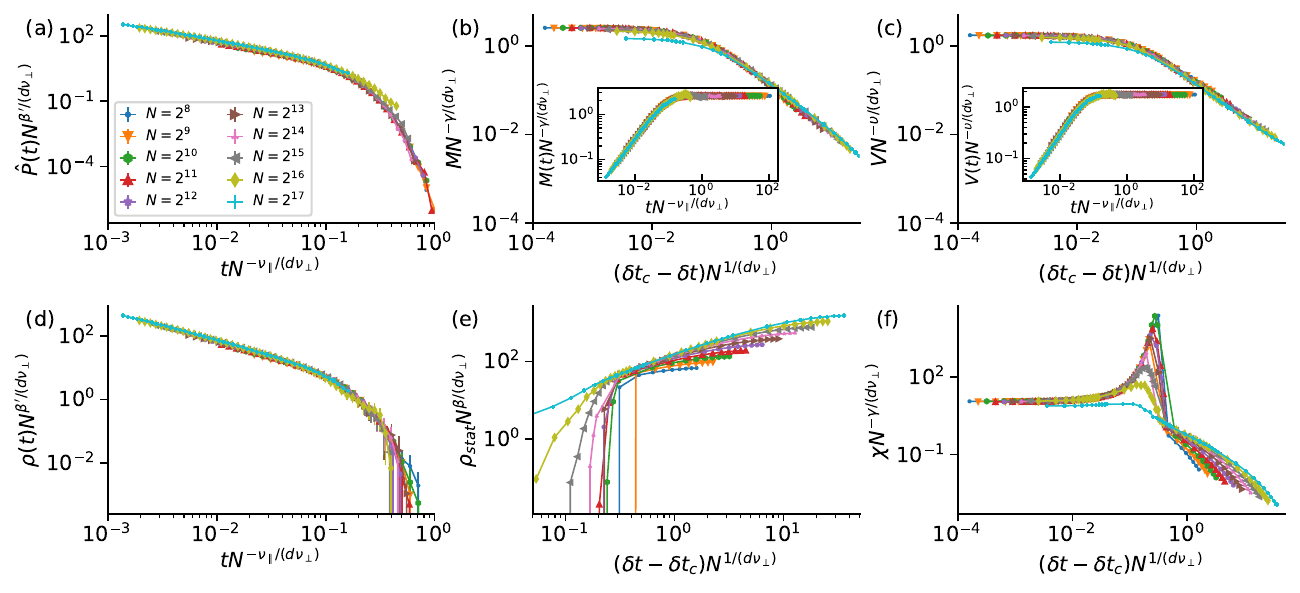}
    \caption{Random 9-regular network with bursty link activation (i.e. power-law inter-event time distribution with minimum cutoff) with mean inter-event time of 1 and inter-event time exponents of 5.2.}
    \label{fig:regular-bursty-scaling-5.2}
\end{figure}

\begin{figure}[htbp!]
    \centering
    \includegraphics[width=\linewidth]{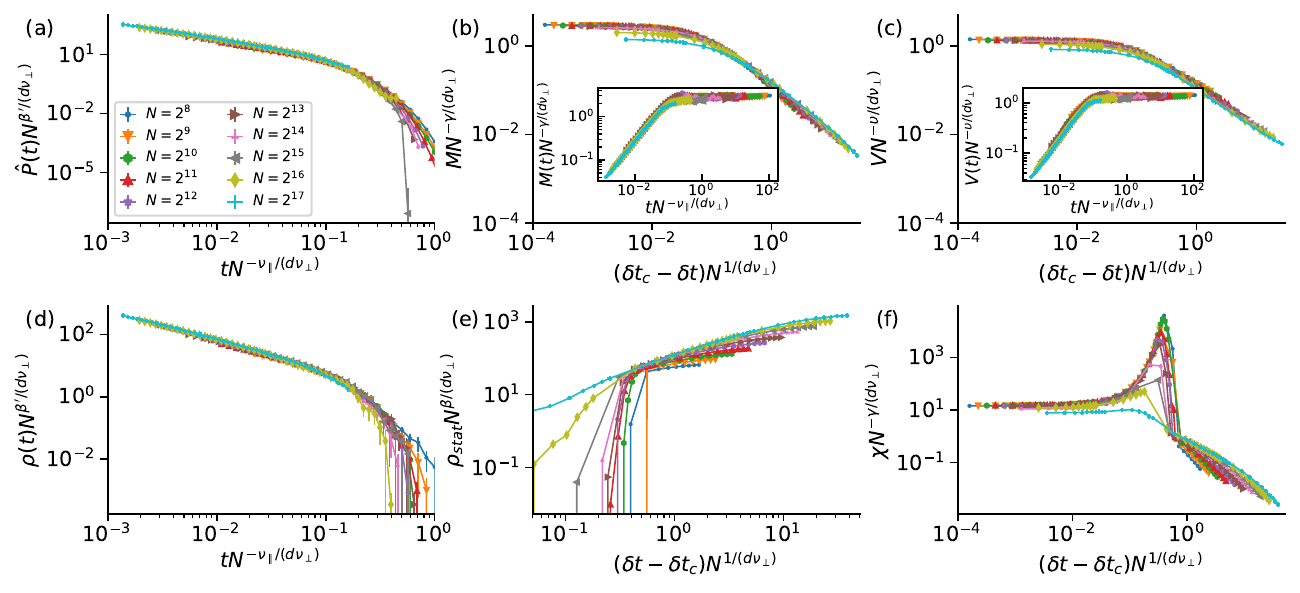}
    \caption{Random 9-regular network with self-exciting Hawkes process with parameters $\mu=0.2$, $\alpha=0.8$ and $\theta=0.50$.}
    \label{fig:regular-self-exciting-scaling-0.2-0.8-0.5}
\end{figure}
\begin{figure}[htbp!]
    \centering
    \includegraphics[width=\linewidth]{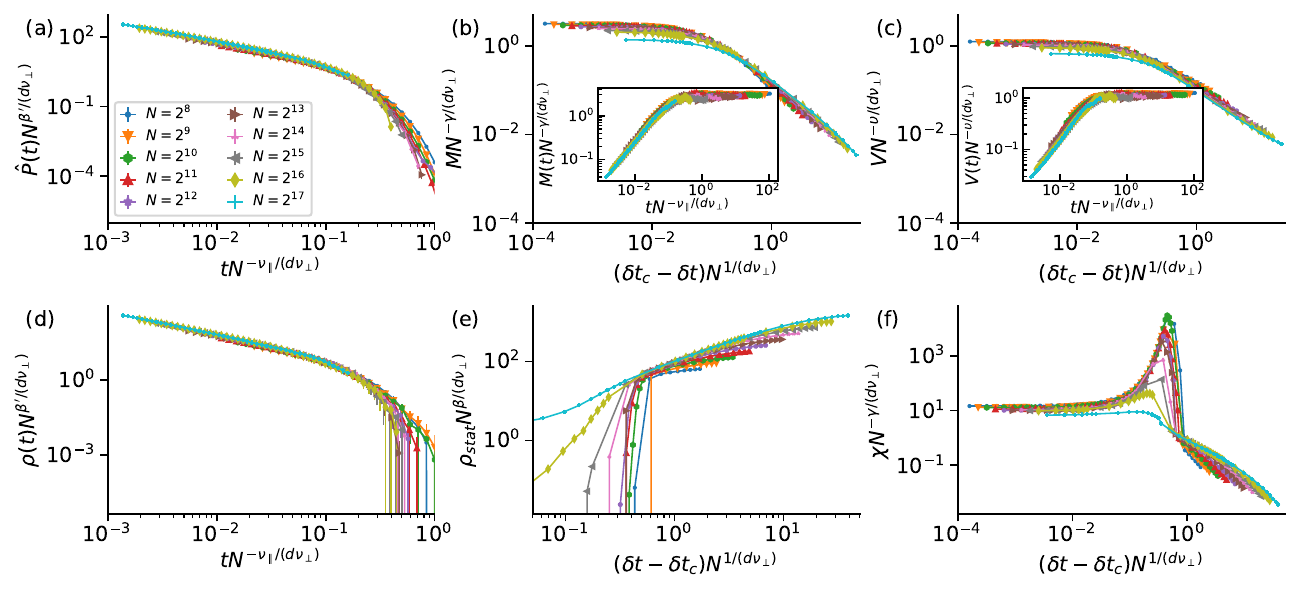}
    \caption{Random 9-regular network with self-exciting Hawkes process with parameters $\mu=0.2$, $\alpha=0.8$ and $\theta=1.0$.}
    \label{fig:regular-self-exciting-scaling-0.2-0.8-1.0}
\end{figure}
\begin{figure}[htbp!]
    \centering
    \includegraphics[width=\linewidth]{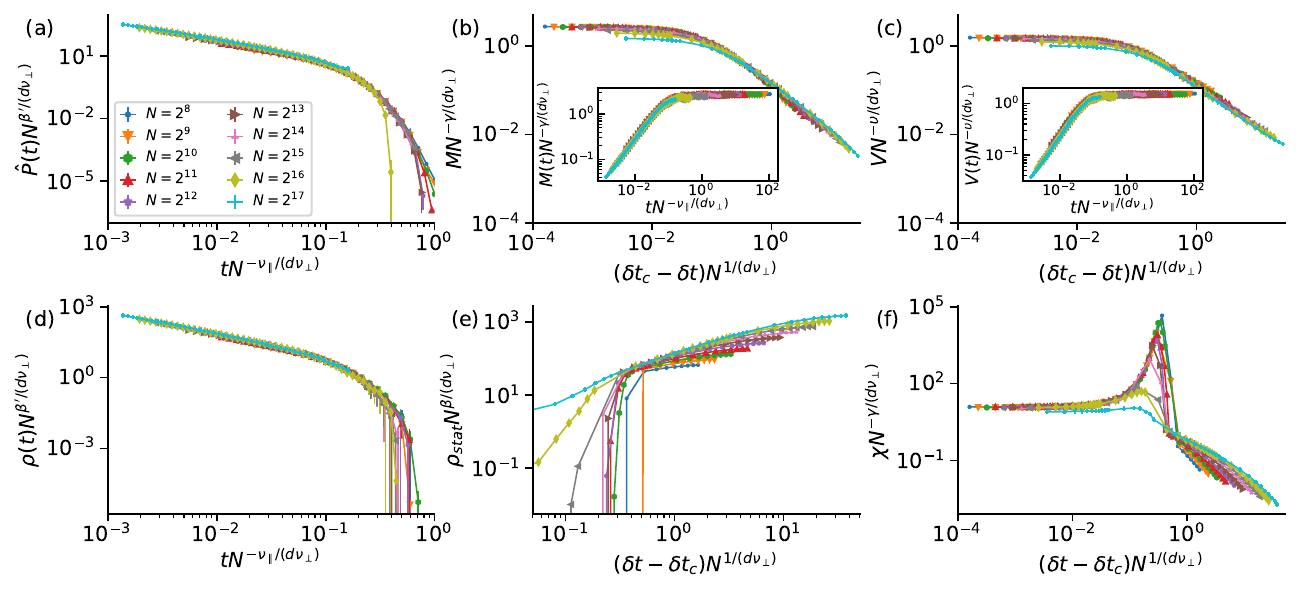}
    \caption{Random 9-regular network with self-exciting Hawkes process with parameters $\mu=0.5$, $\alpha=0.5$ and $\theta=0.5$.}
    \label{fig:regular-self-exciting-scaling-0.5-0.5-0.5}
\end{figure}
\begin{figure}[htbp!]
    \centering
    \includegraphics[width=\linewidth]{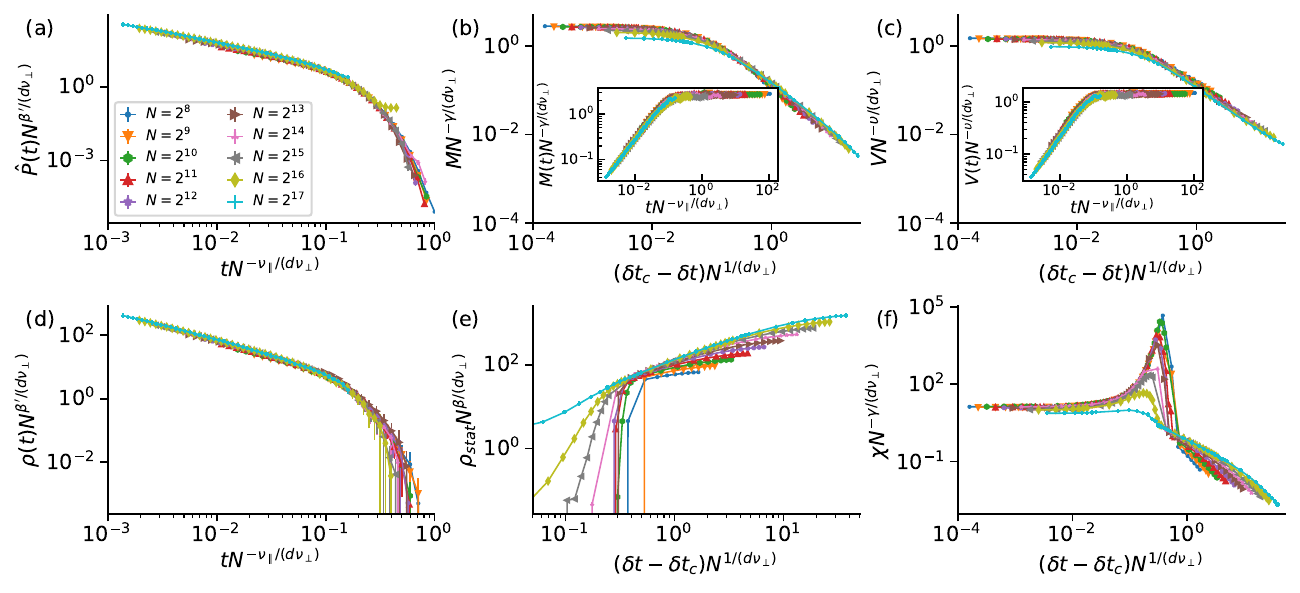}
    \caption{Random 9-regular network with self-exciting Hawkes process with parameters $\mu=0.5$, $\alpha=0.5$ and $\theta=1.0$.}
    \label{fig:regular-self-exciting-scaling-0.5-0.5-1.0}
\end{figure}
\begin{figure}[htbp!]
    \centering
    \includegraphics[width=\linewidth]{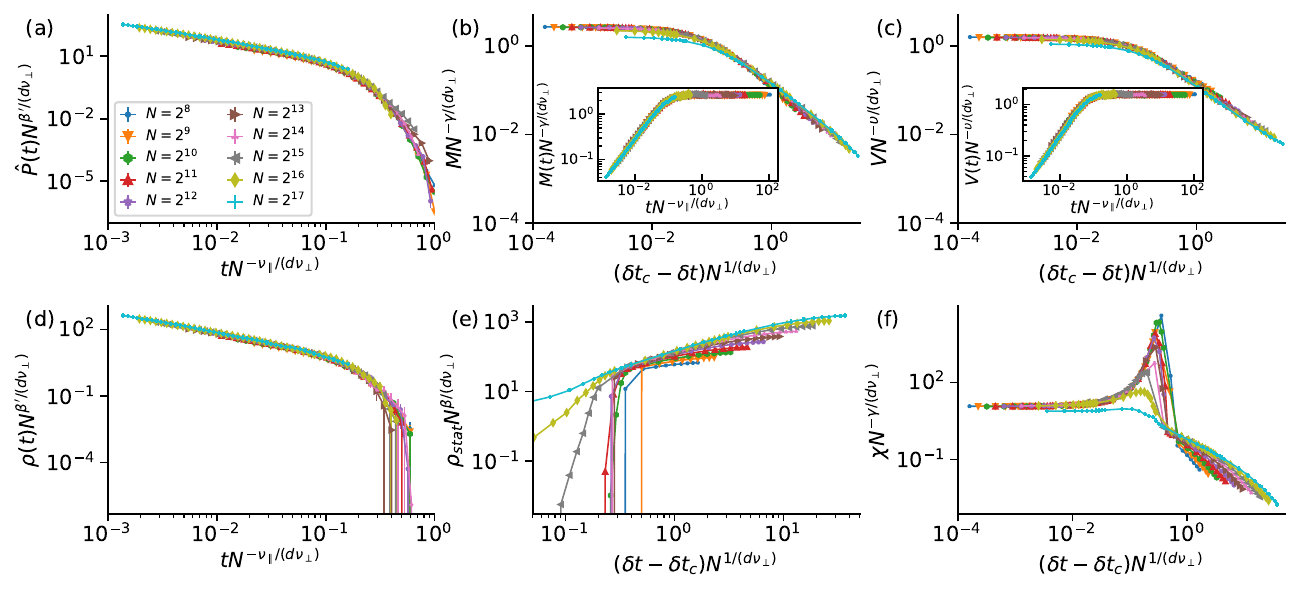}
    \caption{Random 9-regular network with self-exciting Hawkes process with parameters $\mu=0.8$, $\alpha=0.2$ and $\theta=0.5$.}
    \label{fig:regular-self-exciting-scaling-0.8-0.2-0.5}
\end{figure}
\begin{figure}[htbp!]
    \centering
    \includegraphics[width=\linewidth]{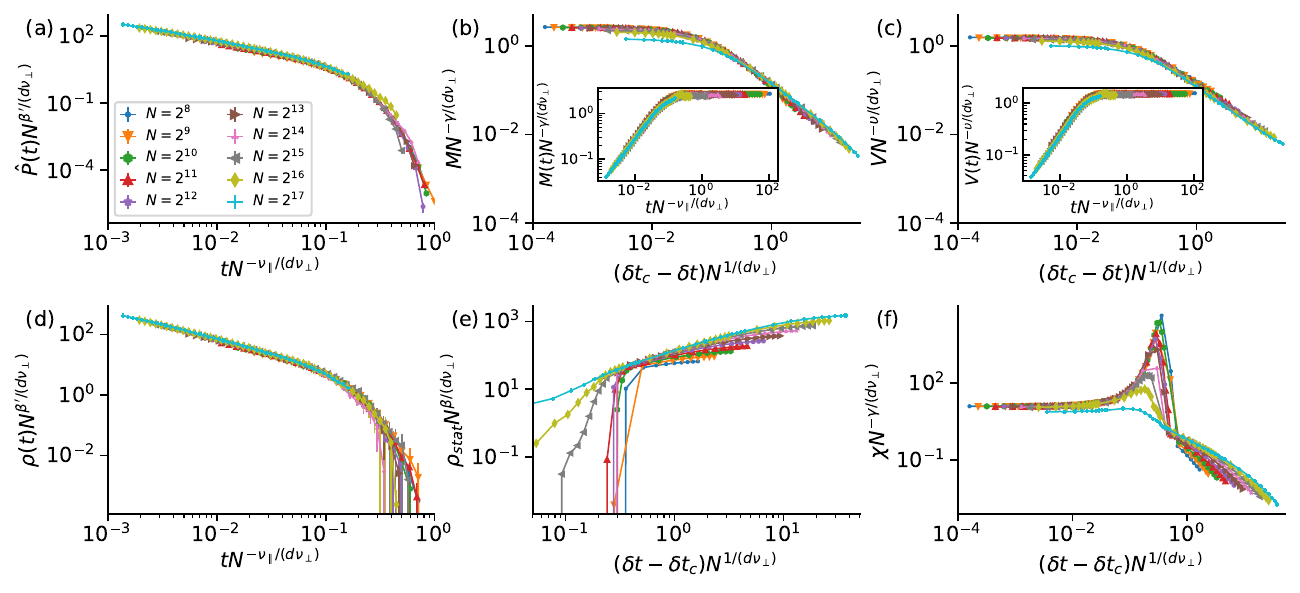}
    \caption{Random 9-regular network with self-exciting Hawkes process with parameters $\mu=0.8$, $\alpha=0.2$ and $\theta=1.0$.}
    \label{fig:regular-self-exciting-scaling-0.8-0.2-1.0}
\end{figure}

\section{Full suit of empirical results for grid lattice based temporal networks}
Grid lattices is generated as a line, square-shaped, cubic and hyper-cubic with periodic boundary condition. Each system has same linear size across all dimensions, as presented in Tabs.~\ref{tab:experiment-parameters-grid-1}, \ref{tab:experiment-parameters-grid-2}, \ref{tab:experiment-parameters-grid-3} and \ref{tab:experiment-parameters-grid-4}.

\begin{table}[htbp!]
\caption{\label{tab:experiment-parameters-grid-1} Experimental setup for 1-dimensional grid.}
\begin{ruledtabular}
\begin{tabular}{rrr}
\multicolumn{1}{c}{\textrm{Size $N$}} &
\multicolumn{1}{c}{\textrm{Time window $T$}} &
\multicolumn{1}{c}{\textrm{Realisations}}\\
\colrule
256 & 8192 & 4096 \\
512 & 8192 & 4096 \\
1024 & 8192 & 4096 \\
2048 & 4096 & 4096 \\
4096 & 4096 & 1024 \\
8192 & 4096 & 512 \\
16384 & 2048 & 512 \\
32768 & 1024 & 512 \\
65536 & 512 & 256 \\
131072 & 256 & 256 \\
\end{tabular}
\end{ruledtabular}
\end{table}

\begin{figure}[htbp!]
    \centering
    \includegraphics[width=0.65\linewidth]{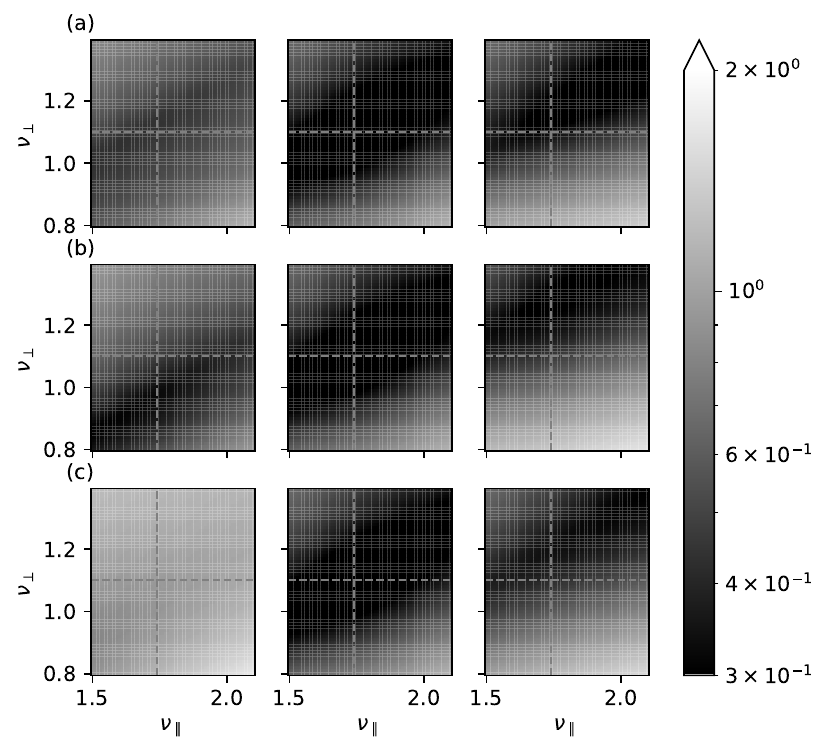}
    \caption{Total error of collapse of $M(t)$, $V(t)$, $\hat{P}(t)$ and $\rho(t)$ universal scaling functions for 1D grid lattice networks $\langle k \rangle = 8$ and Poisson process activation $\lambda = 1$ with $\nu_\perp$ and $\nu_\parallel$ values assuming (a) $\beta \in \{0.12, 0.28, 0.44\}$, (b) $\beta' \in \{0.12, 0.28, 0.44\}$ and (c) $\delta t_c \in \{0.98,0.9919,1.0\}$ which shows a minimum around $\beta = \beta' = 0.28$, $\nu_\parallel = 1.74$ and $\nu_\perp=1.10$}
    \label{fig:grid-1-grid-search-nu-para-nu-perp}
\end{figure}

\begin{figure}[htbp!]
    \centering
    \includegraphics[width=0.65\linewidth]{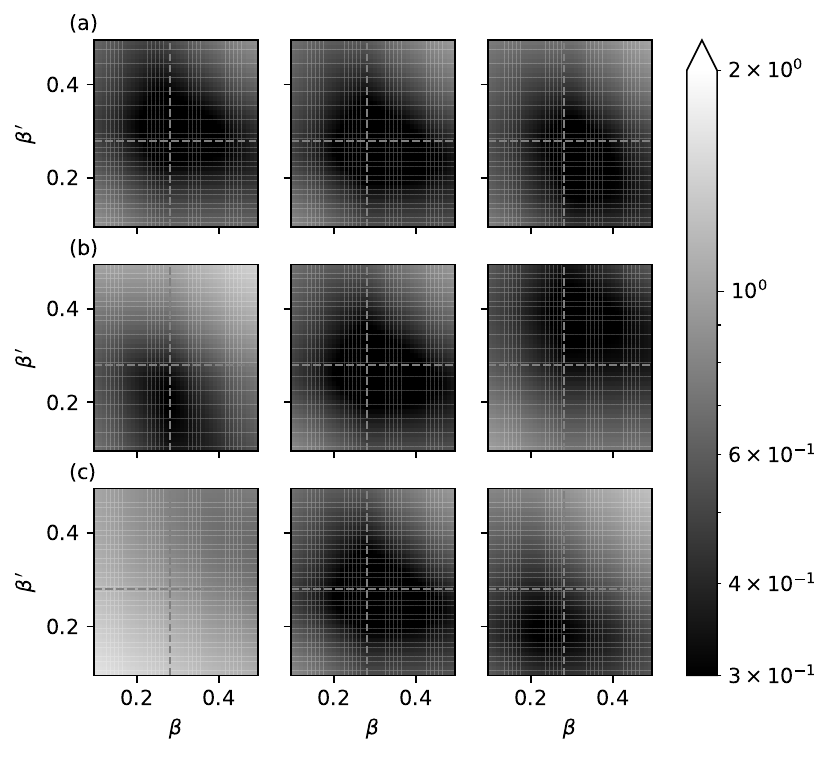}
    \caption{Total error of collapse of $M(t)$, $V(t)$, $\hat{P}(t)$ and $\rho(t)$ universal scaling functions for 1D grid lattice networks $\langle k \rangle = 8$ and Poisson process activation $\lambda = 1$ with $\beta$ and $\beta'$ values assuming (a) $\nu_\perp \in \{0.94, 1.10, 1.26\}$, (b) $\nu_\parallel \in \{1.58, 1.74, 1.90\}$ and (c) $\delta t_c \in \{0.98,0.9919,1.0\}$ which shows a minimum around $\beta = \beta' = 0.28$, $\nu_\parallel = 1.74$ and $\nu_\perp=1.10$}
    \label{fig:grid-1-grid-search-beta-beta-prime}
\end{figure}

\begin{figure}[htbp!]
    \centering
    \includegraphics[width=\linewidth]{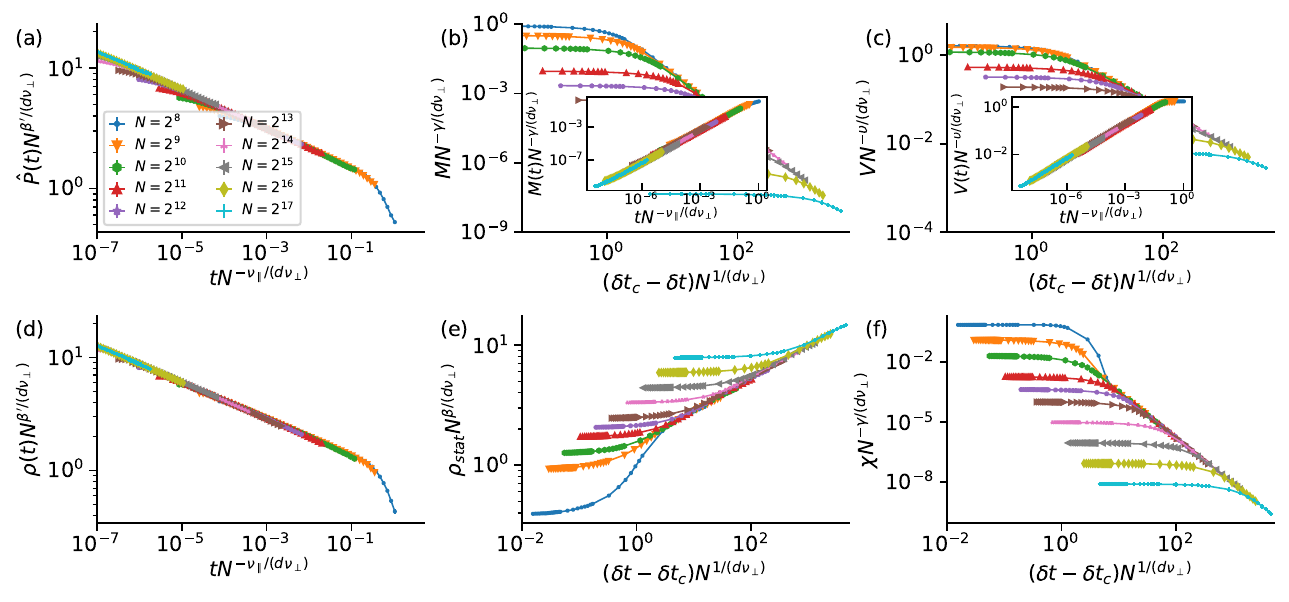}
    \caption{1-dimensional square grid lattice with periodic boundary condition network (i.e.~a ring of nodes) with Poisson link activation with mean inter-event time of 1.}
    \label{fig:grid-1d-poisson-scaling}
\end{figure}

\begin{table}[htbp!]
\caption{\label{tab:experiment-parameters-grid-2} Experimental setup for 2-dimensional square grid.}
\begin{ruledtabular}
\begin{tabular}{rrr}
\multicolumn{1}{c}{\textrm{Size $N$}} &
\multicolumn{1}{c}{\textrm{Time window $T$}} &
\multicolumn{1}{c}{\textrm{Realisations}}\\
\colrule
256 & 4096 & 4096 \\
529 & 4096 & 4096 \\
1024 & 4096 & 4096 \\
2025 & 2048 & 4096 \\
4096 & 2048 & 1024 \\
8281 & 2048 & 512 \\
16384 & 1024 & 512 \\
32761 & 512 & 512 \\
65536 & 256 & 256 \\
131044 & 128 & 256 \\
\end{tabular}
\end{ruledtabular}
\end{table}

\begin{figure}[htbp!]
    \centering
    \includegraphics[width=0.65\linewidth]{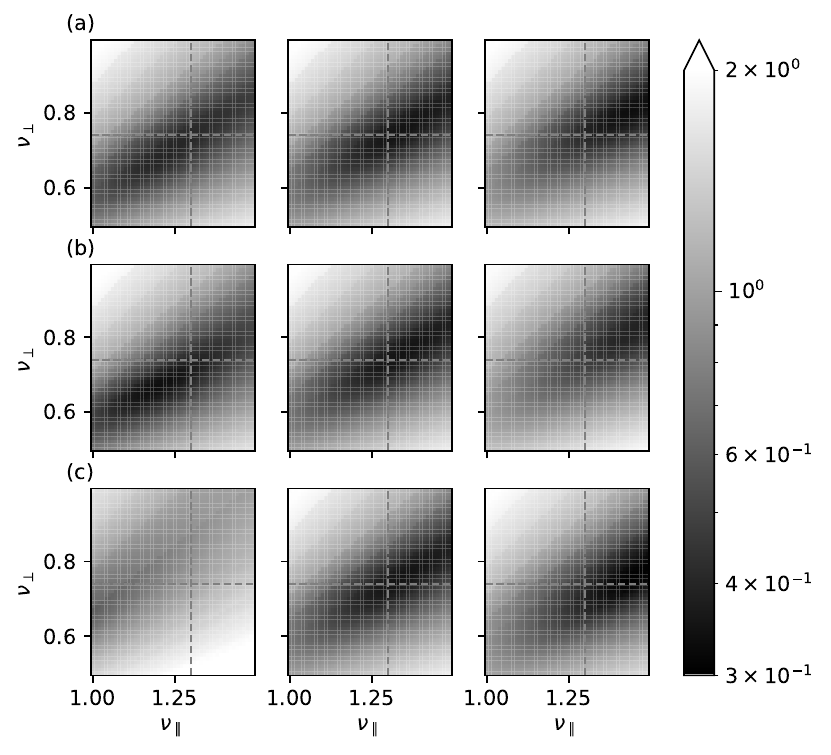}
    \caption{Total error of collapse of $M(t)$, $V(t)$, $\hat{P}(t)$ and $\rho(t)$ universal scaling functions for 2D square grid lattice networks $\langle k \rangle = 8$ and Poisson process activation $\lambda = 1$ with $\nu_\perp$ and $\nu_\parallel$ values assuming (a) $\beta \in \{0.47, 0.58, 0.69\}$, (b) $\beta' \in \{0.47, 0.58, 0.69\}$ and (c) $\delta t_c \in \{0.2835,0.28428,0.2845\}$ which shows a minimum around $\beta = \beta' = 0.58$, $\nu_\parallel = 1.30$ and $\nu_\perp=0.74$}
    \label{fig:grid-2-grid-search-nu-para-nu-perp}
\end{figure}

\begin{figure}[htbp!]
    \centering
    \includegraphics[width=0.65\linewidth]{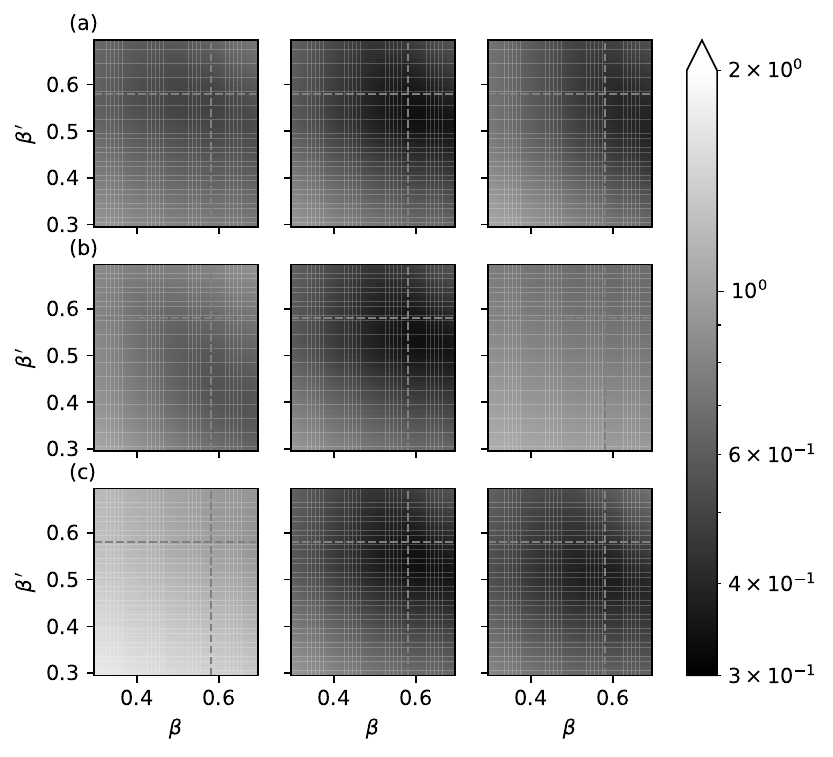}
    \caption{Total error of collapse of $M(t)$, $V(t)$, $\hat{P}(t)$ and $\rho(t)$ universal scaling functions for 2D square grid lattice networks $\langle k \rangle = 8$ and Poisson process activation $\lambda = 1$ with $\beta$ and $\beta'$ values assuming (a) $\nu_\perp \in \{0.63, 0.74, 0.85\}$, (b) $\nu_\parallel \in \{1.19, 1.30, 1.41\}$ and (c) $\delta t_c \in \{0.2835,0.28428,0.2845\}$ which shows a minimum around $\beta = \beta' = 0.58$, $\nu_\parallel = 1.30$ and $\nu_\perp=0.74$}
    \label{fig:grid-2-grid-search-beta-beta-prime}
\end{figure}

\begin{figure}[htbp!]
    \centering
    \includegraphics[width=\linewidth]{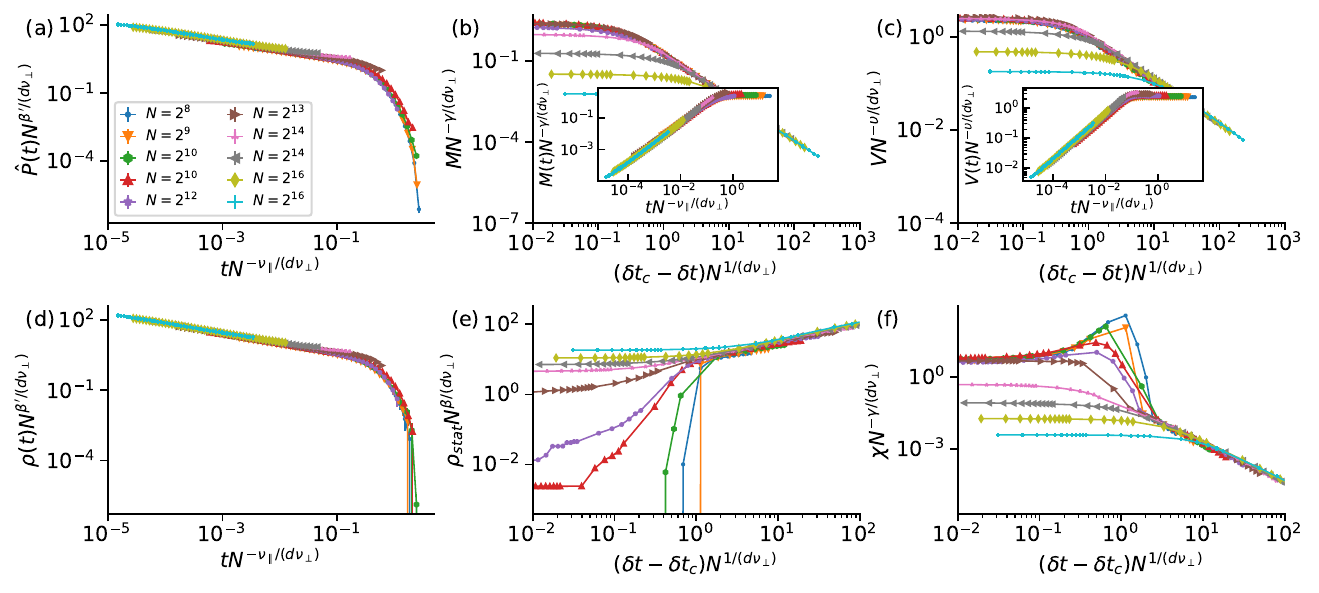}
    \caption{2-dimensional square grid lattice with periodic boundary condition network with Poisson link activation with mean inter-event time of 1.}
    \label{fig:grid-2d-poisson-scaling}
\end{figure}

\begin{table}[htbp!]
\caption{\label{tab:experiment-parameters-grid-3} Experimental setup for 3-dimensional nearest-neighbour square grid.}
\begin{ruledtabular}
\begin{tabular}{rrr}
\multicolumn{1}{c}{\textrm{Size $N$}} &
\multicolumn{1}{c}{\textrm{Time window $T$}} &
\multicolumn{1}{c}{\textrm{Realisations}}\\
\colrule
216 & 2048 & 4096 \\
512 & 2048 & 4096 \\
1000 & 2048 & 4096 \\
2197 & 1024 & 4096 \\
4096 & 1024 & 1024 \\
8000 & 1024 & 512 \\
15625 & 512 & 512 \\
32768 & 256 & 512 \\
64000 & 128 & 256 \\
132651 & 64 & 256 \\
\end{tabular}
\end{ruledtabular}
\end{table}

\begin{figure}[htbp!]
    \centering
    \includegraphics[width=0.65\linewidth]{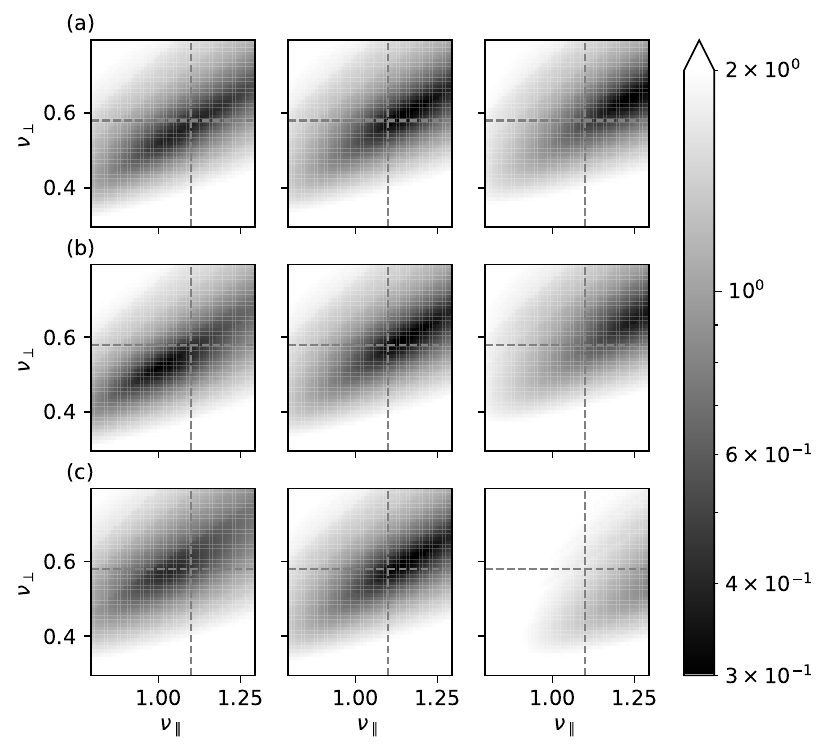}
    \caption{Total error of collapse of $M(t)$, $V(t)$, $\hat{P}(t)$ and $\rho(t)$ universal scaling functions for 3D cubic grid lattice networks $\langle k \rangle = 8$ and Poisson process activation $\lambda = 1$ with $\nu_\perp$ and $\nu_\parallel$ values assuming (a) $\beta \in \{0.66, 0.82, 0.98\}$, (b) $\beta' \in \{0.66, 0.82, 0.98\}$ and (c) $\delta t_c \in \{0.1535,0.15375,0.1544\}$ which shows a minimum around $\beta = \beta' = 0.58$, $\nu_\parallel = 1.30$ and $\nu_\perp=0.74$}
    \label{fig:grid-3-grid-search-nu-para-nu-perp}
\end{figure}

\begin{figure}[htbp!]
    \centering
    \includegraphics[width=0.65\linewidth]{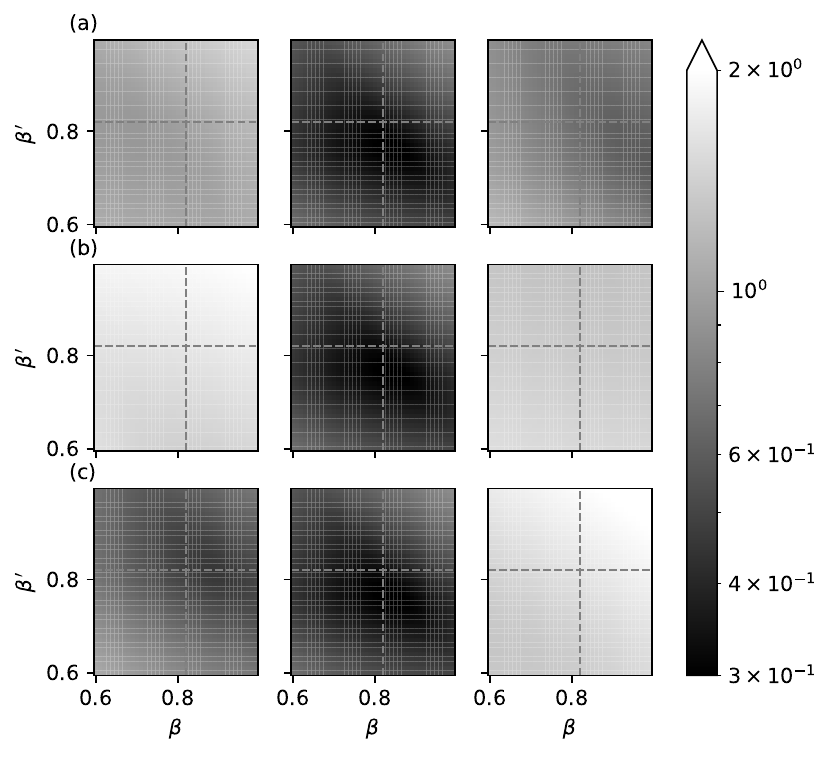}
    \caption{Total error of collapse of $M(t)$, $V(t)$, $\hat{P}(t)$ and $\rho(t)$ universal scaling functions for 3D cubic grid lattice networks $\langle k \rangle = 8$ and Poisson process activation $\lambda = 1$ with $\beta$ and $\beta'$ values assuming (a) $\nu_\perp \in \{0.42, 0.58, 0.74\}$, (b) $\nu_\parallel \in \{0.94, 1.10, 1.26\}$ and (c) $\delta t_c \in \{0.1535,0.15375,0.1544\}$ which shows a minimum around $\beta = \beta' = 0.58$, $\nu_\parallel = 1.30$ and $\nu_\perp=0.74$}
    \label{fig:grid-3-grid-search-beta-beta-prime}
\end{figure}

\begin{figure}[htbp!]
    \centering
    \includegraphics[width=\linewidth]{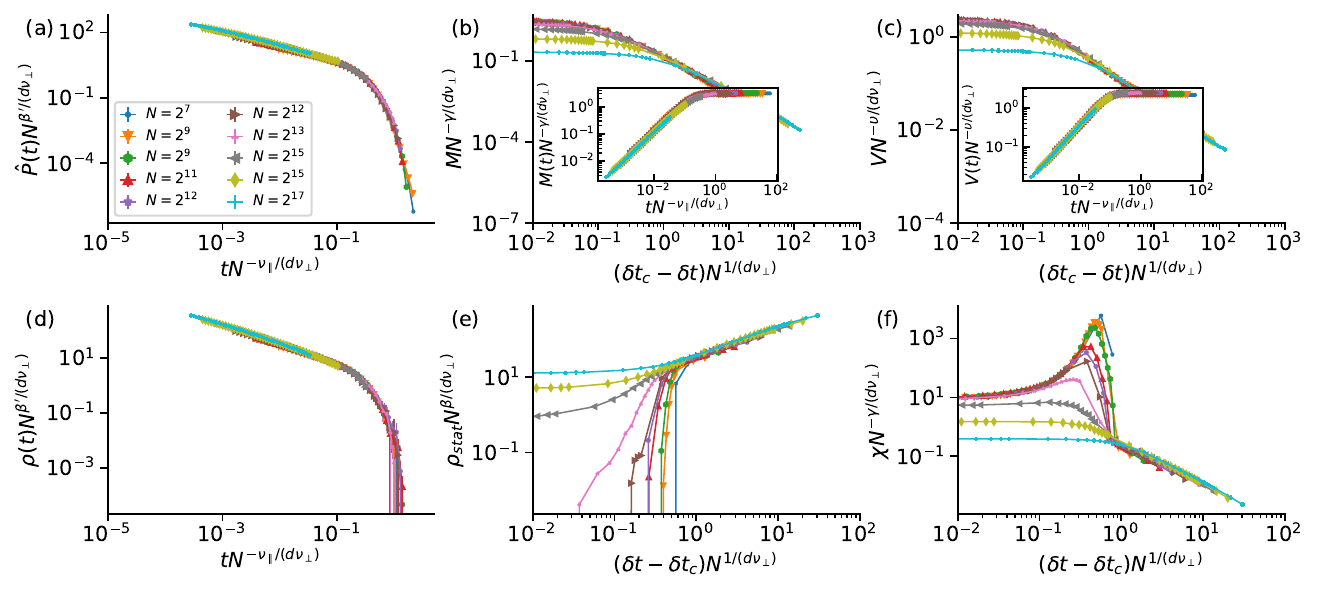}
    \caption{3-dimensional square grid lattice with periodic boundary condition network with Poisson link activation with mean inter-event time of 1.}
    \label{fig:grid-3d-poisson-scaling}
\end{figure}

\begin{table}[htbp!]
\caption{\label{tab:experiment-parameters-grid-4} Experimental setup for 4-dimensional nearest-neighbour square grid.}
\begin{ruledtabular}
\begin{tabular}{rrr}
\multicolumn{1}{c}{\textrm{Size $N$}} &
\multicolumn{1}{c}{\textrm{Time window $T$}} &
\multicolumn{1}{c}{\textrm{Realisations}}\\
\colrule
256 & 2048 & 4096 \\
625 & 2048 & 4096 \\
1296 & 2048 & 4096 \\
2401 & 1024 & 4096 \\
4096 & 1024 & 1024 \\
6561 & 1024 & 512 \\
14641 & 512 & 512 \\
28561 & 256 & 512 \\
65536 & 128 & 256 \\
130321 & 64 & 256 \\
\end{tabular}
\end{ruledtabular}
\end{table}

\begin{figure}[htbp!]
    \centering
    \includegraphics[width=0.65\linewidth]{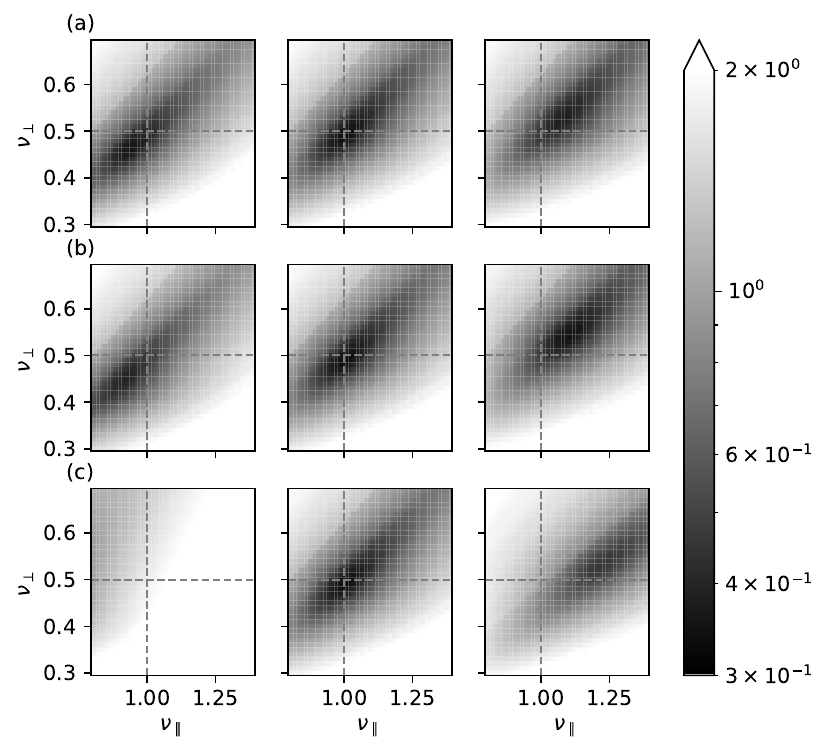}
    \caption{Total error of collapse of $M(t)$, $V(t)$, $\hat{P}(t)$ and $\rho(t)$ universal scaling functions for 4D cubic grid lattice networks $\langle k \rangle = 8$ and Poisson process activation $\lambda = 1$ with $\nu_\perp$ and $\nu_\parallel$ values assuming (a) $\beta \in \{0.84, 1.00, 1.16\}$, (b) $\beta' \in \{0.84, 1.00, 1.16\}$ and (c) $\delta t_c \in \{0.103,0.1045,0.105\}$ which shows a minimum around $\beta = \beta' = \nu_\parallel = 1.00$ and $\nu_\perp=0.50$}
    \label{fig:grid-4-grid-search-nu-para-nu-perp}
\end{figure}

\begin{figure}[htbp!]
    \centering
    \includegraphics[width=0.65\linewidth]{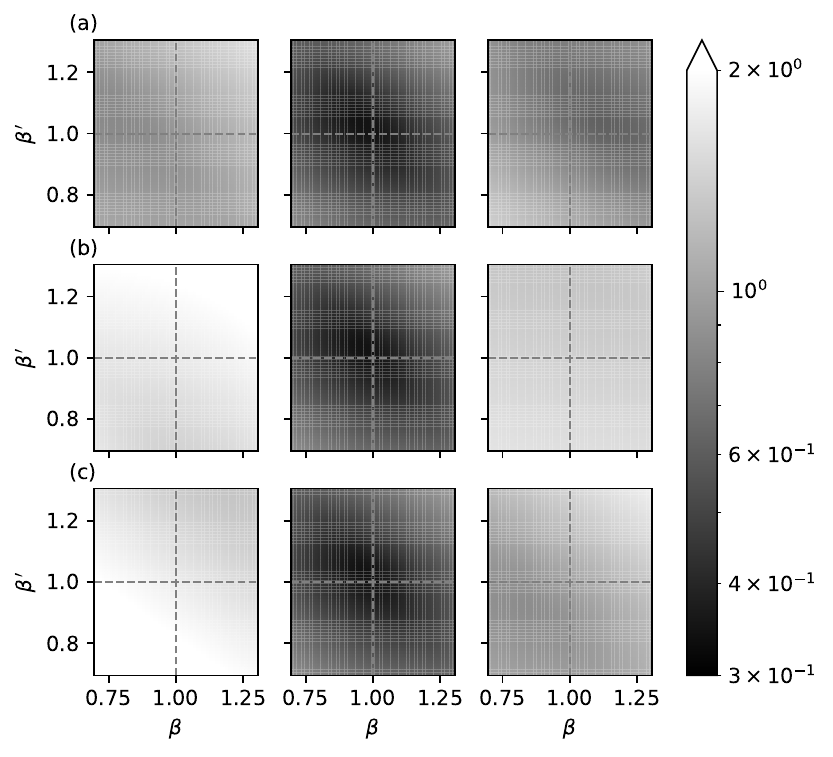}
    \caption{Total error of collapse of $M(t)$, $V(t)$, $\hat{P}(t)$ and $\rho(t)$ universal scaling functions for 4D cubic grid lattice networks $\langle k \rangle = 8$ and Poisson process activation $\lambda = 1$ with $\beta$ and $\beta'$ values assuming (a) $\nu_\perp \in \{0.84, 1.00, 1.16\}$, (b) $\nu_\parallel \in \{0.84, 1.00, 1.16\}$ and (c) $\delta t_c \in \{0.1535,0.15375,0.1544\}$ which shows a minimum around $\beta = \beta' = \nu_\parallel = 1.00$ and $\nu_\perp=0.50$}
    \label{fig:grid-4-grid-search-beta-beta-prime}
\end{figure}

\begin{figure}[htbp!]
    \centering
    \includegraphics[width=\linewidth]{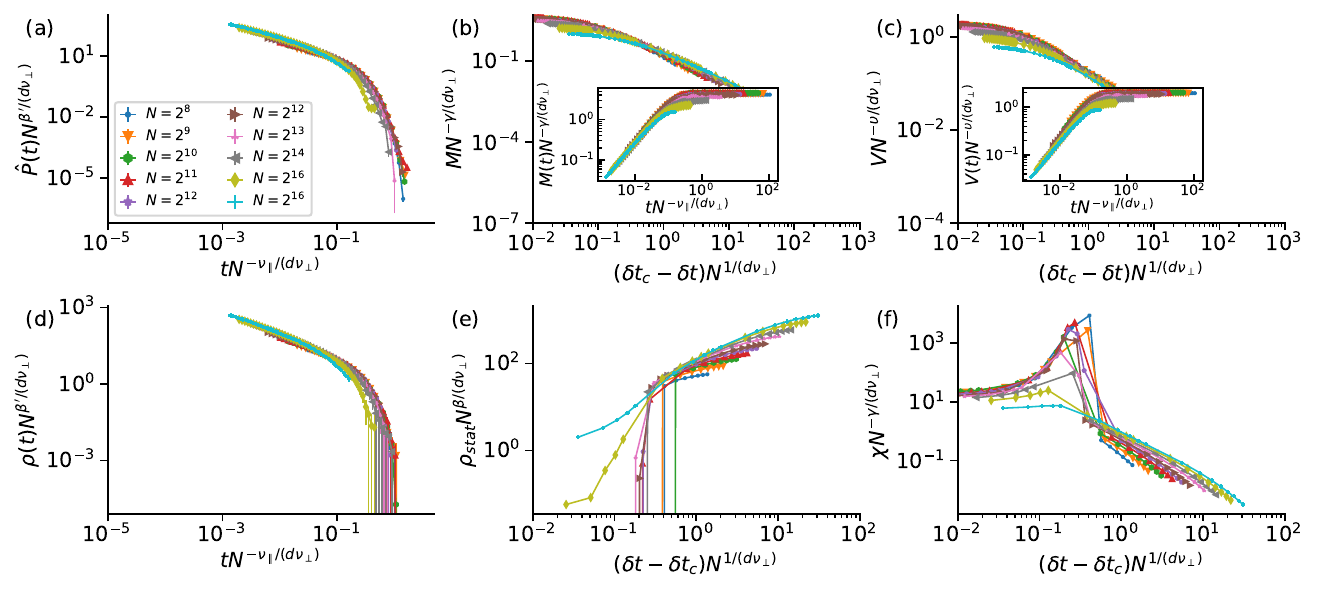}
    \caption{4-dimensional square grid lattice with periodic boundary condition network with Poisson link activation with mean inter-event time of 1.}
    \label{fig:grid-4d-poisson-scaling}
\end{figure}

\bibliography{citations}